# Magnon confinement and trapping at the nanoscale

J. Chen,[a*] H. Yu,[a,b*] R. Gallardo,[c] P. Landeros[c] and G. Gubbiotti[d*]

[a] *International Quantum Academy, Shenzhen, 518048, China*
[b] *Fert Beijing Institute, MIIT Key Laboratory of Spintronics, School of Integrated Circuit Science and Engineering, Beihang University, Beijing 100191, China*
[c] *Departamento de Física, Universidad Técnica Federico Santa María, Valparaíso 2390123, Chile*
[d] *Cnr-Istituto Officina dei Materiali, Via A. Pascoli, 06123, Perugia, Italy*

Magnon confinement and trapping refer to the localization of magnons—quasiparticles that represent collective spin-wave excitations in magnetic materials—within specific regions or structures. This concept is essential in magnonics, a subfield of spintronics that leverages spin waves for processing and transmitting information. Compared to conventional electronics, magnonics offers lower power consumption and faster operation, making it a promising technology for future devices. Magnons can be confined using both static and dynamic methods, often relying on potential wells and barriers to restrict their free propagation and trap them in designated locations. In this review, we will explore the main strategies for magnon confinement and trapping, including: magnetic field inhomogeneities, spin textures (i.e. domain walls, vortices, skyrmions) nanostructured materials (i.e. nanowires, disks, and magnonic crystals), topological states, chiral magnons and flat band formation, induced by dipole-dipole interactions and Dzyaloshinskii-Moriya interaction. Microwave cavities and resonant magnetic fields, as well as spin-torque effects and Bose-Einstein condensation contribute to magnon localization. Furthermore, spin-wave edge and cavity modes have been observed in two-dimensional magnetic materials and twisted moiré superlattices at a specific twist angle. Magnon trapping has broad applications in computing and data processing, particularly in the development of magnonic crystals, waveguides, and memory elements. Additionally, magnon systems are being explored for quantum computing, where confinement can enhance the coupling between magnons and other quasiparticles in hybrid quantum systems. Precision control of magnons could lead to next-generation spintronic devices, offering improved efficiency and scalability.

*Author to whom correspondence should be addressed:
chenjilei@iqasz.cn
haiming.yu@buaa.edu.cn
gubbiotti@iom.cnr.it

**Contents**

# 1. Introduction

Magnon confinement and trapping refer to the process of confining magnons—quasiparticles associated with the collective spin-wave (SW) excitations in a magnetic material—within a specific region. This concept is particularly important for the emerging field of magnonics or magnon spintronics. Magnonics involves the manipulation and utilization of SWs to process and transmit information, offering a promising alternative to conventional electronic approaches due to its potential for low power consumption and fast operation.

Although **magnon confinement** and **magnon trapping** are closely related, they describe fundamentally different physical situations. While both involve spatial localization of magnons, they differ fundamentally in terms of wave propagation and dynamics.

*Magnon confinement* refers to the spatial restriction of SWs to a narrow region or channel, typically on the micrometer or nanometer scale. Despite being confined laterally, SWs are still free to propagate along the channel. In other words, they are guided as light in an optical fiber. This confinement is typically engineered through lithography patterning, magnetic anisotropy modulation, or geometrical constraints and is often used for designing magnonic circuitry where information is carried by SWs moving along such defined and controlled paths.

In contrast, **magnon trapping** annotates the localization of magnons in a closed or cavity-like region from which they cannot escape. This condition arises due to potential wells created by magnetic inhomogeneities, shape anisotropies, or external fields. Trapped modes exhibit discrete eigenfrequencies and form standing wave patterns that do not propagate outside the region. Such trapping often results in the accumulation of a large number of magnons in a small volume, which can be exploited for coherent magnon states or Bose–Einstein condensation, if the magnons can occupy the same quantum state.

Magnon confinement has been explored from multiple perspectives in numerous review articles and edited volumes, collectively defining the state of the art in nanoscale magnon control and manipulation. To help readers follow the evolution of magnon confinement and trapping effects from both experimental and theoretical perspectives, as well as the emergence and establishment of the field, we provide a comprehensive list of key publications from the past two decades that focus on this aspect of magnetization dynamics.

Seminal contributions to this field are the comprehensive reviews about laterally-induced magnon confinement effects observed by Brillouin light scattering (BLS) [1,2, 3] These reviews cover various confinement scenarios, including SW quantization due to lateral boundaries and the influence of static and dynamic coupling in planar magnonic crystals.

In 2012, foundational background on confined SW modes was provided by S. Demokritov's edited books, particularly *Spin Wave Confinement*, [4] *Spin Wave Confinement, Propagating Waves* [5] and *Magnonics: From Fundamentals to Applications* (2012, co-edited with Andrei Slavin). [6] These volumes systematically treat the physics of localized SW eigenmodes in confined geometries such as stripes, dots, and multilayers, bridging theoretical models with experimental observations. They remain widely cited reference works for understanding how confinement alters dispersion relations, mode quantization, and nonlinear dynamics, and they laid the groundwork for much of the subsequent development reviewed in later articles.

The review articles titled *Magnonics* [7] *and The Building Blocks of Magnonics* [8] offer a thorough overview of the magnonics research field. These reviews describe engineering magnetic structures, such as waveguides, antidot lattices, patterned nanostructures, or regions with locally modified magnetic potentials, magnons can be spatially restricted to well-defined areas. Furthermore, the article *Review and Prospects of Magnonic Crystals and Devices with Reprogrammable Band Structure* provides a comprehensive overview of magnonic crystals—artificially structured magnetic materials designed to control the propagation of SWs through spatial modulation of magnetic properties. [9] The review also emphasizes the reconfigurability of magnonic band structures, enabling dynamic control over magnon confinement. This tunable confinement is crucial for the development of functional devices such as waveguides, filters, and logic gates, where controlled localization of SWs enhances energy efficiency and miniaturization.

The 2021 *Magnonics Roadmap* by Barman *et al.* [10] and the 2024 *Magnonics Roadmap* by Flebus *et al.* [11] emphasized the SW geometrical confinement and trapping in patterned structures and periodic arrays. More recent overviews, including *Recent advances in magnonics*, [12] have updated this picture with a focus on experimental techniques for probing confined modes and on-chip architectures for integrating them into functional devices.The review article *Review on Magnonics with Engineered Spin Textures* provides comprehensive examinations of how nonuniform magnetic structures—such as domain walls, skyrmions, and vortices—can be utilized to control and manipulate magnons for information processing applications. [13] A central theme of the review is the confinement and trapping of magnons within these engineered spin textures, which offers a versatile approach to guiding and controlling SW propagation at the nanoscale.

A third body of reviews, grouped under the labels of cavity magnonics and spin cavitronics, redefines confinement in terms of hybridization with resonant electromagnetic fields. The comprehensive review *Cavity magnonics*, [14] and the associated *Roadmap for spin cavitronics* [15] analyze how placing magnetic elements inside microwave or optical resonators imposes spectral selectivity and enables strong or even ultrastrong coupling between magnons and photons. In this framework, magnon confinement is as much about mode hybridization and coherence as about spatial localization. The focus here is more on how resonator geometries and quality factors determine the selectivity, coherence, and potential for quantum applications.

In addition, complementary reviews on hybrid and quantum magnonics place materials and coherence properties at the center of the discussion. Articles such as *Hybrid magnonics: physics, circuits, and applications* [16] and *A review of common materials for hybrid quantum magnonics* [17] compare the suitability of materials ranging from YIG to metallic ferromagnets and antiferromagnets for achieving long-lived confined states, particularly when magnons are embedded in hybrid circuits involving superconducting qubits, optical cavities, or phonons. The review *Quantum magnonics: When magnon spintronics meets quantum information science* [18] further extends the scope by treating confined modes as genuine quantum excitations, discussing the generation of non-classical magnon states, squeezing, and entanglement, all of which require high-fidelity confinement and readout.

Taken together, these reviews and books converge on a common picture: confinement in magnonics can be realized through geometry, magnetic textures, or cavity coupling, with each route offering distinct advantages and challenges. Geometrical approaches are technologically mature and straightforward to integrate; texture-based strategies promise reconfigurability and adaptive functionality; cavity-based methods deliver coherence and quantum-level control. Materials and fabrication constraints cut across all of these approaches, setting the ultimate limits on linewidth, coherence, and scalability. What emerges is not a single dominant pathway to confinement but a complementary set of strategies, each suited to different application spaces, from classical signal processing to quantum information technologies.

The existing review articles and books on magnonics have primarily focused on broad aspects such as SW dynamics, magnon-based logic devices, and fundamental materials properties. While these works provide valuable overviews of the field, they often treat magnon confinement and trapping without systematically addressing the mechanisms, design strategies, and technological implications of spatially controlling magnons at the nanoscale. Moreover, novel emerging approaches, including topological magnons confined at edges, magnonic circuits based on confined magnons in domain walls, and nanomagnonic cavities, are notably absent from these reviews. Therefore, a dedicated review article focused specifically on magnon confinement and trapping would fill a critical gap by integrating scattered knowledge into a unified framework, highlighting design strategies, and guiding the development of practical magnonic technologies—an area inadequately covered in previous reviews and textbooks. This review consists of seven Sections, including the introduction, that presents and illustrate the main strategies/mechanisms to confine and trap magnons, as summarized in Fig. 1.1.

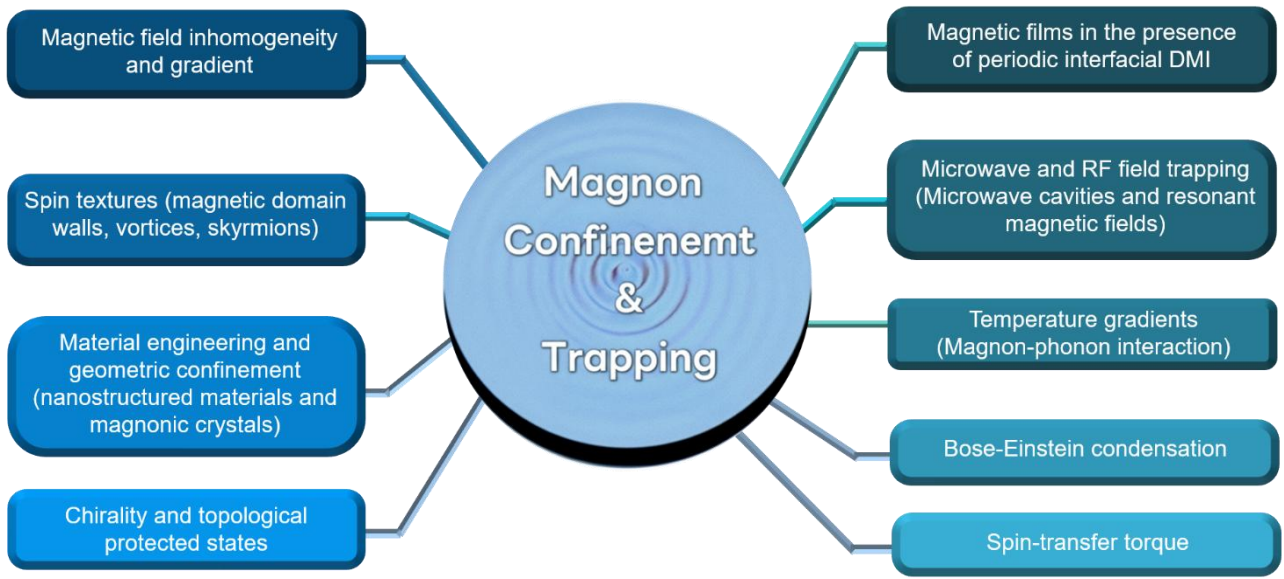

**Fig. 1.1** Illustration of the main strategies/mechanisms useful to confine and trap magnons.

*Section 2* provides a brief introduction to the SW properties of magnetic thin films and discusses the effects of geometrical confinement and trapping arising from physical boundaries, lateral dimensions, potential wells, spin textures, and artificially introduced periodicities. In confined magnetic nanostructures—such as nanowires, nanodisks, and nanorings—magnons can be trapped due to geometric boundaries that reflect SWs and restrict their free propagation, leading to localized modes. Edge magnons are also observed in regions where the internal magnetic field is nonuniform, particularly near element boundaries. The finite width of these nanostructures enforces a quantization of magnon modes, with the lowest-energy states typically localized at the edges. The degree of magnon localization depends on the shape, size, and material properties of the nanostructure. This section also reviews magnetic spin textures, including vortices and skyrmions with nonuniform magnetization, where magnons can be confined. Both vortices and skyrmions host internal modes that localize magnons in specific regions on the nanometer scale. Owing to their high stability and robustness against external perturbations, skyrmions are particularly promising for next-generation information technologies.

*Section 3* reviews various nanomaterial designs that enable the manipulation of SW dispersion to achieve nonreciprocal responses, unidirectional propagation, and the formation of magnonic flat bands. Particular focus is placed on magnetic field gradients, magnonic crystals, moiré superlattices, artificial superlattices, and periodic modulation of the interfacial Dzyaloshinskii–Moriya interaction (DMI). Band flatness implies low group velocity and enhanced magnon density of states, which strengthens interaction effects and can be exploited to realize novel magnetic phases or improve SW functionalities in magnonic circuits. At sufficiently high magnon densities, condensation and even amplification may occur, with a large number of magnons occupying the lowest-energy state. Such amplification provides a mechanism to counteract natural damping in magnetic materials, thereby sustaining or enhancing magnon intensity.

In *Section 4*, chirality and topological effects for controlling magnon dynamics in magnetic systems are presented and discussed. Edge magnons, arising from the nontrivial topology of SW bands, can propagate preferentially along the edges of the material. In topologically nontrivial systems, these chiral magnons are protected from backscattering by defects or impurities, making the edge states robust and stable. Chiral magnon propagation is often induced by the DMI, which breaks inversion symmetry and enforces a preferred rotational sense of spin precession, enabling directional SW transport. Additionally, non-Hermitian phenomena—arising from the interplay of gain, loss, and dissipation—can modify magnon lifetimes, generate exceptional points, induce edge magnon localization, and produce unconventional propagation behaviors.

*Section 5* reviews the coupling of magnons with microwave photons, utilizing microwave cavities or resonators engineered with specific resonant frequencies aligned with magnon energy levels, enabling coherent coupling and efficient energy transfer between the two systems. The co-localization of magnons and photons at the nanoscale may enable ultrastrong magnon-photon coupling, a critical factor for advancing future quantum information processing and conversion technologies. Furthermore, resonant magnetic fields provide a method to enhance magnon trapping by stabilizing localized states, and tuning these fields to match the frequency of the localized magnons can significantly enhance coupling strength, resulting in nonlinear magnonic effects. Magnons can also interact with phonons (lattice vibrations), and by creating temperature gradients, their behavior can be controlled or trapped through thermal influences.

*Section 6* delves into macroscopic quantum effects and nonlinear magnonic phenomena. One notable example is the Bose-Einstein condensation (BEC) of a large number of magnons into a low-energy state, utilizing parametric pumping to excite magnons in a YIG thin film, which subsequently thermalize into the low-frequency state. The section also explores the potential for creating magnon condensates within nanoconduits and discusses the achievement of magnon trapping through spin-torque injection from a charge current. Two primary topics are analyzed here: the localization achieved by electric currents in ferromagnetic layers, and the exploration of bullet modes and their relevance for magnon localization in nanoscale systems.

Finally, *Section 7* presents and discusses diverse platforms and key mechanisms for magnon trapping and confinement, emphasizing their potential technological applications. Geometrically confined magnetic nanostructures allow precise control of magnon dynamics, producing localized modes and on-chip magnon cavity resonances. These cavity magnonic systems provide promising platforms for quantum information processing by interfacing magnons with qubits, spins, or photons, enabling hybrid magnonic devices. Nanomagnonic waveguides, including flat-band architectures, support controlled SW propagation and interference, which can be harnessed for logic operations, signal routing, and magnon interferometry. Confinement in flat-band systems enhances magnon-magnon interactions, offering opportunities to study nonlinear SW dynamics and magnon BEC. Integration with spin-torque mechanisms enables active magnon control and amplification, paving the way for spin-wave–based logic circuits and in-memory computing. Together, these advances position magnons as versatile carriers for next-generation information technologies, bridging classical and quantum functionalities.

**2. Magnon confinement in nonuniform magnetic backgrounds**
Magnons hold great potential for developing future low-power devices and circuits. Magnon confinement enables diverse functionalities such as filtering and signal enhancement in magnonic applications. This chapter reviews the principles and mechanisms of magnon confinement within nonuniform magnetic backgrounds. Confinement across various dimensions, nanostructures, and spin textures offers a versatile catalog of methods for precise magnon control, thus important for nanoscale magnonic applications.

*2.1 Spin wave properties in ferromagnetic materials*

A **spin wave** is a collective excitation of the electron spins in a magnetic material. It represents a wave-like disturbance that propagates through the ordered arrangement of magnetic moments (spins) in a material, typically occurring in ferromagnetic or antiferromagnetic systems. Spin waves arise from the exchange interactions between neighboring spins and can be visualized as a coordinated precession of spins around their equilibrium directions. The quantized version of a SW is known as a **magnon**, which is a quasiparticle carrying energy and momentum associated with the traveling spin disturbance.

A **uniform spin wave (Kittel mode)**, is the simplest type of SW existing in a magnetic material. It describes a collective precession of all the spins **in phase** and with **the same amplitude** across the entire sample Fig. 2.1 (a). [19]

This configuration corresponds to an **infinite wavelength** ($\lambda \rightarrow \infty$) and **zero wave number** (k=0) and this mode corresponds in the experiments to the **ferromagnetic resonance (FMR)**, particularly when driven by a uniform radio-frequency (RF) magnetic field.

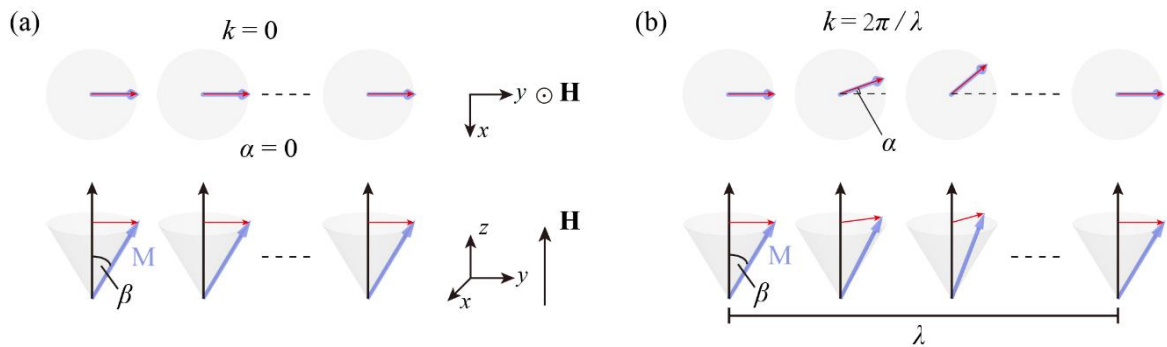

**Fig. 2.1** Schematic representation of the precessional modes of magnetization (**M**) with opening angle b measured from the external field (**H**): (a) Uniform precession with wave number $k = 0$, all magnetic moments are precessing in-phase. (b) SW with wavelength $\lambda$, where neighboring spins are tilted by an angle $\alpha$. [20]

However, when considering real experimental conditions, nonuniform and time-dependent magnetic fields often arise. These may result, for instance, from dipolar interactions at the boundaries of the sample or from inhomogeneous excitation fields. Such perturbations lead to spatially nonuniform magnetization distributions within the sample. The collective excitations that emerge in this context, characterized by a well-defined spatial correlation among spins, are known as SW, where a small, constant tilt $\alpha$ between neighboring spins, is introduced along with a slight deflection $\beta$ of each spin from its equilibrium position. With a large number of spins, this results in quasi-continuous spin states, forming a SW with a finite wavelength $\lambda = 2\pi/k$, as shown in Fig. 1(b).

In SW theory for an infinite (unconfined) ferromagnetic material, the SW vector **k** can assume any value and direction; that is, $k_x$, $k_y$, and $k_z$ are independent continuous variables. However, when the dimensions of the magnetic system are reduced, the SW spectrum is modified due to the boundary conditions at the surfaces. [21]

In a thin-film geometry—where the film thickness $d$ is comparable to the SW wavelength—confinement effects arise due to the finite dimension along the thickness (*z*-axis). In this case, the wavevector component $k_z$ is no longer a continuous independent variable but becomes quantized, taking only discrete values determined by the separation between the film surfaces. This phenomenon is known as SW quantization, and it results in a spectrum comprising an infinite set of dispersion branches, each characterized by a distinct spatial distribution of the dynamic magnetization across the film thickness. [4,5]

To describe dipolar SWs in uniformly magnetized thin ferromagnetic films, a dispersion relation can be derived by solving Maxwell's equations under the magnetostatic approximation. In the thin-film geometry, magnetostatic SWs exhibit anisotropic behavior due to the inherently anisotropic nature of dipolar interactions, which affect the total dipolar energy. For an in-plane magnetized film, two distinct SW modes can be identified, [22] depending on the angle between **k** and the applied field (**H**). In absence of magnetic anisotropy and neglecting the exchange interaction between adjacent magnetic moments, waves propagating along the applied field are called backward-volume magnetostatic waves (BVMSWs), while waves propagating transverse to the applied field are called magnetostatic surface waves (MSSWs, also known as DE waves). [23] The complete calculation of the SW spectrum, which includes both dipolar and exchange interaction, was developed by Kalinikos and Slavin. [24]

*2.2 Geometric confinement and trapping in planar 1D and 2D magnetic elements*

In one (1D) and two-dimensional (2D) magnetic systems, the geometry of magnetic nanostructures shapes the internal magnetic field, leading to separated magnon modes at different positions. The magnonic edge mode, localized at the structure boundaries, is the most common example. This section offers a comprehensive review of magnon confinement in planar magnetic structures, including analytical, numerical, and experimental approaches.

*2.2.1 Magnetic stripes and dots*

In general, the internal magnetic field is given by the sum of the external field (Zeeman term) and both the static and dynamic dipolar fields due to static magnetic charges and the precessing magnetization close to the edge of the elements, respectively. Let us now consider the case of infinite ferromagnetic stripe with thickness $d$ and width $w$ (see Fig. 2.2). As a consequence of the lateral confinement, the internal magnetic field felt by the precessing spin is set by the direction of the externally applied magnetic field and could be uniform in longitudinally magnetized stripes (**H** parallel to the stripe length, $x$-direction) or nonuniform in transversely magnetized stripes (**H** parallel to the stripe width, $y$-direction).

In the former case, the situation is rather simple since, considering magnetized stripes with infinite length, no surface magnetic charges ($\sigma = M.n$) are induced by the uniform magnetization distribution. In the latter case the internal magnetic field landscape is more complicated because, due to the static magnetization perpendicular to the stripe edges, magnetic charges are formed leading to the appearance of the static demagnetizing field (shape anisotropy). To minimize this field, the magnetization vector rotates and becomes parallel to the stripe edges leading to a nonuniform magnetization configuration. At the same time, exchange interaction tends to align the adjacent magnetic moments in parallel. Fig. 2.2 shows the magnetization configuration in transversely magnetized stripes (a) without and (b) with shape anisotropy. Calculations of the ground state were performed in the presence of a transverse magnetic field by using the OOMMF micromagnetic code. [25] The external magnetic field is applied perpendicular to the stripe with the value of $\mu_0 H = 80$ mT. In the former case, magnetic momentsare orthogonal to the stripe lateral edges while in presence of shape anisotropy they rotate and tend to orient parallel to the stripe edge to avoid the formation of magnetic charges and the consequent increase of magnetostatic energy.

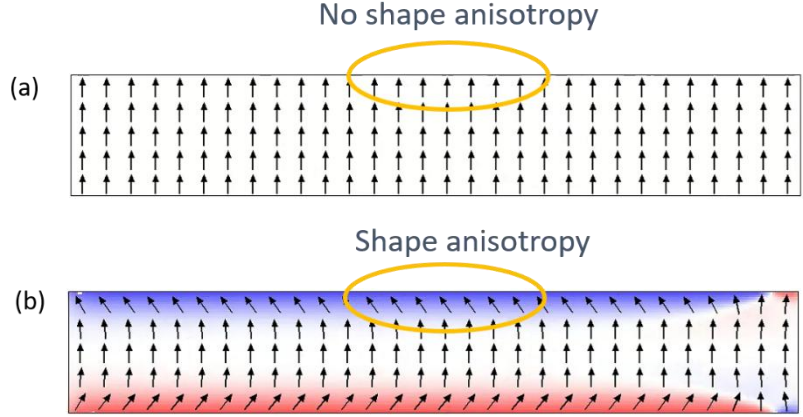

**Fig. 2.2** Magnetization equilibrium configuration calculated by the OOMMF code for a transversely magnetized stripe (a) without and (b) with shape anisotropy. The white backround regions indicates the SZ while the red and blue regions, close to the stripe edges, represent the FFBZ.

This situation generates what is called shape anisotropy, meaning that the magnetic moments confined at the lateral edges do not maintain parallel alignment but deviate with respect to the center of the elements trying to align parallel to the edge of the element.
This deviation from the aligned state produces significant effects on the nature of the spin modes that exist within the system. Considering, for example, a transversely magnetized stripe with width *w*=3 μm and thickness *d*=30 nm in presence of an applied magnetic field along its width (*y*-axis, the axis of difficult magnetization). Two distinct regions can be defined within the stripe. The region near the lateral edges, consisting of spins that are affected by the edge effects, is called the 'field-free boundary zone' (FFBZ) and has a width *s* inversely proportional to the external field. The internal region, composed of spins aligned with the external magnetic field, is instead called the saturated zone (SZ). This description takes into account the dipolar interaction but remains valid even in the presence of exchange interaction (in this case, the transition region becomes narrower because exchange favors the parallel alignment of spins). [26]
The presence of edge modes is justified by the fact that the spins at the edges tend to orient themselves in such a way as to create a dipolar field that opposes the externally applied magnetic field, and their confinement in the FFBZ region can be explained through the creation of a potential well.

$$H_{pot}(y) \approx H_{equ}(y) + M_s \frac{d}{s+d} \sin^2(\phi) \qquad (1)$$

where $s = \frac{dM_s}{\pi H_0}$ is the width of the FFBZ that is inversely proportional to the strength of the applied magnetic field. The first term, $H_{equ}(y)$, is the static dipolar field, which tends to localize the modes near the edges of the element; it is maximum at the center and decreases at the edges of the confined element, indicating the tendency of the system to minimize the overall dipolar energy. The profile of the first field term can be calculated numerically and it is well reproduced by the following analitically expression and depends on the *w/d* ratio:

$$H_{equ}(y) \approx H_0 - 4M_s \left( tan^{-1} \frac{d}{2y+w} - tan^{-1} \frac{d}{2y-1w} \right) \qquad (2)$$

For larger *w/d* ratios the width of the edge regions decreases and $H_{pot}$ becomes more uniform in the center of the stripe.
The second term represents the dipolar dynamic field, which tends to confine the modes in the internal region and which arises due to the precessional motion of the magnetic moments located at the edge of the element (it is maximum at the edges and zero at the center of the element and it represents, in some way, an obstacle to the magnetic moments located at the edges, preventing them from large spin precession amplitude close to the element edges and outside the boundaries of the system).

We use OOMMF micromagnetic simulations with an applied field $\mu_0 H = 80$ mT to calculate the micromagnetic potential (see Fig. 2.3) in transversely magnetized stripes. The values of $H_{equ}$ and the dynamic dipolar field on a short segment perpendicular to the edge were extracted from the relaxed state (see inset). These two components of the micromagnetic potential were then plotted as functions of $y$ (open and solid circles in Fig. 2.3). $\phi$ is the angle between the equilibrium magnetization direction and the field direction ($y$- axis), and it is obtained from micromagnetic simulations.

The competition between these static and dynamic demagnetizing fields leads to the formation of a SW well (SWW) potential in the internal magnetic field, represented by the blue curve. The eigenvalue equation for the in-plane oscillating magnetization $m_\phi(y)$ is given by:

$$\left(\frac{\omega}{\omega_M}\right)\frac{d+w}{s} m_\phi(y) = -\lambda_{ex}^2 \frac{\partial^2}{\partial y^2} m_\phi(y) + \frac{H_{pot}(y)}{M_S} m_\phi(y) \tag{3}$$

where $\lambda_{ex} = \sqrt{2A/(\mu_0 M_s^2)}$ is the exchange length, $\omega = \gamma M_s$ with $\gamma$ as the gyromagnetic ratio, $M_S$ as the saturation magnetization, and $A$ is the exchange constant.

By solving Eq. (3) using the finite difference method, discrete eigenvalues and corresponding eigenfunctions that describe quantized SW modes within the SWW potential are obtained. The power spectrum of Fig. 2.3 (c) exhibits several peaks in the frequency range between 4 and 12 GHz. The associated eigenfunctions, plotted above each line, illustrate the spatial mode profiles of standing SWs localized at the edge. They exhibit an increasing number of nodes as the mode frequency increases (Fig. 2.3 (a)).

These results demonstrate that edge modes are indeed confined within the SWW potential, and exhibit a characteristic quantization: with increasing mode frequency, the number of nodal lines increases, consistent with higher-order quantum states. Because the width of the potential well is relatively small, the exchange interaction dominates over dipolar contributions, particularly for higher-frequency modes with short wavelengths. This leads to a mode character primarily governed by exchange dynamics, rather than long-range dipolar interactions.

Fig. 2.3 (d) displays the corresponding two-dimensional spatial mode profiles associated with each peak. Modes A, B, and C are identified as edge modes, characterized by the localization of the spin precession amplitude near the stripe edges. Here symmetric edge modes with in-phase precession amplitude are excited, only due to the symmetric excitation magnetic field. These modes exhibit an increasing number of nodal lines parallel to the edges with rising frequency, consistent with quantization along the transverse direction. In contrast, mode D represents a quasi-uniform mode, extending throughout the entire width. Its amplitude reaches a maximum at the stripe center, indicating minimal confinement and a relatively uniform excitation across the sample.

We also calculate the evolution of the SW potential in panel (e) of Fig. 2.3 for different values of the applied magnetic field ranging from 50 to 100 mT. Importantly, the depth and width of the SWW potential are sensitive to the strength of the effective field: as the applied field increases, the well becomes narrower and deeper, enhancing the localization of edge modes.

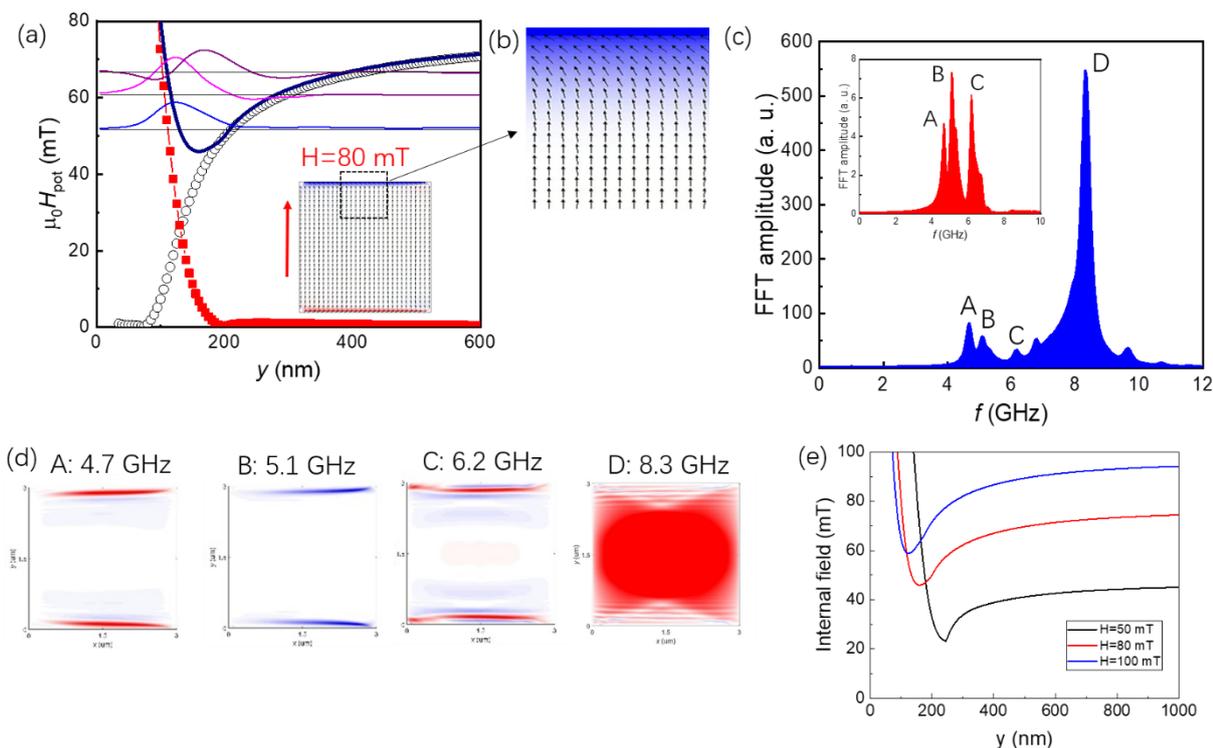

**Fig. 2.3** (a) Micromagnetic simulation of the total internal magnetic field (dark blue curve), sum of the static (open points) and dynamic (full red point) dipolar fields—for a transversely magnetized NiFe stripe with a width of 3 μm. The inset shows the magnetization configuration of the stripe: blue and red regions correspond to the forward-folded Brillouin zone (FFBZ), while the white region denotes the spin-zero (SZ) area. (b) Magnified view of the upper FFBZ region, highlighting its spatial characteristics. (c) Fourier power spectrum of the stripe, with distinct peaks labeled A, B, C, and D. Peaks A–C appear in the low-frequency range (below ~8 GHz), while a dominant peak at 8.3 GHz indicates a higher-frequency excitation. The inset shows the Fourier spectrum performed only in the edge region. (d) Spatial distribution across the stripe width of edge-localized (A, B, and C) and central (D) SW modes. (e) Evolution of the internal magnetic field versus the applied field for values of 50, 80, and 100 mT.

To better understand the role of the inhomogeneous internal magnetic field, we performed additional micromagnetic simulations on stripes with a reduced width of $w = 500$ nm, while keeping the thickness at 5 nm. The results are presented in Fig. 2.4. In this narrower geometry, the internal field profile is significantly altered compared to the previous case of wider stripe. This is due to the dominant influence of the stripe edges, which now play a much more pronounced role in shaping the internal magnetic landscape. The spatial profile of the internal magnetic field as a function of the transverse $y$-coordinate is shown in Fig. 2.4(a).

It is particularly noteworthy that the internal field is now markedly more inhomogeneous. A uniform internal field is only maintained within a narrow central region of approximately 120 nm, located at the center of the stripe. This uniform region corresponds to the white area observed in Fig. 2.4(c), where the equilibrium magnetization remains relatively unaffected by edge effects. Additionally, it can be observed that the FFBZ region is more extended and penetrates deeper into the stripe's interior, reflecting the increased influence of edge-induced inhomogeneities in the narrower geometry. Peaks appearing in the FFT of Fig. 2.4(b) are less separated in frequency and have spatial profile that are mainly concentrated in the central portion of the stripe with significant oscillations. In contrast, no clear localization close to the stripe edges is observed. This is due to the strong inhomogeneity of the internal field. We also present Fig. 2.4(d) the 2D spatial profile of the modes showing the oscillating character parallel to the field direction and extended profile parallel to the stripe edges. Mode A and B are very similar and precess in phase. This is opposite to mode C. Mode D exhibits a larger number of oscillations and nodal lines perpendicular to the field direction.

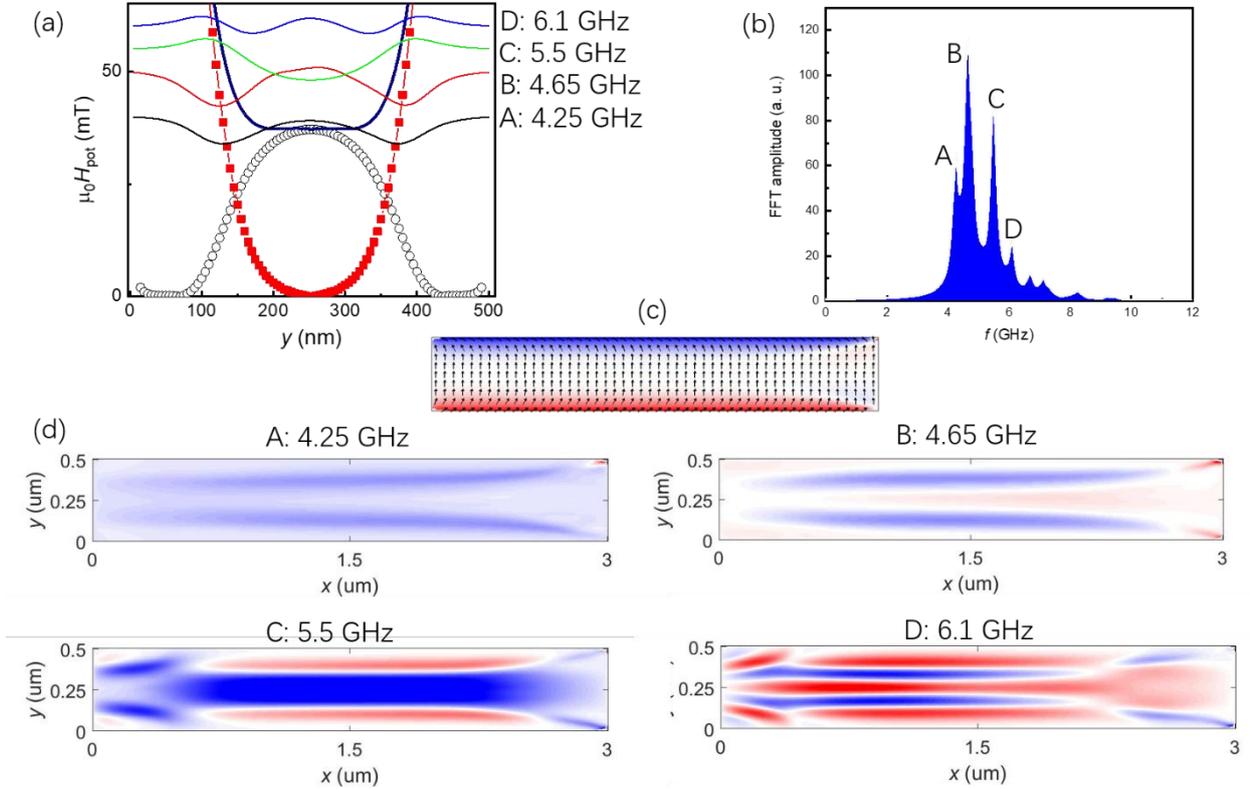

**Fig. 2.4** (a) Same analysis as in Fig. 2.3, but for a stripe width of 500 nm. (b) FFT spectrum showing peaks that are now closer in frequency, without clear separation between modes localized at the edges and those in the central region of the stripe. (c) Magnetization configuration indicating that the FFBZ extends toward the central region, thereby reducing the SZ area. (d) Spatial distribution across the stripe width of all modes identified in panel (b). (e) Intensity map of the SW modes corresponding to the peaks shown in panel (b).

We also consider the effect of lateral confinement of spin waves in longitudinally magnetized stripes. Micromagnetic simulations were performed on stripes with the same geometrical parameters as those considered in Fig. 2.5, in the presence of a magnetic field H= 80 mT applied parallel to the stripe length. The system is excited by an out-of-plane field pulse, and the power spectra and mode profiles are calculated. The magnetization distribution shown in panel (a) is uniform and parallel to the field direction in the central region of the stripe, except in the edge regions where the magnetization begins to deviate and tends to align with the stripe boundaries. This is the consequence of the fact that the simulated stripe does not have an infinite length. The system is then excited by a uniform out-of-plane field pulse. The Fourier power spectrum shown in Fig. 2.5(b) contains a set of well-resolved peaks at frequencies higher than those obtained for the transversely magnetized stripes. The corresponding 1D and 2D spatial profiles are presented in panels (c) and (d) and correspond to stationary modes resonating across the stripe width. In this case, no edge modes are observed at the stripes edge in the direction parallel to that of the applied magnetic field. The resonant modes exhibit symmetric spatial profiles—with an even number of nodal lines—with respect to the center of the stripe (x=1.5 μm). Mode A is the quasi-uniform mode without nodal lines, while modes B, C, and D display two, four, and six nodal lines parallel to the field direction, respectively.

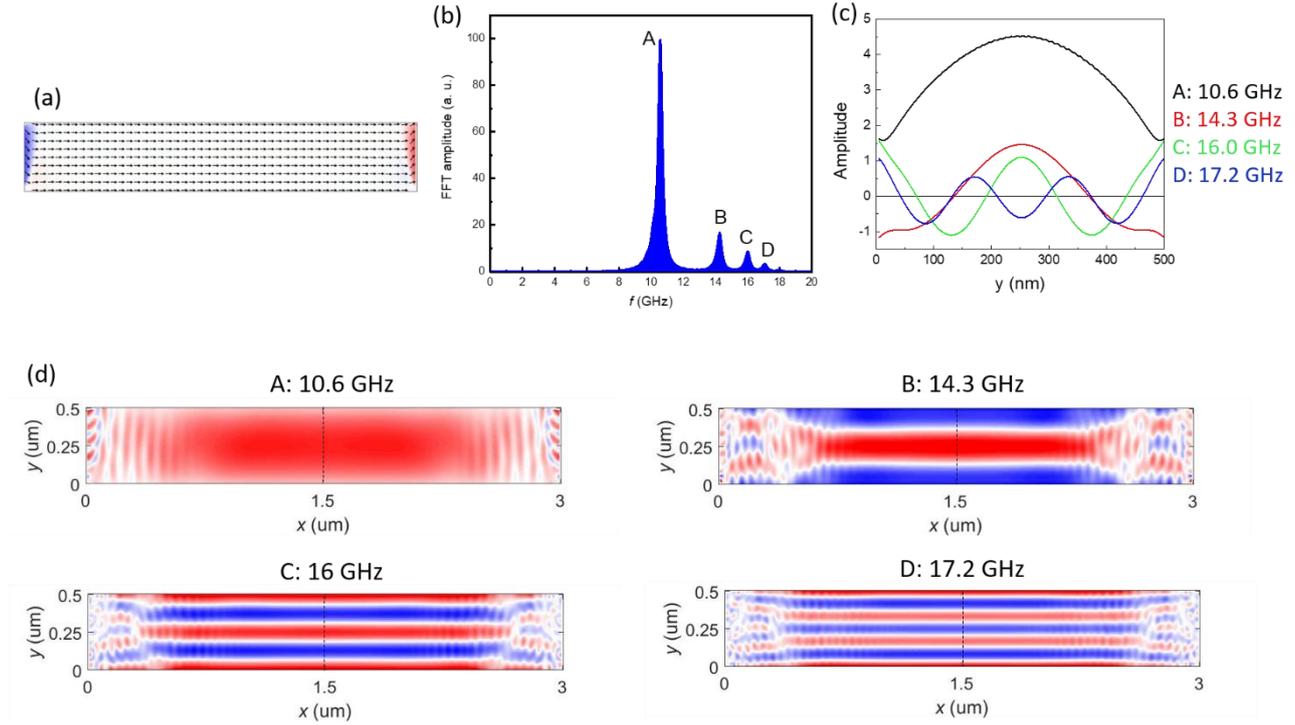

**Fig. 2.5** (a) Magnetization distribution, (b) Fourier powers spectrum and (c) 1D and (d) 2D spatial profiles of the SW modes calculated by micromagnetic simulations for stripes shaving a width of 500 nm, thickness of 30 nm, in presence of a magnetic field of H=80 mT.

To further investigate the role of lateral confinement and internal magnetic field inhomogeneity on SW properties—particularly on the in-plane wavevector components $k_x$ and $k_y$, we consider a d=30 nm thick Permalloy (NiFe) stripes and w=600 nm wide. [27] BLS spectroscopy was employed in four different field and wavevector configurations, either parallel or perpendicular to the stripe's length and width. The results show that due to lateral confinement, SW modes behave as stationary (non-propagating) across the stripe width (y-axis) and as propagating along the stripe length (x-axis). This is especially evident from the 2D mode profiles of narrower 500 nm stripes. These stationary modes are dispersionless, meaning their frequency remains constant regardless of wavevector. They arise from interference between counterpropagating waves reflected at the stripe edges.

In one configuration of Fig. 2.6(a), the external magnetic field is aligned with the stripe length, minimizing demagnetization and maintaining a uniform internal field. Here, the wavevector selected by the BLS scattering geometry ($k_x$) is perpendicular to both the field and stripe length, resulting in discrete, quantized standing-wave modes observed in the BLS spectra. The detected peaks are identified as MSSW stationary modes without any frequency variation with $k_x$, confirming their dispersionless nature. In Fig. 2.6(c), both the magnetic field and wavevector are aligned along the stripe length, leading to a dispersive spectrum with multiple peaks that shift with angle of incidence (i.e., varying $k_y$). The transverse wavevector $k_x$ remains quantized due to confinement, while $k_y$ varies continuously and the mode frequency decreases on increasing $k_y$ as expected for BVMSWs.

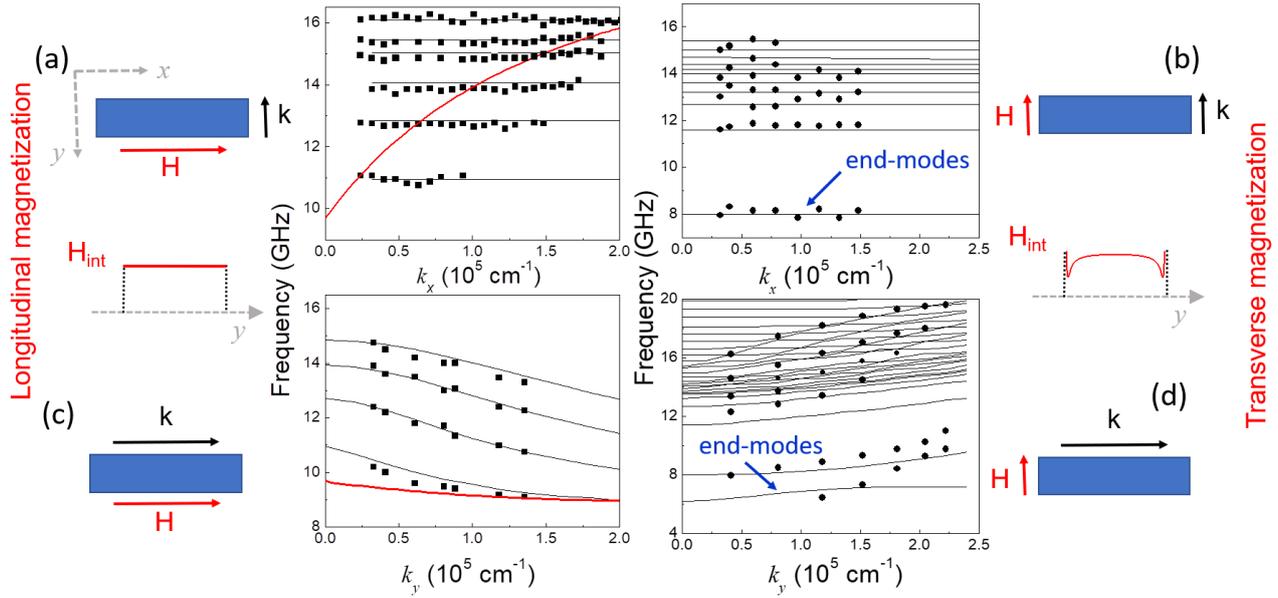

**Fig. 2.6** SW frequencies measured by BLS in the four scattering geometries shown in the insets for an array of NiFe stripes. The direction of the applied magnetic field (**H**) and wavevector (**k**), as well as the profile of the internal magnetic field (homogeneous in panels (a) and (c) and inhomogeneous in (b) and (d) with respect to the stripe's length are shown. Continuous curves are calculated frequencies using the theoretical models described in Ref. 11. Adapted with permission from [27].

Measurements with transverse magnetization in Permalloy stripes are presented in Fig. 2.6(b) and (d), using a magnetic field of 2.9 kOe to ensure sufficient magnetization across the stripe. In this configuration, the internal magnetic field is inhomogeneous due to demagnetizing effects, especially near the stripe edges where the static field is reduced, as shown in the inset.

When the wavevector **k** is perpendicular to the applied magnetic field **H** and parallel to the stripe width, all observed SW modes are non-dispersive. The corresponding BLS spectrum, shown in Fig. 2.6(b), consists of a series of standing-wave resonances across the entire stripe width, along with a limited number of low-frequency modes. These lowest-frequency modes, although dominant in the spectrum, are governed primarily by exchange interactions and are localized near the stripe edges, where the internal magnetic field is strongly inhomogeneous due to demagnetizing effects. For this reason, they are commonly referred to as edge modes or end modes. The experimental mode frequencies are compared with theoretical results (solid lines) in Figs. 2.6(b) (d), obtained using the analytical model described in Ref. [27].

For **k** parallel to the stripe length, the BLS spectra shown in Fig. 2.6(d) reveal two distinct families of peaks. The first set appears above 10 GHz and forms a dense band with partially resolved peaks. The second set is observed at lower frequencies, where at least two well-separated peaks can be clearly distinguished. Notably, the group velocity of these modes is positive—that is, the frequency of each peak increases with increasing $k_y$. The inhomogeneity of the internal magnetic field in this configuration leads to the emergence of additional low-frequency modes localized near the stripe edges. Interestingly, the frequencies of the modes at $k_y=0$ in Fig. 2.6(d) coincide with those measured in Fig. 2.6(c), highlighting the correspondence between the two geometries at normal incidence. With these experiments we showed that, depending on the scattering geometry used in the BLS experiment and the corresponding selected wavevector direction, the resonant or propagating (guided modes) nature of these SW modes can be put in evidence.

In the case of transverse magnetization, Demidov *et al*. demonstrated that SW beams with submicron widths can be generated, guided, and manipulated within laterally confined magnetic waveguides. For frequencies smaller than the critical frequency, which corresponds to the propagation of a wide

SW beam centered with respect to the central axis of the stripe, the SW beam splits into two much narrower beams shifted from the center of the stripe toward its edges. These SW beams have the same frequency and are localized in the edge potential wells, and thus, two narrow waveguiding channels are formed. [28] Furthermore, microscopic waveguides with varying widths can be used to induce transformation of the SW mode profiles, leading to mode conversion and interference effects. [29] These transformations depend on the geometry of the waveguide and the SWs can be easily channelized, splitted, and manipulated.

The influence of adjacent magnetic elements on edge SW in magnetic microstripes has recently been investigated through micromagnetic simulations. Specifically, the study examined the propagation of edge modes along the length of transversely magnetized yttrium iron garnet (YIG) microstripes, both in isolation and in the presence of nearby NiFe microstructures. Under a transverse applied magnetic field, introducing a Py stripe adjacent to one edge of the YIG microstripe led to a splitting of the SW dispersion curve. As a result, the edge modes on the two sides of the YIG stripe exhibited different wavelengths, group velocities, and decay lengths—even at the same frequency. This asymmetry is attributed to the inhomogeneous static dipolar field generated by the nearby Py element, which produces an asymmetric internal magnetic field across the two edges of the YIG stripe. [30]

So far, we have discussed the general properties of SWs in 1D systems consisting of longitudinal and transversely magnetized stripes with lateral confinement only in one direction along the stripe width. In 2D confined magnetic structures having lateral dimensions comparable to the SW wavelength, SW quantization arises directly from the finite lateral dimensions of the structure. The physical boundaries act as reflective surfaces for magnons, giving rise to standing SW modes. These standing waves are supported only when the SW wavelength satisfies the resonance conditions of $w = \frac{n\lambda}{2}$ and $l = \frac{m\lambda}{2}$, where $w$ and $l$ are the width and length of the magnetic element, respectively, and $n$ and $m$ are integers representing the number of half-wavelengths (or nodal lines) along each dimension. These standing wave conditions lead to the quantization of the in-plane wavevector components as $k_x = \frac{n\pi}{w}$ and $k_y = \frac{m\pi}{l}$, as seen in Fig. 2.7. In a square rectangular element with lateral dimensions $w$ and $l$, the two-dimensional profile of each mode can be factorized as $m(x,y) = A\cos\left(\frac{n\pi}{w}x\right)\cos\left(\frac{m\pi}{l}y\right)$.

As previously discussed, the energy of the SW modes depends on the nature of the magnetostatic waves and the number of nodes in the confined direction(s). SW confinement can be understood by considering the dispersion of DE and BV waves in a continuous film with discretized in-plane wavevecor components. This discretization results in a discrete and mode-rich spectrum shaped by geometric confinement and magnetization configuration. For instance, as illustrated in Fig. 2.7, in the case of 1D confinement along the width $w$, the energy of MSSWs increases with the number of nodes $n$, while for BVMSWs, it decreases with increasing node number $m$. This behavior stems from the anisotropic dispersion characteristics of dipole-coupled SWs.

In addition, the SW spectrum becomes even more intricate due to the presence of an inhomogeneous internal magnetic field, particularly along the direction of the applied field. [31] This internal field variation modifies both the dispersion relations and the mode profiles, leading to significant deviations from idealized uniform systems. Notably, such inhomogeneities can give rise to exchange-dominated edge modes, which may hybridize with higher-frequency modes of dipolar origin. This hybridization results in mode mixing and frequency shifts that are strongly dependent on the local field landscape.

Furthermore, depending on the boundary conditions and the magnetization orientation, the resulting SW modes can exhibit complex nodal patterns and a hybrid character, especially in regimes where both exchange and dipolar interactions play significant roles. The interplay between geometry, magnetic configuration, and internal field inhomogeneity thus leads to a rich and highly tunable SW mode structure in confined magnetic systems. Dynamic boundary conditions describe how the dynamic components of magnetization behave at the edges during SW propagation. These conditions are governed by dipolar interactions and can lead to either pinned or partially pinned magnetization at the boundaries. Guslienko and Slavin developed generalized dynamic boundary conditions extending the classical Rado-Weertman model, accounting for the finite-size geometry and edge-induced dipolar pinning in nanostructures. [32] Moreover, Kostylev introduced modified dynamic boundary conditions in systems with interfacial DMI, showing how this interaction induces wavevector-dependent pinning, further modifying mode profiles and frequencies. [33]

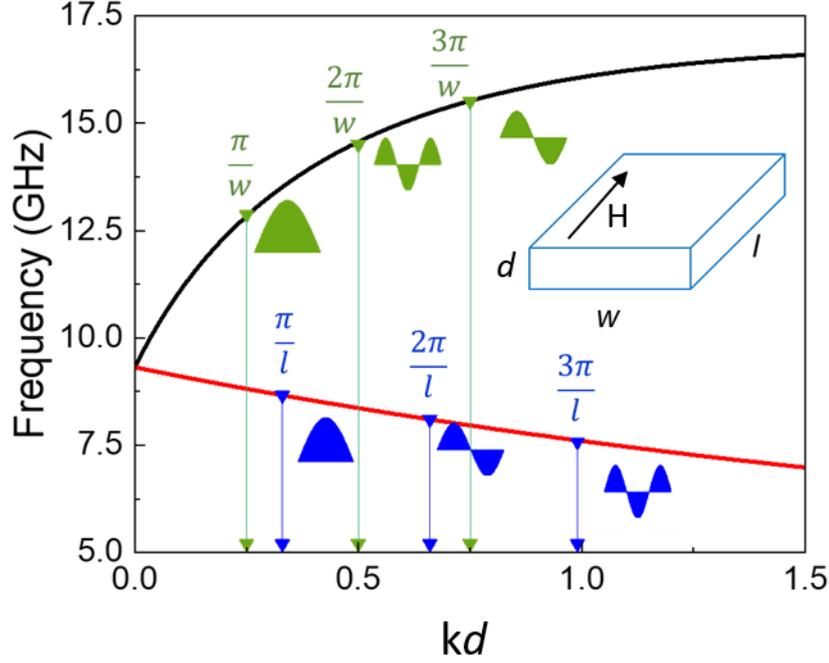

**Fig. 2.7** The dispersion relations of SWs as a function of the in-plane wavevector **k** times for a *d*=30 nm thick NiFe film. The curves correspond to the case of in-plane wavevector *k* perpendicular (MSSW) to and parallel (BVMSW) to the applied field **H**. NiFe magnetic parameters are $\mu_0 M_s = 1$ T; g = 2.11; D = 13 pJ/m. The external magnetic field is $\mu_0$H =100 mT. The curves are degenerate at 9.38 GHz for kt = 0. For small wavevectors the SWs are dominated by the dipolar interaction, while the quadratic contribution from exchange interaction becomes dominant for large **k**.

As an example of SW discretization, we present results obtained by E. R. J. Edwards using scanning Kerr effect microscopy (SKEM) on rectangular NiFe elements. [34] These elements were placed on a gold microstripe transmission line, which served to excite magnetization dynamics. The dynamic magnetic field **h**, generated by the microwave current, was oriented perpendicular to the static magnetic field **H** (110 mT), which was applied along the axis of the microstripe. Fig. 2.8 shows measurements on a square element, with the laser beam focused either at the center (blue) or the edge (red). These measurements probe the out-of-plane component of the dynamic magnetization as a function of microwave excitation frequency. When focused at the center, the spectrum reveals a dominant peak (mode B) associated with the quasi-uniform FMR mode, along with weaker subsidiary modes (C, D, and E) at higher frequencies, corresponding to higher-order SW modes. Mode B can also be considered as the first resonant mode with the element without nodal lines. In contrast, when the laser is focused at the edge, an additional low-frequency mode (~6 GHz, mode A) emerges—absent in center measurements—indicating the presence of an edge-localized mode. To spatially resolve these modes, the microwave frequency was fixed to that of each spectral peak, and the laser beam was scanned across the sample. This enabled the imaging of real-space magnetization eigenmodes. Figure 8 displays the spatial profiles of the dynamic magnetization at the resonance frequencies. Mode A appears as a well-defined edge-localized mode aligned with the direction of the external magnetic field. Mode B represents the quasi-uniform mode, extending across the entire element with maximum amplitude at its center. Higher-frequency modes (C, D, and E) correspond to standing SW modes with nodal plane parallel to the direction of the applied magnetic field (i.e. resonant DE-type SWs), characterized by an increasing number of antinodes.

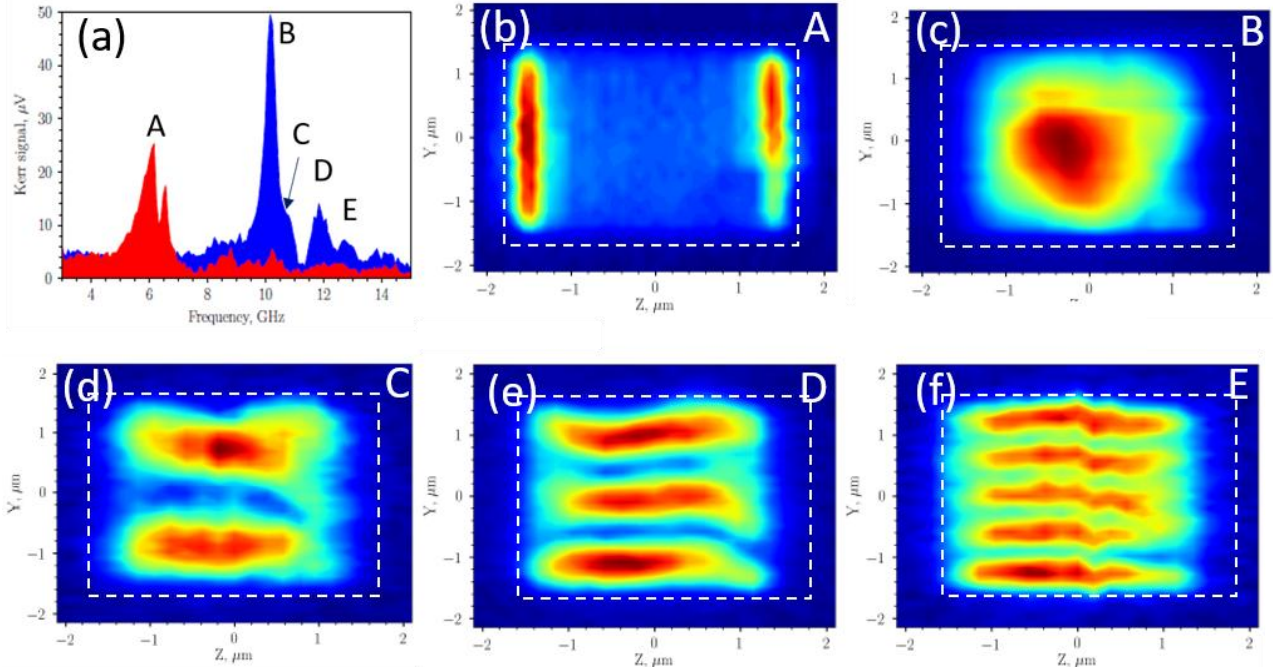

**Fig. 2.8** (a) SKEM spectra measured at the edge (red curve) and center (blue curve) of rectangular dots. (b–f) Intensity maps of the normalized Kerr signal, proportional to the out-of-plane component of the dynamic magnetization in the square element, for different microwave excitation frequencies: (b) $f_{mw}$ = 6.56 GHz (peak A in panel a); (c) $f$ = 10.16 GHz (peak B); (d) $f$ = 11.20 GHz (peak C); (e) $f$ = 11.84 GHz (peak D); (f) $f$ = 12.72 GHz (peak E). Adapted from Ref. [34].

Modifications of standing SW modes in nanostructures—such as stripes and dots of various shapes and sizes—have been observed in the presence of chiral interactions. [35,36] In systems with interfacial DMI (iDMI), interference between counterpropagating SWs no longer produces conventional standing waves, as SWs propagating in opposite directions at the same frequency have different wavelengths. [37] This asymmetry arises because the wavevector magnitudes differ for left- and right-propagating waves at a given energy. As a result, the interference pattern is neither symmetric nor antisymmetric with respect to the central plane of the nanowire, unlike in systems without iDMI where well-defined standing waves can form. Additionally, a strong suppression of the frequency asymmetry between counterpropagating SWs, corresponding to the Stokes and anti-Stokes peaks in BLS experiments, has been observed when the lateral confinement of the magnetic elements is reduced from 400 nm to 100 nm, i.e., below the wavelength of light (i.e., 532nm). [38]

SW edge modes are highly sensitive to both geometric and magnetic parameters. Their frequency, spatial distribution, and field dependence are determined by a combination of shape, size, aspect ratio, edge curvature, and the intrinsic magnetic properties of the material. The aspect ratio influences the demagnetizing field distribution: elongated elements (such as ellipses or rectangles) exhibit a pronounced contrast between edge and center regions, which enhances edge mode localization. In smaller elements, exchange interactions become more dominant, leading to stiffer modes with weaker field dependence.

Edge sharpness or curvature also affects the degree of confinement; sharper edges can lead to stronger field gradients and more pronounced edge mode localization. The intentional introduction of quantifiable asymmetric egglike shape distortion into ideal NiFe elliptical nanomagnets lifts the degeneracy of end modes with concentrated amplitude at the nanomagnet edges. [39] In contrast, modes with concentrated amplitude at the interior are significantly less affected by the distortion. In large arrays of nanomagnets, even slight variations in shape led to significant inhomogeneous broadening of the collective linewidths, especially for edge modes. This broadening arises from the sensitivity of edge modes to geometric imperfections. Magnetic parameters such as the saturation magnetization ($M_s$) and exchange stiffness ($A$) impact the mode profiles and their frequency spacing. Higher saturation magnetization increases the overall internal field, generally pushing the edge mode frequencies upward. Similarly, stronger exchange stiffness tends to delocalize the modes slightly by increasing the characteristic exchange length, thereby influencing the coupling between edge and center modes. [40] Reduced edge magnetization and surface anisotropy decrease the edge saturation field, which is the field needed to align the magnetization at the edge nearly parallel to the applied field. This leadsto a smaller effective demagnetization field along the edges that would cause a significant increase in the edge mode resonance frequency compared to that of an ideal edge, and the shift could be of the order of several GHz. [41, 42]

Furthermore, the spatial extent of both edge and center modes is sensitive to nonlinear effects induced by increasing levels of microwave excitation, as can be seen from inspection of Fig. 2.9(b). In studies of NiFe ellipses using microfocus BLS, it was observed that edge and center modes respond to nonlinearity in qualitatively different ways. [43] As excitation power increases, the center mode maintains an almost constant frequency, whereas the edge mode exhibits a clear positive nonlinear frequency shift. Consequently, the edge mode frequency approaches that of the center mode. However, rather than intersecting, the two modes undergo nonlinear repulsion—a phenomenon in which their frequencies avoid crossing and instead diverge, signaling the onset of nonlinear hybridization. In the strong excitation regime, the spatial profiles of the modes also change markedly: the edge mode begins to acquire features characteristic of the center mode, and vice versa, indicating a mixing of their spatial identities (Fig. 2.9(c)). This mode hybridization is closely correlated with the observed nonlinear frequency behavior, as the reshaping of spatial profiles occurs at the same threshold power level where frequency repulsion becomes evident.

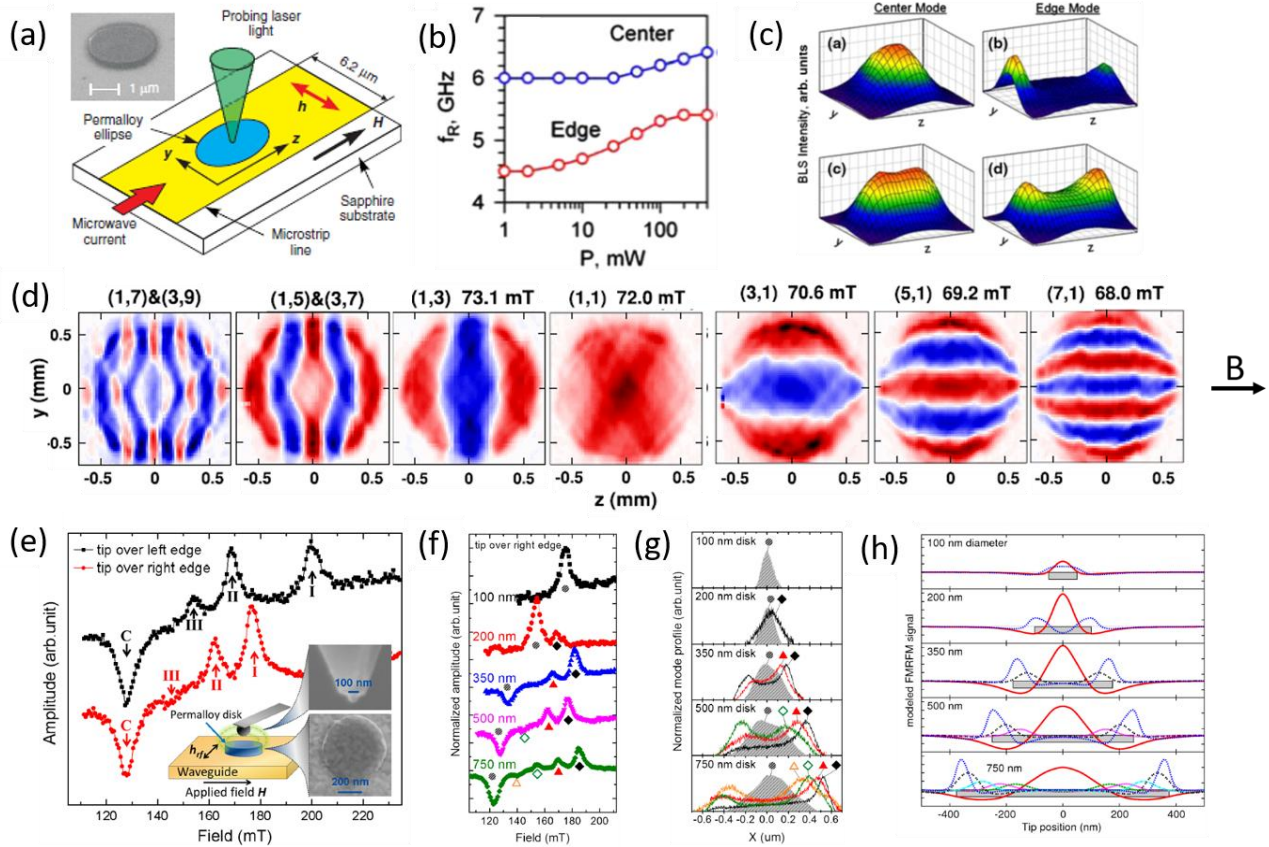

**Fig. 2.9** (a) Schematic of the micro-BLS setup. The inset shows a scanning electron microscope (SEM) image of a NiFe ellipse. (b) Power dependence of the resonant frequencies for the center and edge modes. (c) 2D intensity maps of the center and edge modes measured in both the small-amplitude linear regime (P = 1 mW) and the nonlinear regime (P = 400 mW) at a constant applied magnetic field of H = 350 *Oe*. Adapted with permission from [43]. (d) Experimental spatial profiles of the fundamental (1,1) mode, DE modes (3,1), (5,1), and (7,1), and backward volume modes (3,1), (5,1), and (7,1), measured in a millimeter-sized, in-plane magnetized YIG disk. Adapted with permission from [44]. (e) Measured FMRFM spectra with the tip positioned over the left edge (upper spectrum) and right edge (lower spectrum) of a nominally 500 nm diameter disk, using an excitation frequency of 10 GHz. Insets show a conceptual sketch of the experimental setup and SEM images of the cantilever tip with a cobalt hemisphere and the disk. (f) Measured spectra at the right edges of disks with diameters of 100, 200, 350, 500, and 750 nm, using a 10 GHz excitation frequency. (g) 1D spatial mode profiles measured in the same disks as in (f). (h) Modeled FMRFM tip force for the center (solid curves) and edge (dotted curves) modes in disks of various diameters, each subjected to an applied magnetic field of 0.15 T. Adapted with permission from [45].

The spatial profile and the spectral properties of magnetostatic SW modes in an in-plane magnetized disk-shaped film of YIG were investigated by scanning magneto-optical Faraday microscopy technique. [44] By sweeping the applied magnetic field at a constant excitation frequency, several DE and BV-like modes were detected, as illustrated in Fig. 2.9(d). These are characterized by a series of nodal planes perpendicular and parallel direction with respect to the applied magnetic field, respectively. For backward-like modes the number of antinodes increases with increasing field, as one expects from the negative group velocity of the BV-modes in continuous film. In contrast, for the DE-like mode the number of nodes is increasing with decreasing field. No evidence of localized edge modes has been found. This is ascribed to the large aspect ratio of the disk and to the relatively small YIG saturation magnetization that makes the non-saturated regions in YIG samples were much smaller if compared with the corresponding lateral size. Correspondingly, the spatial localization of the edge mode and the optical signal detected in the optical experiment very small or negligible.

Ferromagnetic resonance force microscopy (FMRFM) was applied to investigate the SW dynamics in individual NiFe nanodisks with diameters ranging from 100 nm to 750 nm. [45] As the microwave frequency was swept and resonance conditions were met, the dynamic magnetization in the disk modulated the force between the sample and the cantilever tip, causing it to oscillate at its mechanical resonance frequency (~8 kHz). These oscillations were then detected using a lock-in amplifier, providing both spectral and spatial information with a lateral resolution of approximately 100 nm. The authors identified multiple edge modes—labeled edge I, II, and III—in disks with diameters of 200 nm or greater, as shown in Fig. 2.9(e). These modes were absent in 100 nm disks, highlighting a strong dependence on element size. The spatial profiles of these modes confirmed their confinement near the edges of the nanodisks, with increasing modal complexity and number of nodal lines as the disk diameter increased. In larger disks (e.g., 750 nm), up to four edge modes were detected, in addition to a central mode corresponding to the quasi-uniform FMR mode. These observations were found to be in good agreement with micromagnetic simulations, validating the experimental interpretations.

Another important finding of the study was the inhomogeneity of the edge modes. The resonance fields of individual edge modes were found to differ between opposing edges of the same disk, and varied with the angle of the applied magnetic field. This behavior suggests that edge imperfections—such as variations in edge roughness or local magnetic anisotropy—can significantly influence the SW dynamics. As such, edge modes serve as sensitive probes of the magnetic and structural properties at the nanoscale, offering a powerful tool for characterizing local inhomogeneities in nanomagnetic devices.

A significant advancement in achieving nanometric spatial resolution in BLS spectroscopy has been demonstrated by the group of Prof. S. O. Demokritov at the University of Münster (Germany). [46] In their pioneering approach, the conventional BLS spectrometer was integrated with an atomic force microscopy apparatus to develop a novel technique referred to as near-field Brillouin light scattering (NF-BLS). To achieve nanometric confinement of the probing light, a nanoscale aperture was fabricated at the apex of an atomic force microscopy tip using focused ion beam milling. This aperture, with a diameter of approximately 50 nm (see inset of Fig. 2.10 (a)), enabled the confinement of the optical field to a volume well below the diffraction limit. This innovative near-field coupling allowed for a lateral spatial resolution on the order of 50–100 nm. The NF-BLS technique was experimentally applied to investigate the magnetic normal of elliptical elements with lateral dimensions of 1.3×2.4 $\mu$m$^2$ fabricated by e-beam lithography and ion etching from a 20 nm thick film of $Ni_{80}Fe_{20}$.

Notably, the width of the edge-localized mode was observed to decrease with increasing magnetic field strength, reaching a minimum spatial extent of approximately 85 nm at higher field values. The width of the localization area of the edge mode has been systematically investigated as a function of the applied magnetic field, showing that it evolves as 1/H, as expected for the size of the FFZB which is inversely proportional to the magnetic field strength. An important practical advantage of this hybrid technique is its ability to simultaneously acquire both magnetic and topographic information. However, the implementation of the NF-BLS technique is not without challenges. A critical requirement is the precise stabilization of both the sample and the atomic force microscopy tip to prevent lateral drift during extended acquisition times—a necessity due to the inherently weak light transmission through the subwavelength aperture. Furthermore, increasing the laser power to compensate for low signal strength is limited by the risk of thermal damage to the tip and sample. As such, long integration times remain necessary to achieve a satisfactory signal-to-noise ratio.

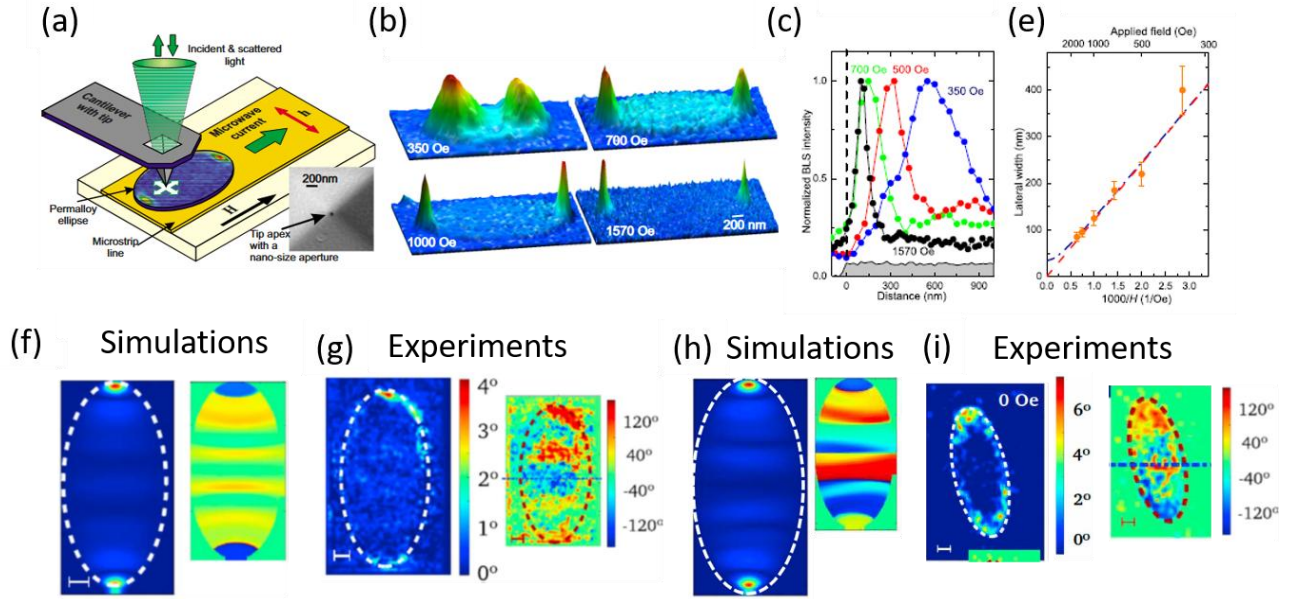

**Fig. 2.10** (a) Schematic drawing of the NF-BLS experimental setup. The sample is illuminated through a nanoscale aperture with a 50 nm diameter, fabricated in a hollow pyramidal atomic force microscopy tip (inset). (b) Normalized 2D maps of the BLS intensity—proportional to the square of the spin excitation amplitude— are shown. 2D intensity maps recorded at different magnetic field values: 350, 700, 1000, and 1570 Oe, corresponding to resonant frequencies of 5.5, 6.5, 7.4, and 9.5 GHz, respectively. (c) Spatial profile of the edge mode, taken along the major axis of the ellipse, as a function of the distance from the edge of the element. The vertical dashed line indicates the position of the edge. (e) Dots represent the experimentally measured spatial width of the localized region, while the dashed curve shows the best fit to the data using a 1/H dependence. Adapted with permission from [46]. (f) Simulation and (g) TR-STXM images of magnetization amplitude (blue background, left) and phase (green background, right), uncapped NiFe ellipse, driven at 2 GHz under zero bias. Scale bar 100 nm. (h-i) Same for panel (f-g) but in the presence of a Ta(3 nm)/Cu(15 nm) capping layer showing mostly antisymmetric precession (odd) phase response of magnetization along the x-direction. Adapted with permission from [47].

Time-resolved scanning-transmission x-ray microscopy (TR-STXM) offers both phase resolution, as a pump-probe technique, and the capability to image through thick conductive overlayers, due to the larger (μm) penetration depth of soft x-rays compared with the optical skin depth. Moreover, it offers spatial resolution higher than near-field optical techniques, in principle to ~15 nm [47] due to the short wavelength of x-rays. Authors employ TR-STXM to image SW eigenmodes in individual NiFe ellipses, each measuring roughly a few hundred nanometers. The research focuses on the small-amplitude, GHz-frequency magnetization dynamics that occur near the vertices (ends) of these nano-ellipses when they are magnetically excited. Using TR-STXM, they achieved a spatial resolution of ~70 nm. The study revealed that the dominant eigenmodes are confined within ~100 nm of each ellipse vertex, as shown in Fig. 2.10(g). These edge or vertex modes exhibit standing-wave patterns, rather than uniform precession across the entire ellipse. A key observation was that the phase relationship between the precessing magnetization at opposite vertices depends critically on sample construction: in ellipses with a conductive overlayer, the vertex modes oscillate out-of-phase (Fig. 2.10(h) and (i)) while in ellipses without an overlayer, the modes remain in-phase (Fig. 2.10(f) and (g)). This points to important interfacial effects—likely Oersted fields generated by the overlayer—altering the local excitation dynamics.

*2.2.2. Magnetic rings*

One of the most interesting nanomagnetic structures with rich spectral features is a magnetic vortex, which is an in-plane circulation of magnetization around a nanometer-scale central core with out-of-plane magnetization pointing either up or down. Neudecker *et al*. examined how the SW eigenmodes in 4 μm diameter NiFe disks evolve under the application of in-plane magnetic fields. [48] The disk samples are prepared in the vortex ground state, and the excitation is performed both via continuous-wave and pulsed in-plane magnetic fields. The response of the system is monitored using time-resolved Kerr microscopy and spatially-resolved FMR techniques, allowing the identification of the spatial character and frequencies of several radial and azimuthal eigenmodes—up to the fifth order.

A central finding of this study is the field-induced splitting of degenerate modes. As the in-plane field increases, the vortex core shifts away from the disk center, breaking the cylindrical symmetry of the structure. This leads to the spatial separation of mode profiles, with some modes localizing near the disk center and others near the edges. Time-resolved imaging technique has been exploited to directly visualize the eigenmodes of similar vortex-state NiFe disks, but under very different experimental conditions. In this case, the system is excited by a short, out-of-plane magnetic field pulse, which disturbs the vortex configuration and initiates spin dynamics. The response is captured using a phase-sensitive pump-probe setup that enables Fourier transform imaging of the spin precession. [49] The primary contribution of this work lies in its ability to reconstruct both the amplitude and phase profiles of the SW eigenmodes across the entire disk, revealing a rich spectrum of dynamic behavior. The authors identify not only radially symmetric (breathing-like) modes but also azimuthal modes with angular nodal structures—evidence of symmetry-breaking dynamic excitations. Particularly noteworthy is the direct observation of core-localized oscillations, a region that is often difficult to probe due to its small size and rapid dynamics.

Curved nanomagnetic devices, such as magnetic rings, have garnered significant interest in magnonics research due to their flux-closure vortex state, which minimizes stray magnetic fields through the circular symmetry of magnetization. Magnon quantization has been observed in magnetic disks and rings, with ring dynamics exhibiting multiple splittings in the SW spectra. Furthermore, the magnetic properties of these curved devices can be readily tuned by adjusting the external magnetic field, enabling transitions between states such as the saturated state, onion state, and vortex state in magnetic ring structures. This tunability offers versatile control over magnonic properties.

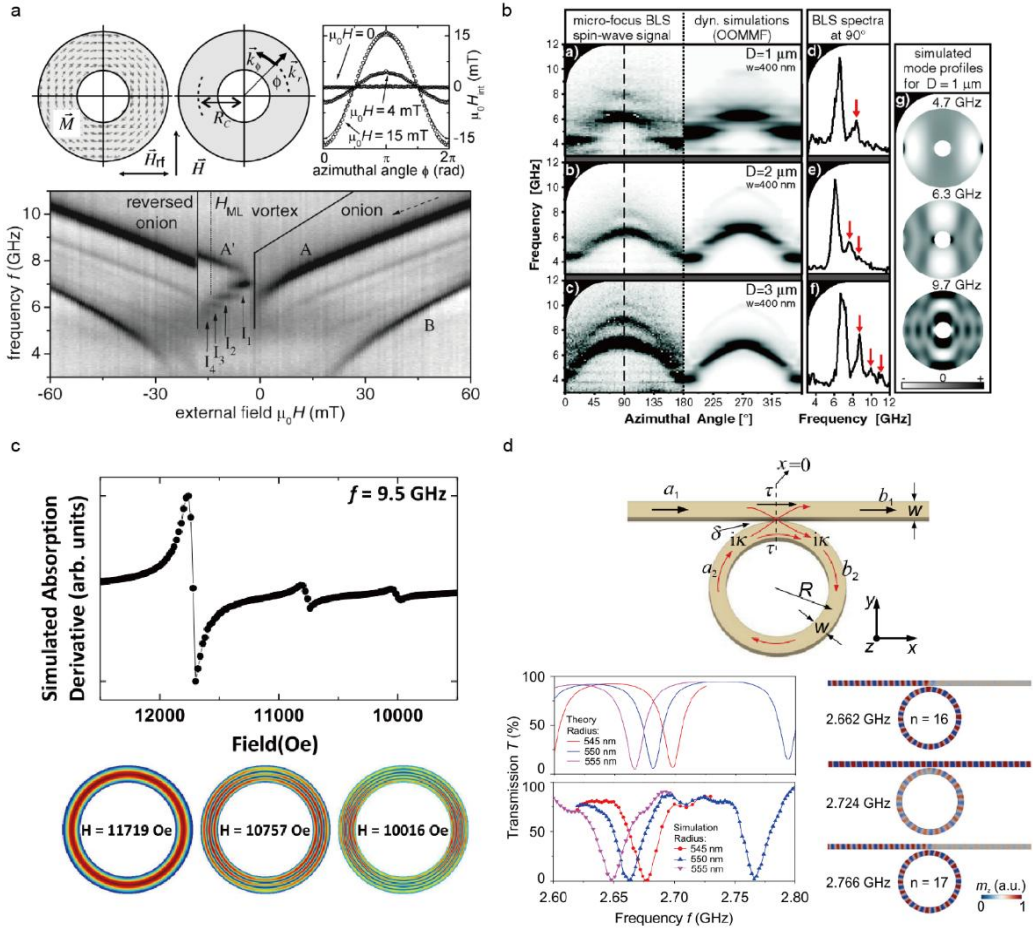

**Fig. 2.11** (a) Schematic illustrating a magnetic ring subjected to an external magnetic field. The ring exhibits distinct magnetic states as the external field is swept from 60 mT to -60 mT. Adapted with permission from [50] (b) Gray-scale maps displaying microfocus BLS spectra (left half) for rings with outer diameters of 1, 2, and 3 μm, respectively, and a constant width of 400 nm, as a function of azimuthal angle. The right half of the panels presents results from micromagnetic simulations, including simulated mode profiles. Adapted with permission from [53]. (c) Simulated FMR absorption spectra for an out-of-plane magnetized nanoring, showing various standing mode profiles. Adapted with permission from [54] (d) Transmission characteristics of the magnonic ring resonator. Snapshots of the out-of-plane SW amplitude in a ring resonator with a 550 nm radius are shown for different excitation frequencies. Adapted with permission from [56].

SW interference has been observed in microscopic magnetic rings using microwave detection techniques. CPWs are patterned on top of NiFe rings, each with a width of 600 nm and an outer diameter of 2000 nm, as illustrated in Fig. 2.11(a). [50] To investigate the collective SW spectra in the ring structure, an array of 750 identical rings was fabricated, with a separation exceeding 2 μm to eliminate dipolar interactions between neighboring rings. As the external magnetic field is swept from positive to negative values, distinct magnetic states, namely the onion state and the vortex state, exhibit unique SW dynamic behaviors. In the onion state, two prominent modes, labeled Mode A and Mode B, are observed. Mode A corresponds to ring segments where the external magnetic field is tangentially aligned to the ring, while Mode B is associated with domain walls. When the external magnetic field is reduced below 2 mT, the system transitions into the vortex state. Unlike magnetic disks, the vortex state in ring structures is less complex due to the absence of a vortex core, allowing radial and azimuthal nodes to serve as well-defined quantum numbers. At least four distinct modes, labeled $n = 1, 2, 3$, and 4, are observed, exhibiting stepwise behavior and minimal dependence on the magnetic field. Semi-analytical calculations reveal that these steps arise from backward volume magnetostatic waves interfering constructively around the ring, inspiring the design of the magnonic ring resonator shown in Fig. 2.11(d). Additionally, increasing the irradiation power reveals nonlinear spin dynamics and a microwave-assisted switching behavior beyond a critical microwave field, providing insights into nonlinear SWs and microwave switching for magnetic memory applications. [51]

The onion state in magnetic rings presents a fascinating phenomenon. Unlike the vortex state, where the circular net magnetization is zero, the onion state exhibits nonzero magnetization aligned with the external magnetic field, while remaining zero in the perpendicular direction. Experimental studies have explored the effects of partial decoherence of SW modes in this state. [52,53] Through micro-focused BLS and micromagnetic simulations, Schultheiss *et al*. have found that the inhomogeneous internal field induces SW confinement in the pole region, forming a so-called SW well. In the intervening regions between two poles, modes display pronounced radial quantization, with a transition from partial to full coherency in the azimuthal direction as the ring size decreases, as shown in the BLS spectra in Fig. 2.11(b). Furthermore, when a strong out-of-plane external magnetic field is applied, sufficient to saturate the nanoring magnetization in the same direction, well-quantized SW modes with varying quantization indices are observed, as shown in Fig. 2.11(c). [54] The SW absorption spectra reveal three prominent modes corresponding to confined radial modes. Thus, the confined SW spectra in different magnetic states of magnetic rings are well characterized and understood.

When light travels from an optically denser to a less dense medium with a sufficiently large angle of incidence, total internal reflection occurs, enabling optical whispering gallery modes (WGMs) at the curved interface of a high-refractive-index medium. In these modes, photons are confined within nearcircular or polygonal microcavities, such as microrings, sustaining stable traveling wave transmission through continuous total internal reflection. Characterized by high quality factors and low mode volumes, these WGMs facilitate quantum information processing and quantum simulation on integrated photonic chips. In magnonics, WGMs have recently been observed in microrings. [55] Wang *et al*. designed a nonlinear nanoscale magnonic ring resonator to enable magnonic logic gates. [56] A straight waveguide, integrated adjacent to the ring, couples with it to inject SWs into the ring. Two confined SW resonances occur in the 2.6-2.8 GHz frequency range, corresponding to the 16th and 17th azimuthal resonant modes, as shown in Fig. 2.11(d). At these frequencies, the output signal diminishes due to destructive interference in the outgoing waveguide between the transmitted and coupled-back spin waves. Moreover, increasing the SW input power shifts the resonance frequency, producing transmission that depending on the frequency, resembles neuronal activation functions or exhibits power limiter characteristics. Thus, the design of the magnonic ring resonator offers a nonlinear building block for unconventional magnonic circuits.

The investigation of the magnetic mode transformation during the onion-to-saturated state in NiFe rings have been investigated by BLS spectroscopy. Similarly to what was observed in ferromagnetic disks, a clear field-induced splitting in both radial and azimuthal eigenmodes has been detected. This splitting is attributed to symmetry breaking introduced by the external field, as well as to the spatial localization of SW modes at specific regions of the ring—such as inner or outer edges. Interestingly, the observed degeneracy lifting in these modes is not continuous but occurs in discrete steps, reflecting the interplay between geometric confinement and the applied field's magnitude. [57]

*2.2.3 Magnetic antidot lattices*

Antidot lattices (ADLs) are magnetic thin films patterned with periodic arrays of non-magnetic holes, creating artificial magnonic crystals. These engineered structures serve as tunable platforms for controlling SWs by shaping the magnonic band structure. ADLs can be viewed as a network of interconnected magnonic waveguides, where both confined and propagating SWs coexist. The effective waveguide width is defined by the spacing between neighboring holes, while the relative orientation of these paths depends on the lattice symmetry**.** Key geometric parameters that govern the SW behavior in ADLs include the lattice geometry, hole size, filling fraction, and dimensionality of the structure, which collectively determine wave dispersion, confinement, and mode hybridization. Due to their engineered periodicity and the resulting internal magnetic inhomogeneities, ADLs can support a rich variety of SW modes, each characterized by unique spatial distributions and underlying in the inhomogeneity of the internal magnetic field felt by precessing magnetic moment within the different ADL regions. The main types of SW modes observed in ADLs include **extended modes**, **localized modes**, and **edge modes**. Their spatial localization is schematically illustrated in Fig. 2.12 (a).

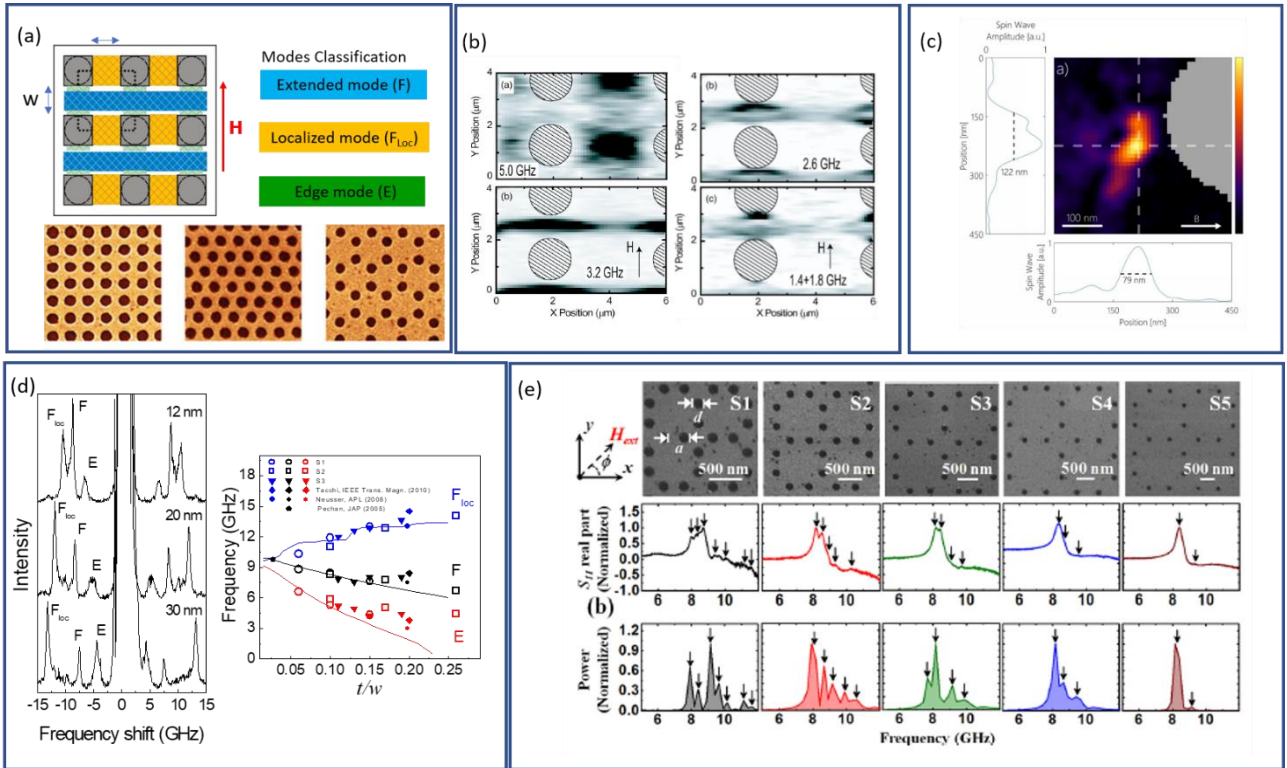

**Fig. 2.12** Schematic representation of an ADL with a square unit cell and holes with either circular or squared shapes (grey area). The direction of the applied magnetic field is indicated by the red vertical arrow while w is the width of the nanowires between different rows and columns. Blue, green and yellow area indicate the regions where the edge, localized and extended mode exist. [58] Lower row of panel (a) shows the SEM images of antidot arrays with square (left), rhombic (center), and honeycomb (right) lattice symmetry. Adapted with permission from [62]. (b) Gray- scale 2D map of spectral power of magnetic modes at 5.0, 3.2, 2.6 and 1.4 GHz showing different spatial localization. Adapted with permission from [59]. (c) 2D distribution of the SW amplitude and phase of the edge mode in squared ADL. Adapted with permission from [60]. (d) BLS spectra show how mode frequencies evolve with the thickness-to-width ratio ($t/w$), identifying edge (E), fundamental localized ($F_{Loc}$), and fundamental (F) modes. Adapted with permission from [72]. (e) SEM images display quasiperiodic octagonal NiFe ADLs with hole diameters of 140 nm and varying lattice spacing (300–700 nm). FMR spectra and simulated SW spectra at a magnetic field of 800 $Oe$ ($\phi = 0°$) identify multiple SW modes in both experimental and theoretical data. Adapted with permission from Ref. [73].

**Extended SW** modes propagate throughout the entire ADLs, traversing the periodic structure in the orthogonal direction to the applied magnetic field. These modes often form band structures similar to those in electronic crystals and can be tuned by adjusting the lattice parameters, such as antidot size, spacing, filling fraction, and film thickness. Extended modes are crucial for SW transport, enabling wave-based logic operations, interference, and waveguiding in magnonic circuits. **Localized SW** modes reside between neighboring holes perpendicular to the field direction, often aligned with the lattice axes, where the magnetic potential is favorable for mode confinement. **Edge SW modes** are localized along the perimeters of the antidots, where demagnetizing fields are strongest and abrupt changes in magnetization and demagnetizing fields of the array holes create unique conditions for SW confinement. As a result, edge modes are energetically and spatially "trapped" in localized zones, typically between antidots or within defect regions and are unable to propagate freely across the entire lattice. Edge modes are usually lower in frequency, and their amplitude rapidly decays outside the edge region. These edge modes are analogous to those observed in single or arrays of 1D and 2D magnetic wires and dots where the inhomogeneity of the internal magnetic field acts like SW potential wells. Edge modes are sensitive to the geometry of the antidots and the orientation of the applied magnetic field, making them highly tunable. For a small lattice period and proper directions of the magnetic field, edge mode can modify their character from localized to propagating waves through the array, giving rise to nanochannel waveguides.

The first experimental evidence of these modes with different spatial localization was provided by Pechan *et al.* by using time-resolved Kerr microscopy TRKM. As can be seen in Fig. 2.12(b), the 5.0-GHz response is a localized mode, while the mode at 3.2 GHz has an extended character through the nanochannels between the hole rows in the direction perpendicular to the applied magnetic field. Its intensity is predominantly localized between vertical holes with a small amount of spectral power seen along the line connecting to the adjacent regions. Moreover, the modes at approximately 1.4 and 2.6 GHz are localized along the edges of the holes in the direction of the magnetic field similar to the edge modes observed in magnetic wires. [59] The 1.4 GHz mode is mostly localized close to the antidot hole, while the mode at 2.6 GHz spreads over a more extended horizontal region.

More recently, Groß *et al.* [60] have found by scanning transmission X-ray microscopy that the edge mode does not have a uniform temporal phase along the antidot edge but exhibits a phase evolution with nodal planes parallel to the applied field (DE-type waves). This indicates that the edge mode inherits its characteristics from a propagating SW mode, which becomes trapped in a potential well formed in the vicinity of the antidot (see Fig. 2.12(c)).

On reducing the ADL constant (i.e., the periodic spacing between adjacent antidots) from 800 nm to 300 nm, the edge modes, confined near the perimeters of circular antidots for large periodicity, become delocalized, giving rise to minibands with finite dispersion and enabling efficient SW propagation. As the lattice constant is reduced, the spatial separation between antidots decreases, enhancing the overlap of the dynamic magnetization at the antidot edges. This leads to stronger dipolar coupling between neighboring edge modes, resulting in broader minibands and significantly increased SW group velocities. [61]

Several experimental, theoretical and numerical studies have been carried out to investigate the dependence of SWs on ADLs symmetry, [62] hole shape [63], and dimensionality. [64] The lattice geometry, including the symmetry and periodicity of the antidot arrangement (e.g., square, hexagonal, triangular), plays a crucial role in determining the allowed SW modes and their dispersion relations. The symmetry of the ADL affects the bandgap formation and the directionality of SW propagation. For example, square lattices typically support more isotropic propagation characteristics, whereas hexagonal lattices may give rise to anisotropic band structures. [65] The periodic boundary conditions imposed by the lattice create Brillouin zones (BZs) in reciprocal space, and the magnonic bands develop as a consequence of Bragg scattering at the zone boundaries. [66,67] Moreover, defects and irregularities in the lattice can introduce localized modes or alter the coherence of SW propagation. Recent studies have shown that even small deviations from ideal periodicity can significantly affect magnonic bandgap formation and SW lifetimes. [68] The diameter of the antidots is directly related to the effective magnetic potential landscape experienced by the SWs. Larger holes increase the non-magnetic volume fraction and can lead to stronger confinement and larger bandgaps. However, excessively large holes may disrupt the connectivity of the magnetic matrix and reduce the overall magnetization, negatively impacting SW propagation. Optimal hole size enables bandgap engineering while maintaining sufficient magnetic connectivity. [69] The shape of the holes also plays a critical role in determining its physical properties of ADLs for tailoring their functionality. Numerical simulations and experimental results have demonstrated that increasing hole size leads to a redshift of SW modes and can enhance mode hybridization effects due to the increased interaction between SWs and the edges of the antidots. Edge-localized modes are especially sensitive to hole diameter, as they are confined in the narrow magnetic regions between adjacent holes. [70]

Dimensionality refers both to the in-plane size of the lattice (e.g., finite vs. infinite arrays) and the thickness of the magnetic film. In thin films, SWs are typically quantized in the in-plane direction, and the interplay between quantization and lateral confinement due to the antidots becomes significant. [71] To elucidate the effect of the ADL geometrical parameters on the SW band structure, Tacchi *et al.* [72] investigated ADL etched into NiFe films, varying film thickness and hole size to systematically alter the thickness-to-width aspect ratio (*t/w*) of the effective nanochannel defined between adjacent antidot rows. Each BLS spectrum measured at the Γ (center of the BZ) point is characterized by the presence of three dominant peaks (see the left panel Fig. 2.12(d)). The low-frequency mode corresponds to the so-called edge mode, confined in the deep wells of demagnetizing field appearing close to the hole edges in the direction of H. At higher frequency, the fundamental mode, characterized by a quasi-uniform spin precession amplitude extending in the vertical channels comprised between rows of holes, and the localized fundamental mode, having the maximum amplitude in the horizontal channels between rows of holes, are found. Their key finding is a universal scaling behavior: the frequencies of the most intense SW modes at the Γ-point—including edge, fundamental, and fundamental-localized modes—depend almost exclusively on *t/w*, regardless of absolute dimensions (see the right panel of Fig. 2.12(d)). Similarly, both the *position* and *width* of the main band gap at the BZ edge scale predictably with this aspect ratio. Detailed simulations reveal that the demagnetizing field within the nanowire regions, which originates from surface charges around the antidots, underpins this scaling relationship. As *t/w* increases, the demagnetizing field and thus the SW frequencies and bandgap characteristics shift in a well-defined manner.

Choudhury et al. systematically investigated SW modes in 20 nm-thick NiFe ADLs with circular holes arranged in an octagonal lattice, varying lattice spacings (a) from 300 nm (S1) to 700 nm (S5). SEM images (Fig. 2.12(e)) and bias-field-dependent SW absorption spectra reveal that SW dynamics evolve with lattice spacing and external field strength. [73] At $H_{ext}$=800 *Oe*, SW modes decrease from seven (S1) to two (S5), with mode counts varying significantly with field strength, peaking below a critical field ($H_c$≈800 Oe) at nine, five, five, and three modes for S1–S4, respectively, and reducing above $H_c$. LLG micromagnetic simulations of magnetostatic field distributions show that the internal field in some region of the ADL decreases with increasing period *a*, while it increases in other regions due to opposing demagnetizing fields, leading to nearly uniform magnetization. This unified relationship enables predictive magnonic design, allowing engineers to tune SW band properties like frequency and bandgap by adjusting film thickness or hole size for desired localized and propagating SW features.

Furthermore, the magnetic field orientation plays a crucial role in dynamically controlling the mode type, directionality, and propagation characteristics of SWs at the nanoscale. Tacchi et al. explored the SW dynamics in a patterned magnetic thin film, specifically focusing on a rhombic ADLs formed in a 30 nm-thick NiFe film. [65] The lattice comprises circular holes with a diameter of 250 nm and arranged with a lattice constant of 400 nm. Using BLS spectroscopy, the researchers investigated how the SW modes in this structure respond to variations in the orientation of an externally applied in-plane magnetic field. A central discovery of this work is the controlled conversion between quantized (localized) and propagating (extended) SW modes. This transition is driven purely by the orientation of the applied magnetic field. Specifically, when the field is rotated by 30 ° with respect to the lattice symmetry axes, the SW modes undergo a dramatic variation of the spatial character: modes that were initially localized become propagating, and vice versa. This demonstrates a highly tunable method for manipulating SW propagation into highly directional and channeled modes.

Similar results have been achieved in artificial anti-spin-ice systems consisting of a 20 nm-thick NiFe film, patterned into periodic square lattices of elliptical antidots. [74] BLS spectroscopy was employed to measure SW frequency dispersion under different in-plane orientations of an external magnetic field. When SWs propagate along the principal lattice directions (aligned with the axes of the square lattice), the spectra reveal the formation of flat bands separated by bandgaps. These features arise due to mode confinement between the elliptical antidots, where the periodic potential landscape restricts the SW propagation and localizes the excitations in discrete frequency bands. In contrast, when the SWs are directed along the diagonal (45°) directions of the lattice, narrow, quasi-1D channels that support SW propagation are formed. These channels effectively act as waveguides, enabling the flow of SW energy across the lattice. As a result, the SW dispersion of these modes acquires a nonzero group velocity. This agreement not only validates the experimental observations but also confirms the micromagnetic origin of the mode confinement and propagation phenomena observed in the system.

Another interesting strategy for tuning the SW properties is to play with the magnetic contrast between magnetic materials in bicomponent ADL. Duerr *et al.* engineered SW properties in a bicomponent ADLs by embedding periodic cobalt (Co) nanodisks into shallow-etched nanotroughs in a NiFe film, creating magnetic contrast between the materials for precise control over SW dynamics. [75] Using micro-focused BLS, they identified two dominant SW eigenmodes propagating perpendicular to the applied magnetic field through parallel nanochannels formed by the stripe configuration. Co nanodisks create localized regions with reduced internal magnetic fields, corresponding to low-frequency guided modes (Fig. 2.13(a)). In contrast, horizontal Py stripes above and below the nanodisk row, where demagnetizing fields align with the external field, exhibit higher internal fields and thus higher SW eigenfrequencies, as shown in Fig. 2.13(a) and (b). This contrasts with pure NiFe ADL, where demagnetizing fields typically oppose the external field, resulting in lower-frequency SW modes extending across continuous Py regions between parallel rows of holes (Fig. 2.13(e,f)).

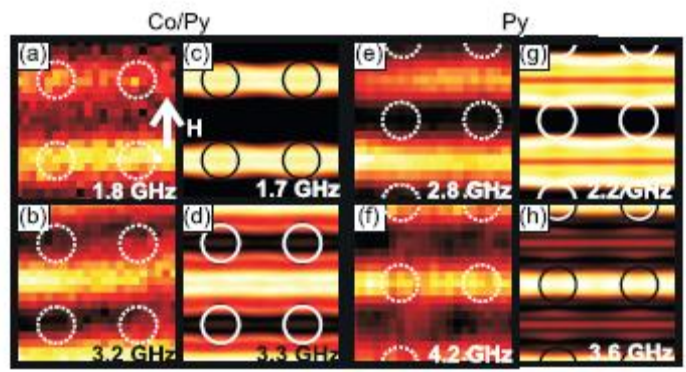

**Fig. 2.13** Measured spin precession profiles (first and third column) and corresponding simulations (second and fourth column). The outer edge of the CPW ground line aligns with the left edge of each graph. Data for Co/Py sample 1 are shown in graphs (a) through (d), and for Py sample 2 in graphs (e) through (h). Relevant frequencies are provided. Dotted circles in graphs (a), (b), (e), and (f) indicate nanotrough positions derived from microscopy. Adapted with permission from [75].

*2.3 Spin-Wave Modes in 3D magnetic nanostructures*
The exploration of SW in 3D magnetic nanostructures represents an emerging frontier in nanomagnetism and magnonics. [76] While two-dimensional systems have long provided a platform for understanding fundamental spin-wave modes and developing prototype magnonic devices, extending these concepts into three dimensions unlocks a wealth of novel physical phenomena and technological opportunities. The additional geometric degree of freedom in 3D nanostructures allows for unprecedented control over mode hybridization, localization, and propagation, enabling functionalities that cannot be realized in planar systems. The novelty of 3D SW modes lies not only in their complex spectral characteristics—arising from curvature, connectivity, and volumetric confinement—but also in their potential to harness topological effects, nontrivial mode coupling, and anisotropic dispersions.

*2.3.1 Magnons in nanotubes and curved shells*

Three-dimensional (3D) magnonics is a rapidly growing research field with significant technological potential, fueled by sophisticated magnetic architectures, including curved shells, [77] nanotubes, [78,79] Möbius strips, [80] buckyballs, [81] nanocaps, [82] and other curvilinear structures. [83,84] Special attention has been given to magnetic nanotubes, [85] which enable the development of analytical descriptions of their behavior. In the case of tubular nanostructures, one can distinguish between axial, azimuthal, and radial modes, where the latter are similar to the perpendicular standing spin waves in flat films. Fig. 2.14(a) illustrates the frequency of radial modes as a function of the tube thickness, [86] where one can see that the radial modes decrease in frequency, as observed also for planar films. The azimuthal modes (Fig. 2.14(b)) also decrease in frequency with the thickness and are quantized due to the periodic boundary conditions of closed nanotubes along the azimuthal directions. The dynamic profiles of the purely radial modes, that is, the dynamic magnetization along the thickness, are shown in Fig. 2.14(c-e), which are also shown in panels (f) and (h), where nodal points are shifted from their symmetric positions along the thickness. [86,87] This leads to a modification of the quantization condition for the radial spin waves along the thickness of a curved shell, which is due to the symmetry breaking imposed by curvature. In other words, the wavevector of the radial waves is not symmetrically distributed along the shell thickness, as occurs for a planar film. [86,87]

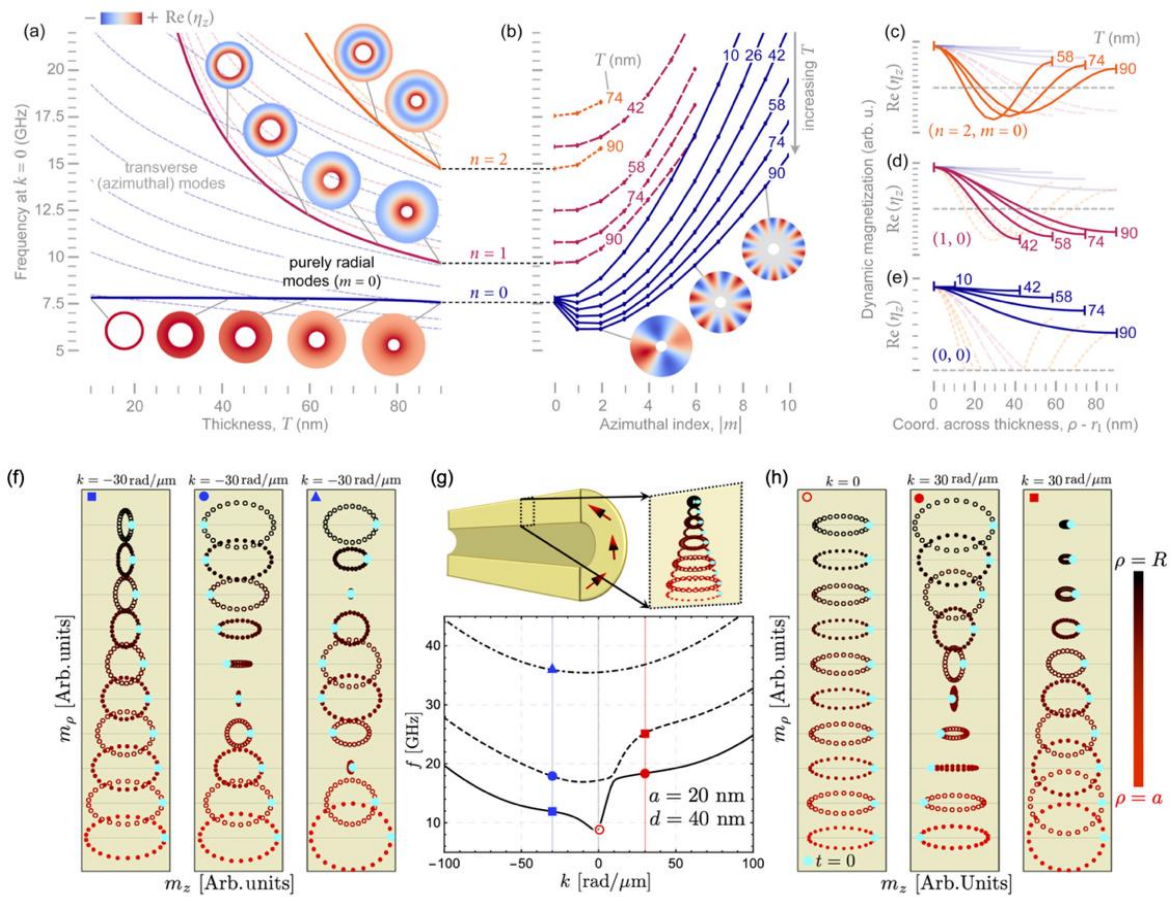

**Fig. 2.14** Spin-wave modes in thick magnetic nanotubes. (a) Radial modes at $k = 0$ vs. tube thickness. (b) Azimuthal modes vs. the azimuthal index. (c) Dynamic magnetization profiles along the thickness of the closed

tube. Adapted from [86]. (f-h) Nonreciprocal spin-wave dispersion for thick tubes, where the mode profiles are illustrated for selected wavevectors. Adapted from [87].

For a flat shell, the SW modes are either even (fully mirror symmetric) or odd (antisymmetric), as shown in Fig. 2.15(c), which depicts the frequency vs. the transverse field. Nonetheless, such parity symmetry disappears with increasing shell curvature due to a chiral exchange contribution in a curvilinear frame of reference. [88] This can be seen in Fig. 2.15(d,e), where the expectation value of the parity operator was calculated following Ref. [89], and is represented by the color code in each mode. As can be seen, there is parity lost for low transverse fields, where the azimuthal modes become heterosymmetric, while parity is preserved for larger fields. For the fully closed tubes (Fig. 2.15), the odd modes disappear due to the lack of edge along the azimuthal direction. [88]

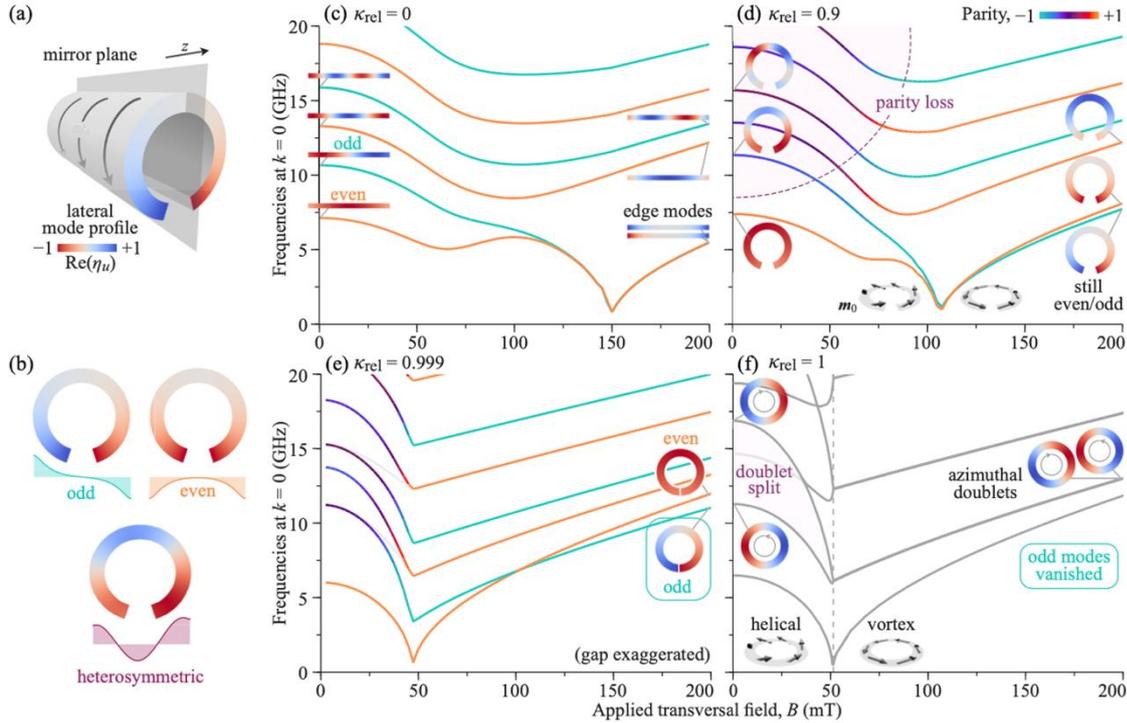

**Fig. 2.15** Mode parity and ferromagnetic resonance for different flat and curved shells. The expectation value of the parity operator serves as a tool for analyzing the magnetization profiles within the local coordinate systems corresponding to each mode, with respect to the mirror plane, as illustrated in panel (a). In (b), different types of modes are described. FMR frequencies for (c) flat shells, (d-e) curved shells, and (f) closed nanotubes. Here, the parity for each mode is calculated and represented by the color code. Adapted from Ref. [88].

*2.3.2 Spin waves in complex three-dimensional nanomagnonic networks*

Recent research in three-dimensional (3D) magnetic nanostructures and nanomagnonic networks has revealed rich SW dynamics, with both localized and propagating modes emerging as tunable features of complex magnetic architectures. The specific resonance frequencies can be controlled by altering the geometry or magnetic state, thereby reconfiguring the spectrum and demonstrating that 3D nanowire networks represent promising candidates for tunable and reprogrammable magnonic devices. Such devices might function as frequency-selective filters, switches, or elements within volumetric magnonic circuits, all controlled without removing the physical structure—only through changes in field, configuration, or design parameters. [76]

Sahoo *et al*. have experimentally detected coherent SW modes in a nanoscale 3D artificial spin ice (3D-ASI) lattice for the first time. This 3D-ASI structure, shown in the green boxed panel of Fig. 2.16, composed of interconnected magnetic nanowires arranged in a diamond-bond network, was fabricated using two-photon lithography followed by thermal evaporation. [90] Using BLS spectroscopy, two distinct SW resonances, labeled $M_1$ and $M_2$, were observed. These resonances showed frequency shifts that varied almost monotonically with changes in an in-plane external magnetic field. To support their findings, the team performed micromagnetic simulations, which closely matched the experimental data. Spatial analysis revealed that mode M1 is localized at the bipod junctions—points where the nanowires intersect—while mode M2 is distributed throughout the nanowire network. Both modes exhibit quantized standing wave behavior, with nodal structures and phase patterns dependent on the mode index, similar to wave modes in a 3D crystal. This work demonstrates that coherent, high-frequency magnons can propagate in complex 3D magnetic structures, a crucial insight for the development of 3D magnonic and spintronic devices, potentially enabling more advanced and efficient information technologies. At the same time, SW modes can be localized at the junctions between adjacent segments.

Kumar *et al.* have conducted a groundbreaking study demonstrating how magnetic charge configurations in a nanoscale artificial spin ice (3D-ASI) can be used to control magnon dynamics, offering a promising path toward reconfigurable magnonic devices. [91] The main challenge addressed is how to dynamically tune SW behavior without altering the physical structure—by instead manipulating internal magnetic states. The researchers created a 3D network of magnetic nanowires arranged in a diamond-bond lattice, using advanced fabrication techniques. They measured the SW modes using BLS spectroscopy and used micromagnetic simulations to analyze the data and visualize SW behavior. Their central discovery is that the magnetic charge configuration—specifically, the arrangement of monopole-like magnetic excitations at the lattice vertices—directly impacts the SW spectrum. As an external magnetic field is varied, the internal microstate of the 3D-ASI changes, leading to shifts in SW frequencies and intensities. This shows that different magnetic charge states can selectively excite or suppress specific SW modes, offering both spectral and spatial control over magnon propagation.

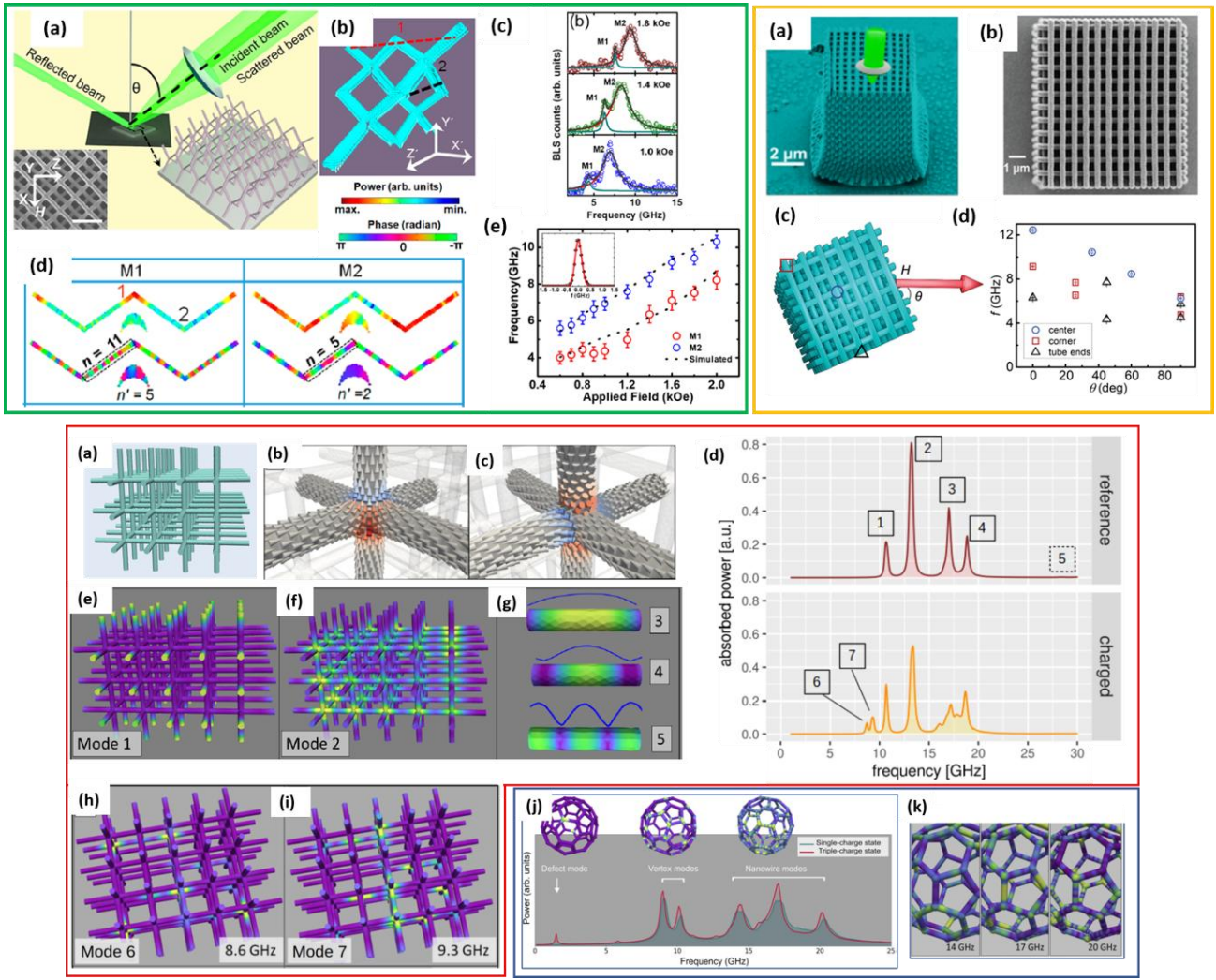

**Fig. 2.16** Green-boxed panel (a) shows schematic of the BLS measurement geometry. The incident laser light is focused normally (θ = 0°) onto the sample. The direction of the applied magnetic field (H) is indicated in the left inset. A scale bar of 2 μm is included for spatial reference. (b) This panel provides a schematic representation of the 3D artificial spin ice (3D-ASI) network, illustrating the complex arrangement of interconnected magnetic nanowires forming a diamond-bond lattice. (c) The BLS spectra are shown for three different values of the magnetic field. (d) SW mode profiles are presented, calculated for different positions in the 3D-ASI at an applied field of H = 1.6 *kOe*. These profiles are extracted from a slice taken along the x′−y′ plane at point 1, providing a spatially resolved view of the SW distribution within the structure. (e) The SW frequencies of mode 1 (M1) and mode 2 (M2) are plotted against the applied magnetic field. Experimental data points are marked with symbols, illustrating how each mode's frequency evolves as the external magnetic field is varied. Adapted with permission from [90].

Yellow-boxed panels (a) and (b) show the SEM image of the 3D Ni nano-network. The green laser and lens represent the *μ*-BLS measurement configuration. (c) The schematic of the measurement configuration. *θ* is the angle between the magnetic field H=150 mT and the tubes of the top layer. (d) Resonance frequencies detected at three different positions when altering the angle *θ* between the magnetic field and the 3D Ni nano-network. Adapted with permission from [92].

Red-boxed (a) Perspective view on the studied sample consisting of three orthogonal sets of equidistant cylindrical magnetic nanowires. The intersection points constitute a regular cubic lattice with a constant of 70 nm. The structure's total size is 420×350×280 $nm^3$. Two qualitatively different vertex configurations were observed: (a) uncharged state of type I, (b) uncharged state of type I and (c) II. Absorption spectra of the nanowire array for the two magnetic states of types I and II. Different peaks are labeled by integer numbers from 1 to 7. The corresponding spatial mode profiles are presented in panel (e-i). Spatial distribution of the oscillation amplitude for the reference state at the frequencies labeled 1-5 in the top frame of panel (d). The mode profile (h) shows a localization of the oscillations at the dangling, free wire ends at the array surfaces, while frame (i) displays the resonance at the intersection points (vertices). Adapted with permission from [93]. (j) Comparison of the power spectrum of the magnetic high-frequency modes in a buckyball geometry at zero static field for two different cases. The dark green line refers to a demagnetized configuration containing only ice-rule-obeying single-charge vertex configurations. (k) Spatial profile of the modes at 14, 17, and 20 GHz. Adapted with permission from [94].

An innovative additive manufacturing approach to fabricate 3D ferromagnetic nanonetworks with unprecedented structural complexity has been demonstrated by Guo *et al*. [92] Their approach involves using two-photon lithography to construct a woodpile-like polymer scaffold, which is then coated with a 10-nm-thick nickel shell through atomic layer deposition. The result is a fully connected 3D ferromagnetic lattice capable of supporting collective spin dynamics at microwave frequencies up to 25 GHz. To study these structures, the team employed μ-BLS, revealing a rich magnonic spectrum consisting of two distinct classes of SW modes: bulk modes, which propagate throughout the entire 3D lattice; surface modes, which are confined to the outer layers of the nanostructure. These two mode types are separated by a consistent ~10 GHz frequency gap, attributed to the different boundary conditions and spatial extents of the bulk and surface regions. The 3D lattice effectively behaves as two interacting magnonic subsystems, each with its own mode spectrum. Further investigation showed that the SW frequencies are highly sensitive to both the orientation of the external magnetic field and the specific spatial location being probed within the structure. By adjusting the field angle and scanning different regions with the BLS laser, the researchers could selectively excite and resolve individual bulk and surface modes. This work demonstrates the feasibility of multi-frequency magnonic signal processing in a single 3D magnetic device. Such capability could enable advanced, low-power functionalities like signal routing, filtering, and logic operations without the need for electric currents, paving the way for next-generation 3D magnonic computing and spintronic systems. Frequency-domain micromagnetic simulations—based on the linearized Landau-Lifshitz-Gilbert equation—have been used to explore the magnetization dynamics of a cubic lattice composed of interconnected magnetic nanowires. [93] Each segment behaves like an Ising-type magnet, i.e., it can be magnetized in one of two possible axial directions. This architecture represents a fully 3D array of cylindrical ferromagnetic wires that intersect at regular points, forming a highly ordered yet geometrically intricate network, see Fig. 2.16 red box (a). To manage the complexity related to the numerous possible magnetic configurations, the authors considered only three qualitatively different states. In the red box of Fig. 2.16, we report the magnetization configuration for two of them: the reference state of type I, in which all wires are uniformly magnetized along their respective axes — *x, y,* and *z* — forming a highly ordered, symmetric configuration. The second configuration (type II) is the remanent state that develops at zero field after saturating the sample in the *x* direction. In this configuration, all nanowire segments oriented parallel to the *x*-axis remain magnetized in the positive *x* direction.

The magnetization in the other two sets of wires, in contrast, is not necessarily uniform, and the corresponding nanowire segments can be either magnetized along the positive or negative *y* and *z* directions, respectively. When subjected to a small-angle oscillating magnetic field, the nanowire network exhibits pronounced resonances at specific frequencies (as shown in panel e-i). These resonances emerge from distinct oscillatory modes, each associated with particular geometric features of the structure or unique micromagnetic configurations within the lattice. In other words, each resonance encodes spatial information about where and how the magnetization is oscillating within the 3D structure. A central insight of the work is that the absorption spectrum of the network is highly sensitive to both the geometry of the array—such as wire diameter, spacing, and connectivity—and its micromagnetic state, including the presence of complex domain structures or topological defects at wire intersections. This highlights how the SW response can be finely tuned by altering structural parameters or adjusting magnetic configurations.

The same research group has extended their micromagnetic simulation studies to explore the high-frequency magnonic behavior of a 3D nanowire network shaped like a "buckyball", technically, a truncated icosahedron. [94] This unique magnetic nanoarchitecture can be built in two formats: a solid structure of interconnected ferromagnetic nanowires, or a hollow version where a nonmagnetic scaffold is coated with a thin (≈4 nm) magnetic shell. In the solid buckyball, micromagnetic simulations show a broad spectrum of SW resonances, each tied to specific structural features (see blue panel in Figure 2.16). Key findings include: low-frequency modes are localized at defect-like vertices, especially at "triple-charge" nodes, where three nanowires intersect. These nodes act like magnetic singularities, concentrating dynamic activity at discrete points. Mid- and high-frequency modes propagate along the nanowires and form collective oscillation patterns spanning the full structure. These modes represent coherent standing-wave-like patterns that traverse the complex wire network. Each resonance effectively forms a "magnetic fingerprint" of the structure, with different modes linked to specific elements: vertices, edges, and segments. When considering the hollow version—where a nonmagnetic framework is uniformly coated with a thin (≈4 nm) magnetic shell—the static equilibrium magnetization remains almost identical to the solid case. However, the dynamic response differs more substantially. The defect-vertex mode now occurs at a higher frequency (~2.7 GHz versus ~1.4 GHz), and this low-frequency resonance becomes a split doublet. (see panel (e) in Figure 2.16). A satellite mode centered around 6.2 GHz appears, localized near the outer surface of the shell rather than at the inner defect regions. This indicates that the spatial distribution of oscillation can shift between the inner and outer surfaces depending on the mode. Furthermore, the hollow architecture exhibits fewer distinct wire modes (two instead of three) because two mid-frequency nanowire resonances in the solid case converge into a single, stronger peak around 15.4 GHz. Above ~17 GHz, additional tube-like oscillations are present, though very weak. Panel (j) represents the calculated power spectrum of the magnetic high-frequency modes in a buckyball geometry at zero static field for two different cases while panel (k) shows the variation of the mode profile between the nanowires, the bulk of the oscillations at these three frequencies can be attributed to the first, second, and third-order standing-wave modes within the nanowires, respectively.
To demonstrate reconfigurability in 3D magnetic nanoarchitectures, focused electron beam-induced deposition was exploited to fabricate a complex 3D tetrapod structure composed of cobalt-iron ($Co_3Fe$) segments meeting at a central node. [95] To probe the magnetic behavior of this intricate geometry, they employ STXM, enabling direct visualization of magnetic configurations at the nanoscale. Complementing these observations, finite-element micromagnetic simulations are used to interpret and predict the structure's magnetization dynamics. The core achievement of this work lies in demonstrating selective, sequential magnetic reversal of the individual arms of the tetrapod. Each leg of the structure can switch its magnetization state independently, enabling a multi-step reversal process. This controlled switching is accomplished simply by varying the orientation of the external magnetic field—meaning that by changing the field angle, the researchers can program the structure to adopt a wide variety of stable magnetic configurations. Micromagnetic modeling reveals how this selective reversal is governed by the interplay of shape anisotropy, magnetostatic interactions, and the biasing effect of internal stray fields within the 3D architecture. The simulations support the experimental findings by reproducing the observed switching order and demonstrating the energy barriers involved in reorienting each arm. This establishes a clear mechanism for field-controlled reconfigurability in 3D magnetic nanostructures.

Recently, Xu *et al* [96] fabricated three-dimensional helical structures using two-photon lithography and subsequently coated these helices with a thin, conformal layer of nickel using atomic layer deposition. Although nickel is a conventional, polycrystalline ferromagnet with no inherent chiral magnetism, the helical architecture of the scaffold forces the spins to adopt a screw-like, spatially twisted configuration. This induces a toroidal magnetic moment whose sign is directly set by the handedness of the helix. In essence, a chiral magnetic state is imprinted onto an otherwise achiral material through the geometry alone. A central result of the study is the observation of strong magnon nonreciprocity at room temperature and in the complete absence of an external magnetic field. Magnons normally propagate symmetrically in opposite directions unless a symmetry between +**k** and –**k** is broken. In the artificial chiral magnets presented here, such symmetry breaking arises from the combination of geometric chirality and the toroidal moment associated with the helical spin texture. As a result, spin waves traveling in opposite directions exhibit markedly different behavior. The degree of nonreciprocity, quantified through a "chiral parameter", exceeds that of many natural chiral magnets by roughly three orders of magnitude, demonstrating that geometrically induced chirality can be far stronger and more tunable than its crystalline counterpart. Importantly, this nonreciprocal magnon response is reprogrammable. By applying a small magnetic field along the axis of the helix, the toroidal moment can be switched between two stable configurations. Once the field is removed, the system retains its new state, meaning that the direction in which magnons preferentially propagate can be rewritten without the need for a continuous external field. This programmability transforms these structures into functional, reconfigurable magnonic elements capable of behaving as isolators or diodes, with the preferred direction of transmission controlled on demand.

*2.4 Spin textures (domain walls, vortices and skyrmions)*

A wide range of spin textures, including domain walls, vortices, and skyrmions, can be stabilized across various magnetic environments. The interaction between magnons and these spin textures is rich in physics and highly compelling. This section reviews magnon confinement within spin textures, which can be further extended to realize applications such as spontaneously formed magnonic nanowaveguides, nonreciprocal magnonic devices, and neuromorphic computing based on magnons.

*2.4.1 Magnon confinement within magnetic domain walls*

Magnetic domain walls are magnetic solitons that separate adjacent magnetic domains with distinct magnetization orientations. [97,98] These magnetic structures can serve as natural waveguides for magnons, with their effectiveness depending on properties such as their width (typically on the nanometer scale), type (e.g., Bloch or Néel), and local magnetic textures, as shown in Fig. 2.17. The confinement of magnons originates from the spatial variation in magnetic potential across different domains, which modifies the magnon characteristics, such as the magnon dispersions, within the domain wall (DW), leading to magnon localization. As magnetic domain walls form spontaneously without requiring nanofabrication, they offer a promising platform for self-assembled nanoscale magnonic networks, facilitating efficient magnon-based information transfer.

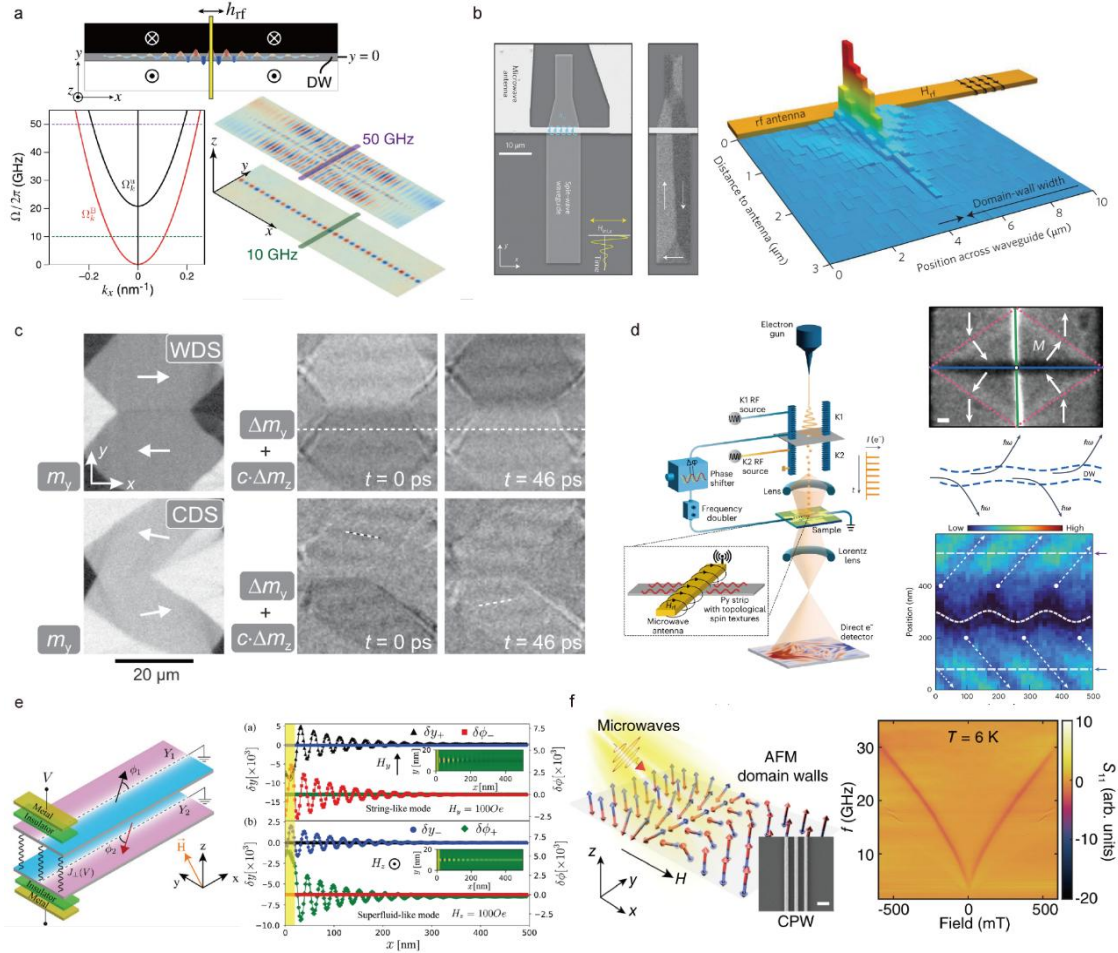

**Fig. 2.17** Illustration of magnon confinement in magnetic domain walls. (a) Micromagnetic simulation of magnon dispersions in the magnetic domain wall, with the magnon confinement along the *y*-direction. Adapted with permission from [99]. (b) Experimental evidence of magnon confinement within a DW waveguide in a patterned NiFe thin film (taken from [104]). (c) Generation of magnetostatic SW in an amorphous CoFeB stripe with a static ferromagnetic domain configuration, excited at 9 GHz. The time evolution depicts the qualitative magneto-optical magnetization response. Adapted with permission from [109]. (d) Magnetic domain-wall oscillations and magnon scattering from the oscillating DW into adjacent domains observed using the laser-free pulser ultrafast electron microscopy system. Adapted with permission from [110]. (e) Schematic of SW excitation in the domain walls of a bilayer van der Waals antiferromagnet via capacitive coupling to electrical gates. The spatial profile of DW mode variables highlights the selective excitation of string- and superfluid-like modes (taken from [111]). (f) Gapless magnon confinement in an antiferromagnetic DW in hematite, an antiferromagnetic insulator. Adapted with permission from [112].

Theoretical analyses have been conducted in order to reveal the mechanisms driving magnon confinement within magnetic domain walls. [99,100,101,102,103] The magnon excitations localized within the wall emerge within the frequency gap of bulk SW modes in the surrounding domains. These confined magnons can be guided along curved geometries and propagate in close proximity to other channels. [99] In the presence of the dipolar interaction, the Bloch-type DW is characterized by magnetic moments that rotate within a plane (*xz*) perpendicular to the wall direction (*y*). In this configuration, the SW eigenmodes can be described by:

$$\psi_k(x,y,t) = \exp[i(\Omega_k^B t - k_x x)]\text{sech}(y/\lambda) \quad (4)$$

which are localized in the direction perpendicular to the DW ($y$) on a length scale $\lambda$, but propagate as plane waves parallel to the DW ($x$). Here, $\lambda = \sqrt{A/K_0}$ represents the DW width, with $A$ being the exchange constant, $K_0 = K_u - \mu_0 M_s^2/2$ is the effective perpendicular anisotropy constant. These modes correspond to exchange-dominated SWs confined in the domain wall, with their dispersion relation is given by:

$$\Omega_k^B = \sqrt{\omega_k(\omega_k + \omega_\perp)} \tag{5}$$

where $\omega_k = \frac{2\gamma K_u}{M_s}$ is the term dominated by the exchange interaction. $\omega_\perp = \mu_0 N_y M_s^2/2$ represents the dipolar interaction due to volume charges at the DW center, and $K_u$ is the uniaxial anisotropy constant along the easy axis of the domain wall. For ultrathin films, the demagnetization factor $N_y \approx d/(d + \pi\lambda)$, where $d$ is the film thickness. These modes are gapless because the effective field associated with the perpendicular anisotropy cancels out at the center of the domain wall, as shown in Fig. 2.17(a). For ultrathin ferromagnetic films with strong spin-orbit coupling, the presence of DMI leads to the formation of Néel-type domain walls. The eigenfrequencies of magnons in these Néel-type domain walls can be expressed as $\Omega_k^{N\pm} = \sqrt{\omega_k(\omega_k - \omega_\perp + \frac{\omega_{D,k}}{k_x\lambda})} \pm \omega_{D,k}$, where $\omega_{D,k} = \pi\gamma D k_x/2M_S$ and $D$ is the DMI parameter. The DMI introduces a linear dependence of the magnon frequency on the wavevector, resulting in an asymmetric spin-wave dispersion. Consequently, a quasi-linear variation appears for $k > 0$, while a more pronounced quadratic behaviour is observed for $k < 0$. This asymmetry in the spin-wave dispersion leads to different spin-wave wavelengths at the same frequency for counter-propagating spin waves within the domain wall. The magnon can also propagate along Néel-type domain walls induced by the DMI, revealing that group velocities can exceed 1 km/s in the long-wavelength limit for specific propagation directions. This ability to guide magnons through curved paths is critical for the development of large-scale magnonic applications. Furthermore, the design principles behind the simplest logic component, a SW diode, are investigated by the chiral bound states of magnons within a magnetic DW influenced by the DMI. A diode, which permits unidirectional signal transmission, is a foundational element of logic architectures and underpins modern information systems. Realizing a magnonic diode is a critical step toward the practical implementation of magnonic devices and circuits.

The confinement of magnons has also been experimentally explored across various systems. Domain walls with nanometer-scale dimensions were engineered in a patterned NiFe thin film, enabling the observation of magnon propagation within the DW using a μ-BLS microscope, as shown in Fig. 2.17(b). [104] The SWs were found to be channeled within a width of approximately 40 nm, exhibiting resonance frequencies below 4 GHz. By applying a small magnetic field, the propagation path of these SWs could be shifted over distances of several micrometers. Additionally, a magnetic vortex pattern was employed on top of a NiFe thin film, resulting in the spontaneous formation of a magnetic DW along the length of the NiFe rectangle. [105] Using a BLS microscope, short-wavelength SWs are detected, with the wavelength as short as 80 nm and the frequencies reaching up to 15 GHz, propagating along the domain wall nanochannel under zero magnetic bias field. These exchange-dominated SWs, traveling along the DW, are crucial for the development of nanoscale magnonic devices and circuits that operate at high frequencies. Building on this approach, magnon propagation was further demonstrated in a 2D vortex network. [106] The magnons were observed to travel through nanochannels formed by Bloch- and Néel-type domain walls, with a propagation decay length extending several micrometers. This technique opens opportunities for the design of low-energy, reprogrammable, and miniaturized magnonic devices.

STXM is an advanced imaging technique that integrates the high spatial resolution of X-ray microscopy with the capability to investigate specific material properties, such as magnetic textures and spin dynamics. It has been employed to study the emission and propagation of 1D and 2D SWs with nanoscale wavelengths in anisotropic spin textures confined within $Ni_{81}Fe_{19}/Ru/Co_{40}Fe_{40}B_{20}$ multilayers. [107] These multilayers are structured into disk- and square-shaped elements with lateral dimensions spanning several micrometers. Moreover, through spin-texture nanoengineering, spatially tailored nanoscale SW wavefronts can be generated and manipulated. [108] These confined SWs in different spin textures exhibit fascinating properties, including resilience to back reflection, which naturally emerges from the nonreciprocity of SWs in synthetic antiferromagnets, as observed using STXM.

Another mechanism for the interaction between magnons and domain walls involves using magnetic domains to excite SWs. Direct observation of MSSWs, generated by magnetic domain walls in a CoFeB stripe with varying domain-wall configurations, was achieved using magneto-optical Kerr effect (MOKE) microscopy, as shown in Fig. 2.17(c). [109] At an excitation frequency of 9 GHz, the SWs are directly visualized. Unlike elastic waves, the propagation direction of these SWs aligns with the magnetization direction, in both the wide domain state and the canted domain state. However, the resolution of MOKE microscopy is limited to detecting SWs with wavelengths above the micrometer scale, specifically magnetostatic waves. To further explore SW excitation by confined domain walls, advanced techniques are required.

Laser-free ultrafast Lorentz transmission electron microscopy (LTEM), equipped with a microwave-driven electron pulser, enables high spatiotemporal resolution and provides a powerful platform for directly imaging the interactions between magnetic domain walls and SWs. Compared to conventional LTEM, the system developed by Liu *et al*. incorporates a laser-free ultrafast RF electron pulser, which preserves the intrinsic beam quality of the field-emission electron source and maintains the original performance of the TEM. [110] By implementing a synchronized RF excitation scheme via a custom-designed sample holder and employing a direct electron detector optimized for capturing weak signals and minimizing noise, the authors successfully visualized SW dynamics. Fig. 2.17(d) presents the excitation of SWs by domain walls in the dipolar-exchange regime. SWs with wavelengths as short as 200 nm are directly observed in a NiFe stripe hosting spin vortices and anti-vortices. Furthermore, DW oscillations are captured, accompanied by the spontaneous emission of SWs. In the space-time intensity maps shown in Fig. 2.17(d), the DW positions oscillate as SWs are emitted on both sides and propagate outward. Although the emitted SWs on either side of the DW share the same frequency as the wall oscillation, they exhibit a phase difference of 180°. This observation provides a microscopic picture of the SW emission and propagation driven by the domain wall.

Compared to ferromagnetic magnons, antiferromagnetic (AFM) magnons offer distinct advantages, as they are minimally affected by stray fields from the environment and exhibit higher velocities. However, their operational frequencies, which can reach the terahertz range, pose challenges for applications in microwave technologies. In particular, van der Waals antiferromagnets, characterized by weak interlayer exchange coupling compared to conventional antiferromagnets, are promising for studying GHz SW dynamics. Due to spin-charge coupling in easy-axis van der Waals antiferromagnets, confined AFM magnons within domain walls can be activated by voltage-induced torques, offering a pathway for low-dissipation, nanoscale excitation. [111] Fig. 2.17(e) presents a schematic of domain-wall magnon excitation in van der Waals antiferromagnets. A bilayer easy-axis van der Waals antiferromagnet is considered, with an antiferromagnetic DW positioned at the center of the simulated structure. An electric gate is used to excite confined magnons within the domain wall. Two independent Goldstone modes, arising from the broken continuous symmetries of the AFM domain-wall ground state, manifest as string-like and superfluid-like modes. These are described by the canonically conjugate pairs $(y_+, \phi_-)$ and $(y_-, \phi_+)$ respectively, where $\phi_\pm \equiv \phi_1 \pm \phi_2$ represents the angle in the plane orthogonal to the easy axis, and $y_\pm \equiv y_1 \pm y_2$ denotes the domain-wall position, as illustrated in the schematic. When an external magnetic field is applied along the $y$-direction, the string-like mode is excited, while the external field is in the $z$-direction, the superfluid-like mode is excited, as evidenced by the dynamic response of the domain walls shown in Fig. 2.17(e). These confined AFM magnons in domain walls are promising candidates for nanoscale routing of spin information, with compatible to the microwave technology.

The AFM insulator hematite ($\alpha$-Fe$_2$O$_3$) is particularly notable for its low Gilbert damping, which results in the long magnon decay length. Hematite exhibits a Morin transition at a temperature around 260 K. Above this Morin temperature, hematite behaves as an easy-plane antiferromagnet, displaying a quasi-ferromagnetic resonance in the tens of GHz range, driven by the presence of the DMI. In contrast, below the Morin temperature, hematite transitions into an easy-axis antiferromagnet, with its antiferromagnetic resonance modes shifting up to the sub-THz regime. Surprisingly, a low-frequency magnon mode is observed in its easy-axis phase, with its frequency approaching nearly zero, as detected by all-electrical SW spectroscopy. [112] The theoretical model identifies this low-frequency magnon mode as magnons confined within AFM domain walls. By considering the weak basal-plane anisotropy, the model predicts two distinct branches of AFM domain-wall magnon modes, which align closely with the experimental findings. The eigenfrequencies of these magnon bound states are:

$$\omega_1 = \gamma\mu_0 \sqrt{\frac{\pi}{4} H_D H}, \omega_2 = \gamma\mu_0 \sqrt{H_{\text{ex}} H_{\text{a}} + H^2 + \frac{\pi}{4} H_D H} \tag{6}$$

where $H_D$ is the DMI field, $H$ is the external magnetic field, $H_{\text{ex}}$ is the exchange field and $H_{\text{a}}$ is the anisotropy field. The two magnons modes in AFM domain walls are directly observed at $T = 250$ K. Furthermore, magnon propagation along AFM domain walls has been observed in phase-sensitive SW transmission spectra. This detection of magnons confined within AFM domain walls may contribute to the development of AFM magnon-based DW circuits, which exhibit higher resilience to external fluctuations compared to their ferromagnetic counterparts.

2.4.2 *Magnon trapping in magnetic vortices and skyrmions*

Magnetic vortices [113,114] and skyrmions [115,116] are topological spin textures characterized by their whirling magnetization configurations. A magnetic vortex features a magnetization that curls around its center, with the core magnetization pointing out-of-plane, resulting in a net magnetization close to zero. In contrast, a magnetic skyrmion is a topological structure composed of a skyrmion core, an outer domain, and a transition region that separates the two. Skyrmions typically emerge in chiral magnets with strong DMI. While magnetic vortices are commonly investigated in patterned magnetic systems, skyrmions can form naturally due to inversion symmetry breaking in magnetic materials. This intrinsic stability has attracted significant interest in their potential applications for next-generation spintronic data storage. Since magnetic vortices typically form in patterned magnetic structures with dimensions ranging from nanometers to micrometers, magnons are inherently confined by the geometry of these structures. Here, we review several recent progresses in utilizing these confined magnons within magnetic vortices, including their application in short-wavelength SW emission, nonlinear whispering gallery magnons and related reservoir computing, and vortex-based magnetic metasurfaces.

In an antiferromagnetically coupled trilayer composed of Co/Ru/Ni$_{81}$Fe$_{19}$, the rotational dynamics of naturally formed nanosized magnetic vortex cores can be driven in the absence of magnetic bias fields. [117] The nanoscale dimensions of the vortex cores facilitate the excitation of exchange SWs with short wavelengths. Meanwhile, the rotation of the driven vortex cores, combined with the relatively linear SW dispersion inherent to the antiferromagnetically coupled trilayer, enables the wavelength of the excited SWs to be tuned linearly by adjustments to the driving frequency. Furthermore, ultrashort dipole-exchange SWs, with wavelengths as small as 80 nm, are observed in a ferromagnetic single-layer system, where they are coherently excited by the driven dynamics of a magnetic vortex core. [118] These magnons display a heterosymmetric mode profile, characterized by regions exhibiting anti-Larmor precession and purely linear magnetic oscillation. The exchange magnons confined within magnetic vortices provide a platform for tunable sources of coherent ultrashort SW in the GHz frequency range. Building on those foundational studies, efficient emission of SWs from magnetic vortex pairs in a synthetic antiferromagnet is achieved by applying a microwave current, [119] as shown in Fig. 2.18(a). The current, with a frequency of $f = 1.1$ GHz and a density of $j = 2.9 \times 10^{10}$ A/m$^2$ generates a clear SW pattern with an average wavelength of approximately 300 nm originating from the vortex core. Micromagnetic simulations reveal that the current-driven Oersted field is the primary excitation mechanism, surpassing spin-transfer torques, with internal Oersted fields exhibiting SW excitation efficiency orders of magnitude higher than conventional stripline antennas. The direction of magnon propagation can be controlled by increasing the excitation amplitude, which alters the underlying magnetization profile through additional anisotropy.

Nonlinear magnons arise in magnetic systems exhibiting nonproportional responses, facilitating phenomena such as wave mixing, nonlinear magnon-magnon scattering, and parametric amplification. In a NiFe disk in the vortex state, whispering gallery magnons with high wavevectors are generated through nonlinear three-magnon scattering. [55] Microwave currents are injected via an Ω-shaped gold antenna encircling the vortex. Observations using BLS microscopy reveal that these nonlinear magnons are strongly localized at the perimeter of the vortex, with virtually no amplitude across an extended region surrounding the vortex core. The underlying three-magnon scattering processes are highly tunable regarding the frequency and spatial distribution of the split modes, and the azimuthal mode numbers are determined by the nonlinear magnon frequencies. Extensive research has been conducted on confined nonlinear magnon modes in magnetic disks, with applications in neuromorphic computing and the quantum regime. For example, pattern recognition is realized through magnon modes in a confined magnetic vortex, where radial mode excitations trigger a three-magnon process, as shown in Fig. 2.18(b). [120] This process excites distinct azimuthal modes with amplitudes corresponding to input sequences, achieving recognition rates up to 99.4%. Such high accuracy holds potential for reservoir computing using the confined nonlinear magnons. Furthermore, a quantum transducer harnesses magnon nonlinearities in a ferromagnetic microdisk to mediate microwave interactions with spin qubits in silicon carbide (SiC), which leverages parametric magnon effects to downconvert microwave frequencies and address off-resonant spin qubit ensembles. [121] Highly confined magnon stray fields can drive these qubits at room temperature. This approach offers a wafer-compatible platform, that enriches quantum engineering with magnon nonlinear physics.

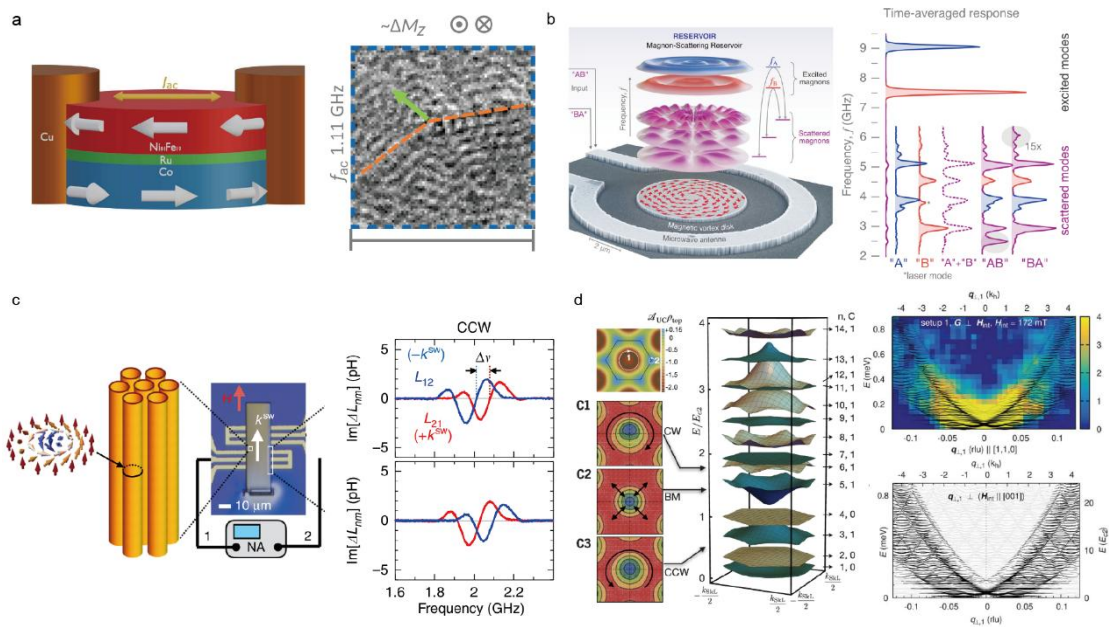

**Fig. 2.18** Illustration of magnon confinement in magnetic vortices and skyrmions. (a) Schematic of a Co/Ru/Ni$_{81}$Fe$_{19}$ microdisk with adjacent, partially overlapping copper leads. A normalized TR-STXM snapshot depicts SW dynamics in response to an AC field. Adapted with permission from [119]. (b) Magnon-scattering reservoir (MSR) based on a magnetic vortex disk. Driving the MSR with two different, temporally overlapping microwave pulses "A" and "B" leads to two separated nonlinear scattered magnon modes. The right panel shows that the experimentally measured output spectra, integrated over time, vary depending on the temporal order of input pulses. Different colors represent contributions from the two input signals: blue peaks arise from input "A" only, red peaks from input "B," and purple peaks from cross-stimulation. Adapted with permission from [120]. (c) Magnon confinement in a skyrmion string, demonstrating counter-clockwise magnon propagation with counterpropagating directions at frequencies ranging from 1.6 to 2.2 GHz (taken from [131]). (d) Calculated magnon bands in the skyrmion lattice of MnSi within the first BZ in units. Magnon spectra of MnSi for momentum transfers perpendicular to the skyrmion lattice tubes are presented experimentally and theoretically. Adapted with permission from [132].

Metasurfaces are planar nanostructured interfaces, that enable precise control over electromagnetic waves by manipulating their amplitude, phase, and polarization. By employing a magnetic metasurface composed of nanodisks in topologically protected vortex states, [122] a reconfigurable spectral filter operating in the GHz range is realized, where confined magnons in the nanodisks couple with those in an adjacent ferromagnetic layer, producing an anticrossing behavior that enhances the efficiency of the spectral filter to 98.5%. Despite covering only 15% of the microwave antenna, this reconfigurable metasurface significantly advances functionality for magnonics and on-chip microwave devices.

Magnon modes in magnetic skyrmions, driven by high-frequency inputs, have been extensively investigated through both theoretical [123,124,125] and experimental approaches. [126,127,128] Typically, three distinct types of magnon modes are observed in skyrmion dynamics: clockwise, counterclockwise, and breathing modes, which generally exhibit frequencies below the ferromagnetic resonance mode, i.e., beneath the magnon gap. In-plane ac fields excite the clockwise and counterclockwise modes, while out-of-plane ac fields induce the breathing mode.

These magnon modes within skyrmion lattices were first detected in the multiferroic material Cu$_2$OSeO$_3$, where microwave input from a coplanar waveguide revealed two fundamental excitations: a counterclockwise circulating mode at 1 GHz, with magnetic field polarization parallel to the skyrmion plane, and a breathing mode at 1.5 GHz, with polarization perpendicular to the [111] plane. Furthermore, magnon modes in helical and conical phases are also observed in Cu$_2$OSeO$_3$. [129] Schwarze *et al*. have explored universal helimagnon and skyrmion excitations across metallic, semiconducting, and insulating chiral magnets, including MnSi, Fe$_{1-x}$Co$_x$Si, and Cu$_2$OSeO$_3$, using all-electrical spectroscopy to probe their collective spin excitations. [127] By considering dipolar interactions, a precise quantitative model spanning the entire magnetic phase diagrams is achieved with just two material-specific parameters, namely quantifying chiral and critical field energies. This universal behavior supports skyrmion-based applications leveraging their resonant responses in chiral magnets, offering benefits such as tunable electrical conductivity and effective integration into electronic systems.

The magnon spectra of Cu$_2$OSeO$_3$, which exhibit distinct resonance modes across various magnetic phases, provide a wide range of reservoir properties and computing performance that can be tuned by external magnetic fields and temperature. [130] Magnetic field cycling is employed to input data, with the magnon spectra measured at each step enabling efficient high-dimensional mapping. The thermodynamically metastable skyrmion phase demonstrates robust memory capabilities due to field-driven skyrmion nucleation, excelling in prediction tasks. This phase tunable nature in Cu$_2$OSeO$_3$, offering strong performance across diverse tasks, allows on-demand access to varied computational reservoir responses for neuromorphic computing. The confined magnon modes mentioned earlier are restricted to the 2D plane of skyrmions. In 3D Cu$_2$OSeO$_3$, skyrmions form tube-like structures, within which magnons are confined along the skyrmion string, as shown in Fig. 2.18(c). [131] The propagation of these magnon modes in the skyrmion string exhibits directional nonreciprocity, with the degree of nonreciprocity, group velocity, and decay length varying significantly depending on the specific excitation modes. These confined magnon modes demonstrate long-distance propagation exceeding 50 µm, highlighting the exceptional long-range order of the skyrmion-string structure. Such nonreciprocal behavior is enterely consistent with the presence of a texture-induced toroidal moment along the skyrmion axis. [88]

The motion of an electrically charged particle perpendicular to a magnetic field generates a Lorentz force, resulting in orbital motion at discrete energy levels known as Landau levels. Similarly, when a spin adiabatically aligns with the local magnetization, the geometrical properties of a smooth magnetization texture produce an emergent magnetic field and a corresponding emergent Lorentz force. In a skyrmion lattice, this phenomenon induces an analogous cyclotron motion of magnons, leading to the formation of Landau levels and topological magnon bands, as illustrated in Fig. 2.18 (c). [132] Polarized inelastic neutron scattering was used to investigate magnon confinement within a lattice of skyrmion tubes in MnSi. For wavevectors perpendicular to the skyrmion tubes, the magnon spectra exhibit finely spaced emergent Landau levels, consistent with a fictitious magnetic field arising from the nontrivial topological winding of the skyrmion lattice. These findings confirm a topological magnon band structure in reciprocal space, driven by the confinement of magnons within the nontrivial real-space topology of the magnetic order.

The confined magnons within skyrmion tubes can enable the control of skyrmion dynamics through the transfer of angular momentum between magnons and the underlying spin textures. Jiang et al. revealed that the orbital angular momentum (OAM) states of magnons are eigenstates of the OAM operator in confined geometries. [133] They theoretically proposed the concept of a "magnetic tweezer", inspired by optical tweezers, in which twisted magnons carrying quantized OAM exert a torque on the skyrmion core, leading to a controllable gyration of skyrmions in chiral magnetic nanodisks. This effect originates from the transfer of magnonic OAM to the topological charge of the skyrmion, offering a purely magnonic approach to manipulate spin textures without involving charge currents.

## 3. Material engineering approaches

Besides magnon confinement arising from nonuniform magnetic textures, another path to manipulate the magnon dispersion in uniformly magnetized systems is to break symmetry through material engineering. The present section discusses various nanomaterial designs that enable manipulation of SW dispersion to produce nonreciprocal responses, unidirectional propagation, and even flat magnonic bands, where the magnonic group velocity is significantly reduced, leading to SW localization and trapping. The section starts with an overview of magnonic architectures incorporating field inhomogeneities, such as graded materials, magnetic bilayers and multilayers. Then, magnetic metamaterials in the form of magnonic crystals and superlattices are introduced, followed by a discussion of chiral magnonic crystals created with a periodic interfacial DMI.

*3.1 Magnetic field inhomogeneity and gradient*
SWs in a uniformly magnetized thick film ($\mathbf{M} = M_s \hat{z}$) exhibit DE surface modes that can localize either on the top or bottom surface of a film, depending on the wavevector direction. These modes present amplitude nonreciprocity since the magnon excitations profile depends on the normal coordinate and concentrates at one of the surfaces, which has been observed in thick films since, for ultrathin films ($d \sim \lambda_{\text{ex}}$, with $\lambda_{\text{ex}}$ the exchange length), the excitation profile is practically uniform across the normal. Therefore, thin films with thicknesses larger than a few tens of nanometers exhibit this inhomogeneity in the amplitude of the surface modes. [118] For thicker films, another group of modes termed perpendicular standing spin wave (PSSW) modes can be excited at GHz frequencies. These standing modes are characterized by a normal quantized wavevector that accommodates the thickness according to $\mathbf{k}_\perp = (n\pi/d)\hat{y}$. Then, the magnon dispersion in the DE configuration for an in-plane wavevector ($\mathbf{k}_\parallel = k\hat{x}$) shows the surface DE mode, which can hybridize with the PSSW modes, as shown in Fig. 3.1(b). [134,135,136] When this happens, the PSSW becomes dispersive if the film is thick enough. Otherwise, for ultrathin films, the PSSW modes occur at much higher frequencies and cannot mix with the surface mode. In any case, flat regions in the magnon spectrum can be observed in specific wavevector ranges, which depend on the thickness and material parameters, allowing localization of SWs in the (positive and negative) $x$ direction [see Fig. 3.1(c-f)]. Nevertheless, breaking symmetry with material engineering approaches is an important strategy that can enable unidirectional SW localization, as discussed in the following.

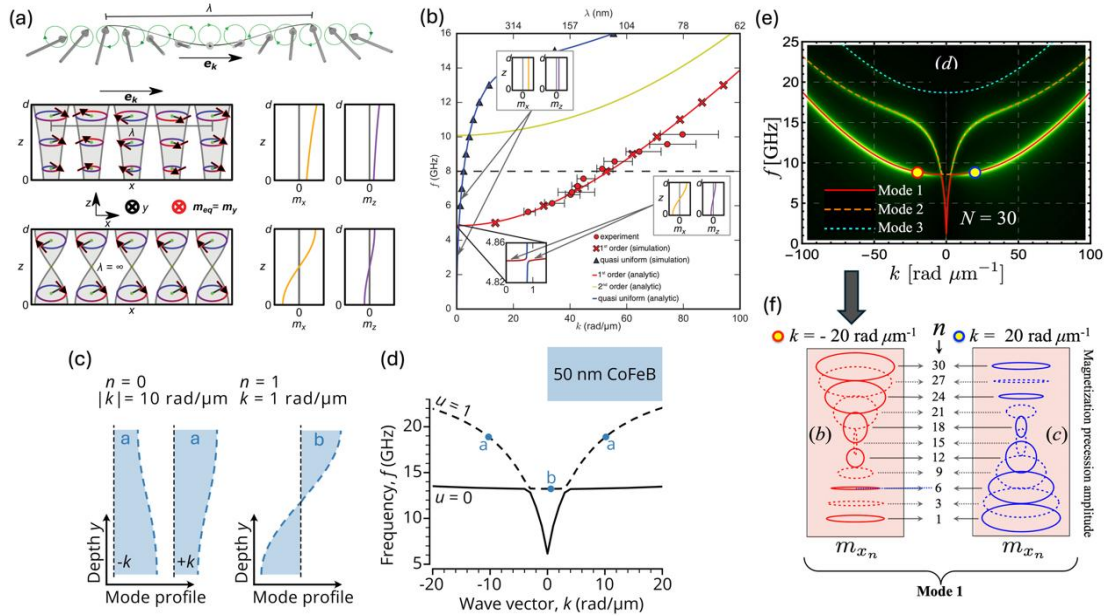

**Fig. 3.1** SW dispersion in thick films with uniform $M_\text{s}$. In (a-b), the dispersion curves were measured, calculated, and simulated, showing the hybridization of the DE with the PSSW modes. (c-d) Reciprocal frequency dispersion and depth-dependent mode profiles, showcasing the amplitude asymmetry of the $n = 0$ mode. (e) illustrates the calculated (lines) and simulated dispersions (color code), where brighter colors indicate a stronger response. In (f), the spatial mode profiles along the thickness are illustrated for two opposing wave vectors. (a-b) Reproduced with permission from [118]. (c-d) Reproduced from [136]. (e-f) Reproduced with permission from [134].

*3.2 Graded magnetic fields in nanometric dimensions*

Spatial manipulation of physical properties is a principle that spans scientific disciplines, particularly those concerned with the transmission of elementary excitations in matter. Optical fibers, for instance, are designed to vary the dielectric constant and the effective refractive index, which decreases gradually from the core center to the periphery, guiding the light more efficiently along the fiber axis by minimizing its dispersion. [137] Periodic spatial variations in refractive index give rise to the development of photonic crystals. [138,139] On the other hand, by adjusting material parameters in a phononic crystal, a periodic acoustic refractive index profile is established, with a graded material design enabling precise control over wave propagation and the generation of unidirectional acoustic waves. [140,141,142] Graded nanomagnets are an engaging class of materials characterized by spatially changing magnetic properties, which can be engineered in a controlled manner to accomplish specific responses, making these systems highly versatile for applications in data storage. [143, 144] The gradient in a magnetic property can be achieved by changing composition or structure, leading to different magnetic behavior not usually observed in amorphous materials. [145] Spatial modulations can be tuned at the nanoscale, enabling the design of nanomaterials with gradual transitions in magnetic properties, such as saturation magnetization, [146,147,148,149,150] exchange constant [151,152,153] magnetic anisotropies, [154,155,156,157,158,159] and DMI strength. [160,161,162,163,164] Exchange-spring media are another class of metamaterials that exhibit magnetic modulation across the normal, where hard and soft layers are strongly exchange-coupled. [165] In magnonics, magnetic graduation entails the manipulation and channeling of SWs. [10,11] The precision in controlling magnetic properties enables the development of graded-index magnonic waveguides for steering SWs, as well as nanodevices with tunable magnetic textures and DW pinning landscapes. [166,167,168] The control over SW propagation can be accomplished via various techniques, including modifying the saturation magnetization, exchange coupling, anisotropies, or the DMI strength. Variations in the magnetic properties of the material affect the internal field distribution, enabling localized propagation of SWs in regions with diminished internal fields. [169,170] Additionally, by modifying the magnetic properties in the normal direction of a thick ferromagnetic film, SW frequency nonreciprocity is induced in the DE configuration due to the symmetry breaking caused by dynamic dipolar fields. [134] The synthesis of graded magnetic properties can be reached through advanced fabrication techniques, such as layer-by-layer deposition with varying compositions [145] or ion implantation. [171,172,173] For instance, vertical graduation across the thickness of a thin film can be realized in epitaxial compositionally graded alloy films by co-sputtering. [149,174] Lithographic masks have been employed to manipulate the structures locally, enabling the creation of lateral gradients within single mesoscopic or nanoscopic structures. [175] Such techniques facilitate the design of complex architectures with precise control over the spatial distribution of magnetic properties.

Developing a SW theory for graded magnetic films is a challenging problem due to the characteristic inhomogeneous field landscapes. It requires considering proper boundary conditions at the surfaces, which can be influenced by magnetic graduation. Additionally, the dipolar coupling depends strongly on the normal coordinate, which consequently varies with each grading profile. Hence, semi-analytical methods have been developed to model 3D nanomagnets, [76] in which the Landau-Lifshitz dynamics is mapped to an eigenvalue problem formulated analytically and solved numerically. [134] A helpful tool is the dynamic matrix method (DMM), [176,177,178] in which the nanomagnet splits into multiple sub-elements connected through dipolar and exchange interactions. [179,180,181] While this method is primarily intended for solving coupled individual layers, it can also be applied to thick films with continuously varying magnetic properties, which can be attained by splitting the film into a computationally manageable number of sublayers. [134,182]

The dynamics of each element in the DMM is determined by the Landau-Lifshitz (LL) equation, which can be mapped into an eigenvalue problem, $i\omega \mathbf{m}(\mathbf{r}) = \gamma\mu_0 \widetilde{\mathbf{A}}\,\mathbf{m}(\mathbf{r})$, where the eigenvalues of matrix $\widetilde{\mathbf{A}}$ are proportional to the frequencies of the SW modes, and the eigenvectors correspond to the dynamic magnetization components. The dynamic matrix can be written as

$$\begin{pmatrix} A^{x_1 x_1} & A^{x_1 y_1} & \cdots & A^{x_1 x_N} & A^{x_1 y_N} \\ A^{y_1 x_1} & A^{y_1 y_1} & \cdots & A^{y_1 x_N} & A^{y_1 y_N} \\ \vdots & \vdots & \ddots & \vdots & \vdots \\ A^{x_N x_1} & A^{x_N y_1} & \cdots & A^{x_N x_N} & A^{x_N y_N} \\ A^{y_N x_1} & A^{y_N y_1} & \cdots & A^{y_N x_N} & A^{y_N y_N} \end{pmatrix}, \quad (7)$$

where the coordinates $x_n$ and $y_n$ are perpendicular to the equilibrium axis $z_n$ axis of the $n$-th sublayer. Number $N$ is chosen based on convergence, in such a way that the eigenvalues are truncated at a given $N$. It is worth mentioning that this method has also been applied to the thick curvilinear shells discussed in Sec. 2.3.1. [86,87] For the graduated films, the matrix elements are obtained from the effective fields and account for interlayer dipolar and exchange interactions. In a simple case of a thick layer without anisotropy, these matrix elements are [134]

$$A^{x_n x_p} = -\left(1 - \delta_p^n\right) ik \frac{2M_{S_n}}{|k|^2 d} \sinh^2\left(\frac{|k|d}{2}\right) e^{-|k||n-p|d} \operatorname{sgn}(n-p)$$

$$A^{y_n y_p} = -\left(1 - \delta_p^n\right) ik \frac{2M_{S_n}}{|k|^2 d} \sinh^2\left(\frac{|k|d}{2}\right) e^{-|k||n-p|d} \operatorname{sgn}(n-p)$$

$$A^{x_n y_p} = \left(1 - \delta_p^n\right) \frac{2M_{S_n}}{|k|d} \sinh^2\left(\frac{|k|d}{2}\right) e^{-|k||n-p|d} + \frac{J}{dM_{S_n}} \left(\delta_p^{n-1} + \delta_p^{n+1}\right)$$
$$- \delta_p^n \left( H_0 + \frac{J}{dM_{S_n}} \left(\delta_p^{n-1} + \delta_p^{n+1}\right) + M_{S_n} \lambda_{ex}^2 k^2 + \frac{2M_{S_n}}{|k|d} \sinh\left(\frac{|k|d}{2}\right) e^{-|k|\frac{d}{2}} \right)$$

$$A^{x_n y_p} = \left(1 - \delta_p^n\right) \frac{2M_{S_n}}{|k|d} \sinh^2\left(\frac{|k|d}{2}\right) e^{-|k||n-p|d} - \frac{J}{dM_{S_n}} \left(\delta_p^{n-1} + \delta_p^{n+1}\right)$$
$$+ \delta_p^n \left( H_0 + \frac{J}{dM_{S_n}} \left(\delta_p^{n-1} + \delta_p^{n+1}\right) + M_{S_n} \lambda_{ex}^2 k^2 + M_{S_n} - \frac{2M_{S_n}}{|k|d} \sinh\left(\frac{|k|d}{2}\right) e^{-|k|\frac{d}{2}} \right).$$

Here, $\lambda_{ex}$ is the exchange length, $d$ is the thickness of the magnetic sublayer, and $H_0$ is the magnitude of the external field. Note that the terms porportinal to $\delta_p^n$ and $(1 - \delta_p^n)$ come from the intralayer and interlayer interaction energies, respectively, with $\delta_p^n$ being the Kronecker delta function (1 if $p = n$ and 0 if $p \neq n$).

Gradual or abrupt variations along the thickness of the saturation magnetization, [134] anisotropies, [159] or exchange constant, can be included in this approach, as these parameters can be discretized per layer, and, thus, labeled by an index $n$, allowing for the modeling of arbitrary grading profiles. For simplicity, free boundary conditions have been assumed at the outer surfaces of the film. In cases involving surface interactions, such as surface anisotropies or interfacial DMI, these effects can be incorporated into the model as averaged bulk energies present only in the magnetic sublayers adjacent to the top and/or bottom surfaces. For sufficiently thin sublayers, this approach is practically equivalent to treating these effects as boundary conditions. [178]

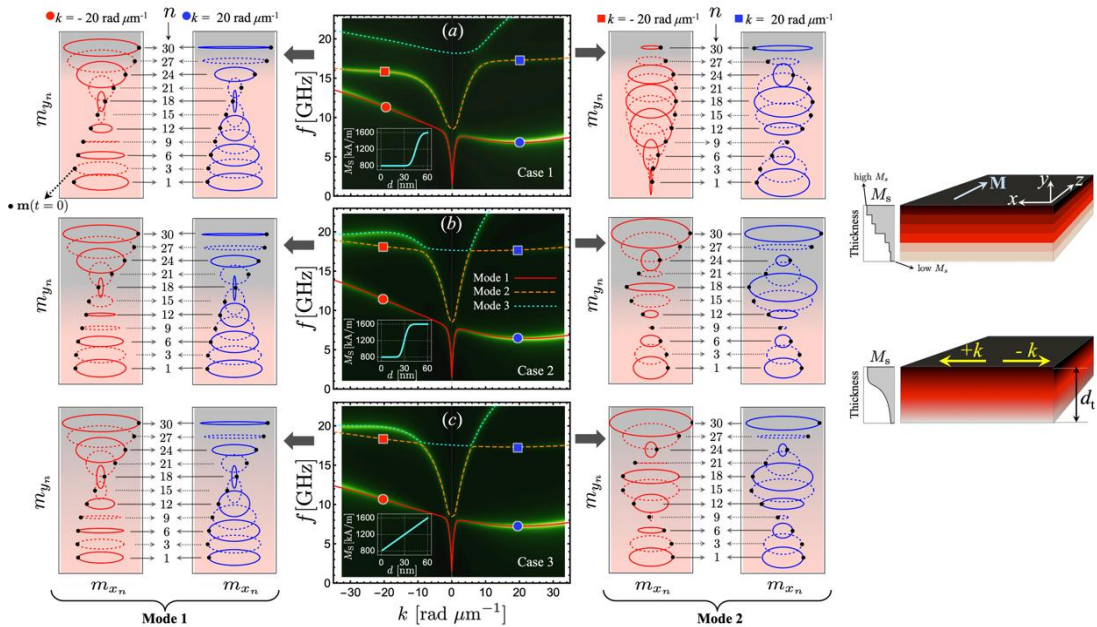

**Fig. 3.2** (a-c) SW dispersion in magnetization-graded films with thickness $d_t = 60$ nm and different spatial profiles $M_s(y)$ along the normal direction as shown in the insets. (a) Case 1 represents an asymmetric gradual profile, while (b) Case 2 depicts a symmetric profile. (c) Case 3 illustrates a linear magnetization profile. The orbits described by the normal component of the dynamic magnetization along the normal direction are illustrated for the first (left panel) and second (right panel) perpendicular standing SW modes for selected wave vectors. In all cases, the color code represents the numerical simulations, with brighter colors indicating a stronger response. The lines correspond to the theoretical results. Reproduced with permission from [134].

A first model case of a graded magnetic system is a layered film, where the saturation magnetization varies along the normal coordinate of the film. Fig. 3.2 depicts the magnon spectra in the DEconfiguration for a 60-nm-thick film with different graded profiles $M_s(y)$ where $M_s \in$ [800,1600] kA/m according to the insets in Fig. 3.2(a-c). The graded films are described with $N = 30$ layers since for larger values, the low-frequency modes in the frequency range of interest do not change considerably, indicating that a graded film with a continuous variation of $M_s$ is well represented. In Case 1 [Fig. 3.2(a)] $M_s$ changes near the top film surface, turning the magnon dispersion asymmetric for the low-frequency modes. Case 2 [Fig. 3.2(b)] represents a symmetric graduation profile in $M_s$, where the first mode is slightly more asymmetric than the corresponding Case 1. However, the second mode is almost symmetric concerning wave-vector inversion for $|k| > 4$ rad/$\mu$m, unlike the second mode in Case 1. For Case 3, $M_s$ changes linearly along the film thickness, as shown in the inset of Fig. 3.2(c). Case 3 is similar to Case 2, where the explanation is related to the localization of the precessional magnetization amplitude, as in both cases, the maxima occur at the bottom and top surfaces of the film, where $M_s$ is the same for Case 2 and Case 3 (not shown). [134] The DMM calculations are compared with micromagnetic simulations (see color code in Fig. 3.2), where an excellent agreement is observed. In the simulations, it is seen that the frequency nonreciprocity is accompanied by an asymmetry in the magnetization precession amplitudes, as indicated by the color code. Therefore, the output signal of a given measurement, sensitive to the dynamic magnetization amplitude, is expected to be asymmetric concerning wave-vector inversion. Spatial symmetry is broken due to the inhomogeneous internal field imposed by the graded magnetization profiles. As in the case of a bare film, the dispersion in all cases exhibits well-defined plateaus where the group velocity is reduced to zero, allowing for the localization and trapping of the SW excitations. Nonetheless, for the graded film with an asymmetric $M_s$-profile, the frequency minimum occurs only for one wavevector orientation, giving a directionally dependent localization (or propagation) of the spin waves.

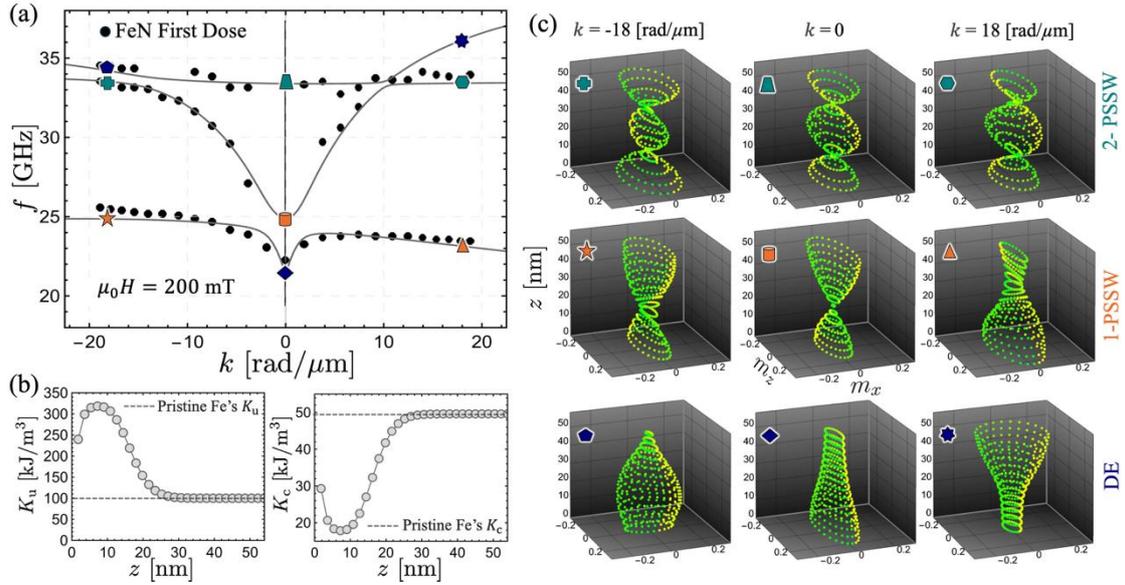

**Fig. 3.3** Magnon dispersion in nitrogen-irradiated iron films. (a) BLS measurements (symbols) and dynamic-matrix calculations (solid lines) showing a nonreciprocal dispersion with marked flat regions. (b) Graduation profiles for the uniaxial (left) and cubic (right) magnetic anisotropies along the film`s normal were obtained to by fitting the theory to the experimental data. The dashed lines correspond to the anisotropy constants for pristine Fe. (c) Layer-resolved precessional orbits of the SW modes calculated at $k = -18, 0$, and $+18$ rad/$\mu$m. Reproduced with permission from [159].

The variation in saturation magnetization is just one example of how SW propagation can be controlled. Although there is currently no experimental evidence of SW nonreciprocity or localization for arbitrary graduation of the saturation magnetization, particular cases have been reported. These include films with different surface anisotropies at the top and bottom, [183,184] as well as parallel magnetized bilayers of different materials and saturation magnetizations, [136,185,186,187,188,189] and antiparallel magnetized bilayers, referred to as synthetic antiferromagnets, [190,191,192,193,194,195,196] which are discussed in the next section.

In irradiated samples, for instance, changes in anisotropy and the exchange constant can also induce inhomogeneous internal fields and asymmetries in the SW dispersion, together with flat regions on the magnon spectrum. A gradual modification of the magnetic anisotropy in irradiated FeN films prepared by low-dose nitrogen implantation of Fe epitaxial thin films has been studied by Christienne *et al.* [159]. Combining ion-implantation techniques, BLS experiments, and the dynamic matrix method, it was shown that nitrogen implantation induces a graded profile in the in-plane and perpendicular anisotropies across the film thickness. Similarly to graded-$M_s$ materials, the graduation in magnetic anisotropy produces a significant modification of the magnon spatial localization and creates a frequency asymmetry in the SW dispersion, as shown in Fig. 3.3. Interestingly, flat regions in the spectrum can be tailored with the film thickness and implantation doses, which are asymmetric respect to wavevector inversion, in agreement with calculations. [159]

To analyze the SW features in different types of graded systems, the SW nonreciprocity induced by magnetization graduation can be studied through the calculation of the toroidal moment ($\boldsymbol{\tau}$), [197, 198] since, if the condition $\boldsymbol{\tau} \cdot \mathbf{k} \neq 0$ is fulfilled, the SWs are nonreciprocal in that direction. [86] For instance, if a thin film with equilibrium magnetization $\mathbf{M}_0$ is graduated along a given direction $\mathbf{g}$, the toroidal moment is proportional to $(\mathbf{g} \times \mathbf{M}_0)$ and its magnitude depends on the symmetry of the graduation profile. [198] Therefore, magnon nonreciprocity is expected in a direction perpendicular to both $\mathbf{g}$ and $\mathbf{M}_0$. In the case described in Fig. 3.2, $\mathbf{g} = \hat{\mathbf{y}}$ (the normal to the film), and $\mathbf{M}_0 \parallel \hat{\mathbf{z}}$, so that nonreciprocity is expected along the *x*-direction [134] since the three vectors, $\mathbf{g}$, $\mathbf{M}_0$, and $\mathbf{k}$, form an orthogonal triad. [198]

Lateral graduation has also been recently explored, [170] where a graduated magnonic waveguide is proposed to channelize waves, and the nonreciprocity condition is not fulfilled since $\mathbf{g} \parallel \mathbf{M_0}$ and the SWs will behave reciprocally. Nevertheless, even though the magnon dispersion is reciprocal in such a case, the magnetic graduation profile can be engineered to mimic the behavior of optical fibers. [166] The graduation of the refraction index is a concept from photonics that has been theoretically explored to propose magnonic fibers based on the grading of a magnetic nanostripe, where different profiles for the saturation magnetization $M_s$ were studied. [170] In particular, for a potential-well-shaped profile in magnetization saturation along the stripe width [see Fig. 3.4(a)], the SWs propagate along the stripe axis, with larger amplitudes in different regions, depending on the frequency. The magnon dispersions are respectively shown in Fig. 3.4(d,f) for graded nanostripes with (d) 200 nm and (f) 1000 nm width. Micromagnetic simulations of SWs excited by an RF field at two selected frequencies (17 and 18 GHz) are shown in panels (e) and (g), corresponding to stripe widths of 200 and 1000 nm, respectively. Here, it is observed that the low-frequency mode is excited at 17 GHz, where SWs are channelized along the nanostrip center (low $M_s$ region), which according to the calculations is the unique mode that can be excited at this frequency since the edges (high $M_s$ region) can be excited at larger frequencies. At a slightly larger frequency, 18 GHz, both the central and edge modes are simultaneously excited, with the central mode having small wavelengths (larger wave vectors) in agreement with the calculated dispersion.

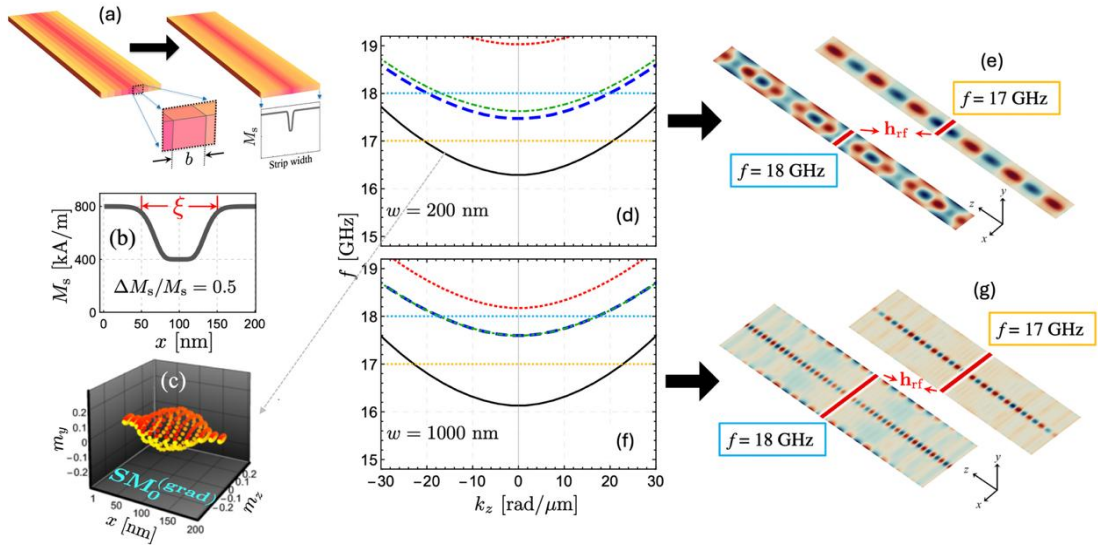

**Fig. 3.4** Magnon modes in laterally graded ferromagnetic nanostripes. Panel (a) illustrates a graded nanostripe, where a (b) potential-well-shaped profile in magnetization saturation was considered. Panel (d) depicts the magnon dispersion for a 200-nm-width graded nanostrip, where (c) the uniform mode is confined in the low-$M_s$ region, while (e) shows micromagnetic simulations at 17 and 18 GHz. Panel (f) shows the SW dispersion for a 1000-nm width graded stripe, and (g) depicts the micromagnetic simulations at 17 and 18 GHz. Adapted from Ref. [170].

*3.3 Formation of localized magnon states in magnetic bilayers.*
A good example of a class of systems hosting inhomogeneous internal magnetic fields is a bilayer stack comprising two different ferromagnetic materials with different $M_s$ and parallel magnetizations. [136,185,186,187,188,189] Here, the saturation magnetization changes abruptly from one layer to another, with a notable reduction in exchange interaction at the interface, giving a similar magnon dispersion to that obtained for graded magnetic materials along the normal. This is illustrated in Fig. 3.5, which shows the dispersion of parallelly magnetized bilayers, demonstrating the nonreciprocal phenomenon and the localization of SWs experimentally. In Fig. 3.5(a-d), the SW dispersions exhibit similar behavior to that shown in Fig. 3.2, which is not surprising, as the graded magnetization example is akin to a bilayer stack with different saturation magnetizations. [134,185]

The work of Grassi *et al*. uses BLS to measure the magnon dispersion in a CoFeB/NiFe bilayer in D configuration, envisioning the concept of a slow-wave-based nanomagnonic diode [188], as shown in Fig. 3.5(b), where the low-frequency mode becomes flat for a broad range of positive wave vectors ($k_z \gtrsim 7$ rad/μm), where SWs only propagate in the negative k-direction, giving rise to the magnonic diode concept. The work of Heins *et al*, also for CoFeB/NiFe bilayers, shows similar dispersions [see Fig. 3.5(d)] with marked flat regions allowing for SW localization. [136] The nonreciprocal dispersions and mode flattening shown in Fig. 3.5(a-d) are a consequence of the dynamic dipolar coupling between the layers. [182] For the case of parallelly magnetized bilayers, these effects are observed when the layers have different saturation magnetizations, which breaks the symmetry along the normal and creates a nonhomogeneous internal field landscape.

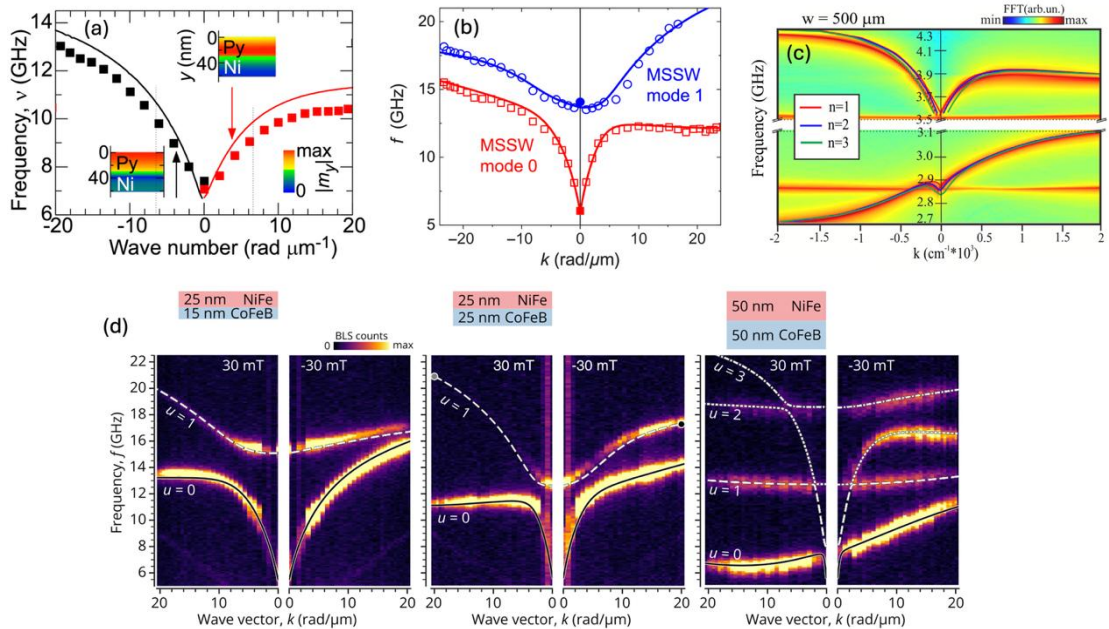

**Fig. 3.5** Magnon dispersions for in-plane magnetized bilayers with different magnetic materials. (a) measured (symbols) and calculated (lines) dispersion for a NiFe film covered with Ni stripes, showing a range of (positive) wave vectors where the dispersion becomes flat. Reproduced with permission [186]. (b) measured (symbols) and calculated (lines) dispersion for a CoFeB/NiFe bilayer, which shows a broader wavevector region with a negligible group velocity. Reproduced with permission from [188]. (c) Calculated dispersion for a bilayer composed of pure and doped YIG with different magnetizations. Reproduced with permission from [187]. (d) Calculated (lines) and measured (color code) dispersions for different CoFeB/NiFe bilayers. [136]

Another class of systems hosting inhomogeneous internal magnetic fields is the antiparallelly magnetized bilayer, colloquially know as a synthetic antiferromagnet (SAF). Such engineered magnetic systems comprising antiparallel-coupled layers separated by a nonmagnetic layer, are designed to emulate the properties of antiferromagnetic materials. [199,200] Unlike natural antiferromagnets, where the magnetic moments of adjacent atomic spins align antiparallel due to intrinsic exchange interactions, SAFs achieve this alignment through interlayer coupling mechanisms. This design enables a highly controlled and reconfigurable magnetic configuration, making SAFs ideal candidates for various technological applications in spintronics and magnonics. [182,182] The antiparallel alignment of the magnetizations is typically mediated by Ruderman-Kittel-Kasuya-Yosida (RKKY) exchange coupling, which oscillates with the thickness of the nanoscale nonmagnetic spacer. [201,202] By controlling the thickness and composition of the spacer, it is possible to achieve ferromagnetic or antiferromagnetic coupling between the FM layers. In the case of antiferromagnetic coupling, the system exhibits a net magnetic moment close to zero, depending on the commensurability of the magnetic characteristics of the coupled layers. The magnetic configuration in SAFs can be manipulated by external magnetic fields or spin-polarized currents, offering pathways to develop low-power and high-speed technologies such as magnetic random-access memory (MRAM). [203]

In magnonics, ferromagnetic bilayers have emerged as a promising platform for the propagation and manipulation of SWs. [204,205] Interlayer dipolar and exchange coupling significantly influence the magnonic dispersion of SAFs, [1900,191,192] modifying the phase and group velocities and leading to nonreciprocal mode localization [182,1888,193,194,195,196,206,207,208] The nonreciprocal feature enables the design of magnonic devices, such as isolators, circulators, diodes, and phase shifters, which are crucial for logic devices and signal processing.[205] The tunable characteristics of SAFs offer a powerful degree of freedom for tailoring the magnonic band structure, which can be controlled by varying the FM material, layer thickness, interlayer exchange coupling, magnetic anisotropies, or the external magnetic field. In the FMR regime, at zero wave vector, SAFs have been demonstrated to be ideal platforms for exploring magnon-magnon coupling at room temperature. [209,210,211,212,213] This coupling is manifested through the anti-crossing of the in-phase and out-of-phase resonance modes, where the frequency branches hybridize near the anti-crossing, creating a gap between the bands. This magnon-magnon coupling can also be enhanced by including an interlayer DMI. [213,214,215,216] Thus, SAFs provide an excellent alternative for exploring unconventional ultrastrong magnon-magnon coupling regimes, paving the way for further exploration of quantum phenomena. [213]

The nonreciprocity in SAFs has been experimentally studied by several groups, demonstrating its high tunability and extending beyond ultrathin magnetic films, in contrast to the nonreciprocity induced by interfacial DMI. BLS measurements [108,182,188,193,206] and time-resolved scanning transmission X-ray microscopy [182,196] have established the magnon nonreciprocity in SAFs, as shown in Fig. 3.6. These experiments have shown that a reconfigurable nonreciprocal frequency difference arises in interlayer exchange–coupled SAFs such as: (a) CoFeB(5.7 nm)/Ir(0.6 nm)/NiFe(6.7 nm), [182] (d) NdCo(64 nm)/Al(2.5-10 nm)/NiFe(10 nm), [206] and CoFeB(15 nm)/Ru(0.6 nm)/CoFeB(15 nm) [193] (not shown). Further, magnon nonreciprocity has been studied in the spin-flop transition region in symmetric and asymmetric CoFeB/Ru/CoFeB SAFs, where the magnetizations are neither fully antiparallel nor parallel, and a nonmonotonic dependence of the frequency nonreciprocity is observed, with a maximum frequency shift around the spin-flop critical point, where the modes change their phases. [207] Even nonreciprocal surface acoustic waves have been observed in CoFeB(20 nm)/Ru(0.46-0.77 nm)/CoFeB(20 nm) [217] and NiFe(20 nm)/Au(5 nm)/CoFeB(5 nm) [218] piezoelectric-ferromagnetic devices, with transmission relying on the propagation direction, which may occur when the two ferromagnetic layers are oriented antiparallel [217] or parallel. [218] Recently, SW asymmetry has been reported in CoFeB/Ru/CoFeB spin-valve (SAF) devices, where a considerable frequency asymmetry enables the unidirectional transfer of acoustical SW modes. [219]

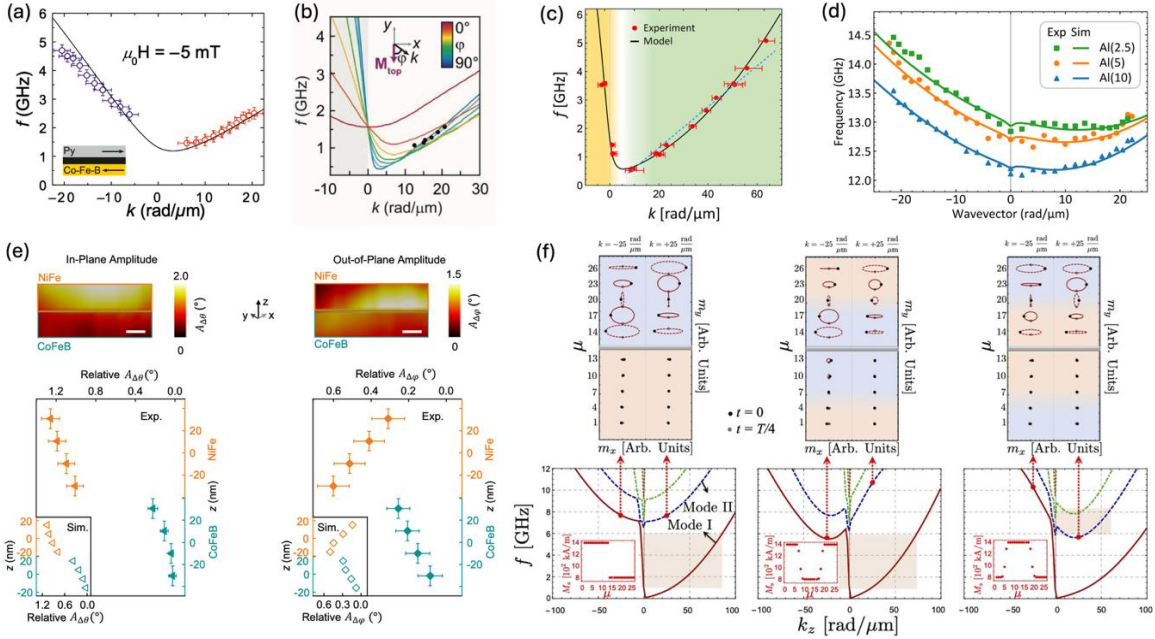

**Fig. 3.6** (a-d) Nonreciprocal magnon dispersions measured for synthetic antiferromagnets (SAFs) in DE configuration. (e-f) Mode-dependent SW localization across the normal direction. (a) Adapted from Ref. [182] (b) Adapted from Ref. [108]. (c) Adapted from Ref. [196]. (d) Adapted from Ref. [206] (e) Adapted from Ref. [208]. (f) Adapted from Ref. [195].

For the SAFs shown in Fig. 3.6(a-d), the dispersion curves are asymmetric, exhibiting a flattened (positive) wavevector window with almost zero group velocity, allowing for localized magnon excitations. In Fig. 3.6(e-f), it is possible to see that the SW excitations are nonhomogeneously distributed across the thickness, which strongly depends on the bilayer characteristics and on the mode frequency.

*3.4 Magnonic crystals and magnetic superlattices*
Magnonics is a rapidly developing area of nanomagnetism and nanoscience, focusing on the manipulation and transport of information using spin rather than charge. A wide variety of magnetic nanostructures play a significant role in magnonics. These include thin films, multilayers, curved nanostructures (such as nanotubes and nanowires), magnonic crystals (MCs), and magnetization-graded nanodevices. [10] Among them, magnonic crystals (MCs) and magnetic superlattices (MSLs) offer unique and powerful platforms for engineering SW behavior through structural and material modulation at the nanoscale. [220,221,222,223]

Magnonic crystals are artificial media where one or more magnetic parameters are periodically modulated in space. These parameters include saturation magnetization, exchange stiffness, magnetic anisotropy, and layer thickness. [222] This periodicity leads to the formation of magnonic band structures, analogous to electronic band structures in semiconductors or photonic bands in optical crystals. The resulting bandgaps, where SW propagation is suppressed due to interference effects, provide precise control over magnon dispersion, group velocity, and density of states. These concepts are critical for envisioning energy-efficient magnonic devices.

By introducing intentional deviations into the perfect periodicity of a magnonic crystal, such as by adding or removing magnetic material, or altering intrinsic properties like saturation magnetization or anisotropy, defects are created. When these defects are inserted periodically, the system forms a magnonic superlattice (MSL). Extensive research on such structures has shown that defects act as magnon traps, capable of confining SW modes within the bandgaps of the host crystal. [224,225,226,227,228,229] These localized defect modes do not propagate freely but are instead confined to specific regions of the structure, offering new dynamic behaviors not present in defect-free crystals. For example, in layered magnetic structures, inserting a defect layer with a different thickness or saturation magnetization induces localized SW modes within the bandgaps.[224] Similarly, carefully designing the defect's anisotropy and geometry can create multiple localized modes that appear either within or below the magnonic bandgap. [225,226] Experimental studies using BLS spectroscopy, and corroborated by numerical simulations, have directly observed these trapped magnon modes within the bandgap. [227,228] These SW modes exhibit strong spatial localization around the defect region, acting thus as a magnon trap. This localization of energy makes such modes particularly attractive for applications requiring high precision and confinement, such as magnonic logic gates and filters. [9] The following subsections explore two fundamental aspects of MCs and MSLs: (i) the creation of bandgaps for controlling magnon propagation and (ii) the localization and trapping of magnons resulting from periodicity breaking or defect engineering. Together, these phenomena showcase how engineered magnetic order enables precise guidance, confinement, and manipulation of SWs for next-generation magnonic technologies.

The introduction of periodicity into a magnetic medium has profound consequences for the propagation of SWs. Like how periodic atomic potentials lead to electronic bandgaps, periodic magnetic structures result in magnonic bandgaps. The gaps originate primarily from Bragg scattering of magnons at the interfaces between regions with contrasting magnetic properties. When the wavelength of the propagating SW becomes commensurate with the periodicity of the structure, constructive and destructive interference result in the suppression of specific modes, leading to a bandgap.

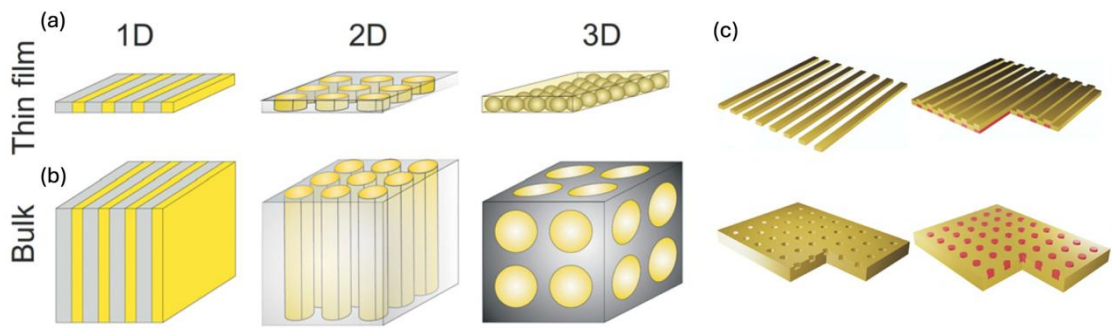

**Fig. 3.7** (a-b) Classification of magnonic crystals based on their dimensionality and geometry. (c) Some examples of 1D and 2D thin magnonic crystals. Adapted from Ref. [9].

According to the dimensionality, MCs can be classified as one-, two-, and three-dimensional, where the parameters vary periodically in one, two, and three dimensions, respectively (see Fig. 3.7(a-b)). Based on the geometry, on one hand, there are MCs with finite thickness (thin-film MCs), where the band structure exhibits confined modes in the vertical direction and extends in the horizontal plane. Examples include arrays of ferromagnetic stripes, continuous films with corrugated surfaces, arrays of holes (antidot lattices), and holes refilled by a different ferromagnetic material (bi-component MCs). [9] On the other hand, bulk MCs are assumed to fill the entire space, meaning the external shape does not significantly influence demagnetization effects. Multilayered thin-film structures can be viewed as the first MCs analyzed and can be approximated as bulk-like MCs when they have a large lateral extent. The position and width of the magnonic bandgaps can be finely tuned by adjusting parameters such as the lattice constant, the contrast in saturation magnetization or anisotropy between layers, and the thickness of the magnetic regions. Moreover, bandgaps can occur not only at the BZ boundaries but also inside the zone due to hybridization between different SW branches, particularly in multilayer or coupled systems. This ability to open and control bandgaps underlies several potential applications. For instance, MCs can function as frequency-selective filters, SW diodes, or waveguides, offering precise control over the flow of magnonic signals. Fig. 3.7(c) shows sketches of thin-film magnonic crystals. At the top, examples of 1D magnonic crystals are presented, including an array of ferromagnetic stripes and a continuous film with one or two corrugated surfaces. The red stripes indicate either the same ferromagnetic material or a different magnetic or nonmagnetic template. At the bottom, 2D magnonic crystals are illustrated, a magnetic antidot lattice formed by an array of empty holes, and a bi-component structure where the holes are refilled with a different ferromagnetic material.

*3.5 Magnon localization in magnetic superlattices*

Defect-induced properties in photonic crystals are crucial for advanced optical technologies due to their unique physics and diverse applications. Engineered defects, which are intentional structural disruptions, enable precise nanoscale control of light, unlocking functionalities not feasible in uniform media. [230,231,232] In magnonic crystals, the magnetic counterparts of photonic crystals, a similar phenomenon occurs with magnons. Specifically, defect modes emerge within the magnonic bandgaps, where propagating SWs are usually forbidden.

The temporal evolution of the SW modes can be obtained from the plane-wave method. [9] This approach is similar to that used in Eq. (7), but with some considerations. First, the temporal evolution of the magnetization dynamics is described by the dynamic matrix $\widetilde{\mathbf{A}}$, assuming that the MC film is sufficiently thin for the dynamic magnetization to be uniform across its thickness, so that $N = 1$. Thus,

$$\widetilde{\mathbf{A}} = \begin{pmatrix} A^{xx} & A^{xy} \\ A^{yx} & A^{yy} \end{pmatrix}, \tag{8}$$

Second, each matrix element is generalized as a new matrix due to the Bloch expansion used in the plane-wave method. Thus, $A^{\eta\eta'} \rightarrow \widetilde{\mathbf{A}}^{\eta\eta'}_{G_i G_j}$ (with $\eta, \eta' = x, y$), where

$$\widetilde{\mathbf{A}}^{\eta\eta'}_{G_i G_j} = \begin{pmatrix} A^{\eta\eta'}_{G_i G_j} & \cdots & A^{\eta\eta'}_{G_{N_p} G_j} \\ \vdots & \ddots & \vdots \\ A^{\eta\eta'}_{G_i G_{N_p}} & \cdots & A^{\eta\eta'}_{G_{N_p} G_{N_p}} \end{pmatrix}. \tag{9}$$

Parameter $N_p$ is related to the plane-wave expansion, and it is chosen to ensure the convergence of the results, and $G_i = 2\pi i/a$, with $i$ being an integer and $a$ the lattice parameter. The matrix elements depend on the effective fields of the system, and periodic dynamic properties can be driven by a periodic magnetic contrast (bi-component MCs), geometric modulation, or spin texture.

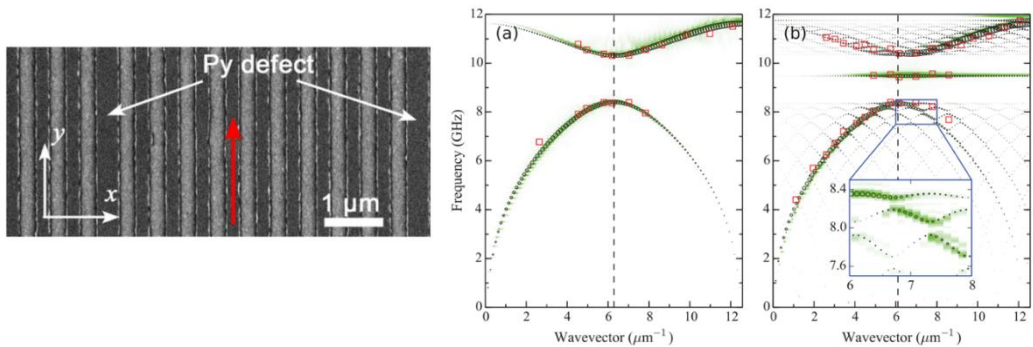

**Fig. 3.8** Left: SEM image of the magnonic crystal featuring 400 nm-wide NiFe defect stripes. The darker

regions correspond to the Py stripes, while the lighter areas represent cobalt stripes. The red arrow denotes the direction of the equilibrium magnetization. Magnon band structures of (a) the perfect MC and (b) the MC with defect stripes (see Fig. 3.7). Measured BLS data are indicated by red squares, while finite-element and micromagnetic simulated data are shown as black circles and green lines, respectively. Adapted from [233].

The left panel of Fig. 3.8 shows an SEM image of a 1D MC composed of alternating NiFe and cobalt (Co) stripes. [233] The perfect MC region consists of 250 nm-wide Py and Co stripes with a periodicity of a=500 nm. In contrast, defects are introduced as wider Py stripes (400 nm) that break the perfect periodicity. The red arrow indicates the direction of equilibrium magnetization. The magnon dispersion relations in DE configuration, shown in Fig. 3.8(a-b), compare the defect-free MC with the defective MC. In the latter, an additional flat band appears within the magnonic gap, distinct from the two primary branches observed in the defect-free case. This flat band signifies the presence of a localized defect mode, where magnons are trapped in the defect region and cannot propagate freely through the crystal. Micromagnetic simulations (green colormap) and frequency-domain finite-element simulations (black circles) confirm this experimental observation. Due to the larger periodicity introduced by the defect-containing supercell, the corresponding BZ is smaller, yielding more band branches. However, the fundamental band structure remains broadly similar to that of the defect-free case, except for the presence of the trapped magnon mode. This behavior is expected, as a low defect concentration has only a limited influence on the collective magnetic excitations.

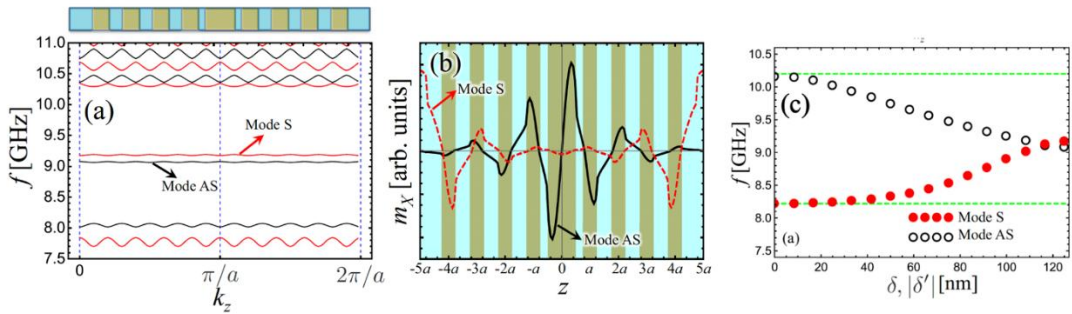

**Fig. 3.9** (a) SW dispersion for a superlattice formed for a defect-free MC with defect zones. The defects correspond to NiFe stripes with a wider width (z = 0), and with a zero width (z = ±5a). (b) SW profiles calculated at $k = 10\pi/a$. Symmetric (S, dashed line) and Asymmetric (AS, solid line) flat modes are illustrated. (c) Evolution of the frequency position of the defect modes as a function of parameter $\delta$ and $\delta'$, which quantify the increase and reduction of the defect stripe width, respectively. Adapted from [229].

As anticipated, the defect mode exhibits strong spatial localization around the defect region. This localized behavior explains the dispersionless, flat-band character of the mode. Indeed, the mode evaluated at $k = 10\pi/a$ shows an asymmetric spatial profile centered on the wider Py stripe (not shown). Interestingly, this asymmetry can be altered by adjusting the defect geometry, as shown in Fig. 3.9, which illustrates a superlattice with two types of defects: wider Py stripes at the center, and the absence of Py stripes at the edge of the supercell. Here, it is demonstrated how the magnitude and character of magnon localization can be tuned through structural engineering. [229] As depicted in Fig. 3.9(a), due to the different nature of the defects, the modes within the bandgaps exhibit distinct localization properties, where a symmetric (asymmetric) mode is excited around the zone with a wider (absent) Py stripe, as illustrated in Fig. 3.9(b). The frequency of the defect modes can be controlled by the geometry of the modified stripe. In Fig. 3.9(c), the frequency of the defect mode is shown as a function of $\delta$ ($\delta'$), which quantifies the increase (decrease) of the defect stripe width, where the case $\delta = 0$ (and $\delta' = 0$) represents a defect-free magnonic crystal.

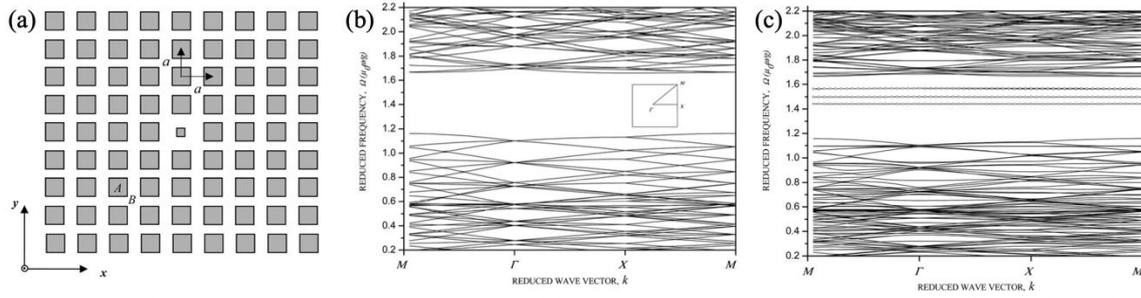

**Fig. 3.10** (a) 2D bi-component magnonic crystal with square scattering centers. The defect is introduced as a smaller square into the lattice. (b) The band structure of the perfect MCs for filling fraction $f = 0.6$. (c) The band structure of MCs with a point defect. Filling fraction is $f = 0.6$ for the zone with large squares, while $f_d = 0.006$ for the defect unit cell. Adapted from [234].

It is worth noting that the localization mechanism originates from confinement in the direction perpendicular to the stripes. In this direction, Bragg scattering arising from the periodic modulation significantly influences the SW dispersion, leading to the formation of bandgaps and the emergence of localized modes in the direction perpendicular to the stripes. In contrast, along the direction parallel to the stripes, the material remains continuous, allowing SWs to propagate in a dispersive way. As a result, the defect modes are spatially confined across the stripes but can still propagate along them. This anisotropic behavior effectively turns the defects into magnonic conduits, guiding SWs in a quasi-one-dimensional manner. However, this situation reverses in 2D magnonic superlattices, where periodicity is introduced in both in-plane spatial directions. In such structures, defect-induced modes become strongly localized in all directions, preventing any propagation. An illustrative example of this behavior is provided in Fig. 3.10, [234] where a 2D crystal is formed by periodically modulating two magnetic materials, A and B. Here, material A has a square cross-section and is embedded in the continuous ferromagnetic matrix B [see Fig. 3.10(a)]. The computed band structures for this system, both with and without defects, are shown in Figs. 3.10(b) and (c), and reveal the presence of defect modes within the bandgap of the pristine crystal. Unlike the 1D case, these defect modes are confined in both directions and do not support propagation, indicating strong magnon trapping at the defect site. [234] For a tiny filling fraction $f$, it has been observed that the defect modes are shape-independent, that is the same dynamic behavior is obtained for circular, square, and rectangular defects. [235] However, at larger filling fractions, the defect modes become highly sensitive to the defect shape.

In engineered defect-based magnonic structures, such as those formed by alternating-width nanowire arrays or aperiodic Fibonacci sequences, magnon localization arises from the disruption of translational symmetry and the inhomogeneous internal field landscape.[236,237,238] These designed variations in geometry, especially the presence of wider and narrower nanowires or irregular spacing, create effective potential wells that act as magnon traps, leading to the formation of confined SW modes with negligible dispersion. In particular, the alternating-width nanowire arrays studied by Goolaup *et al*. [236] show that the differential switching fields of the narrow and wide wires induce a stable antiparallel magnetization configuration over a broad field range. This antiparallel state not only modifies the overall dynamic response of the system but also enhances the conditions for magnon trapping through local demagnetizing field contrasts. The drop in magnetization and the emergence of distinct peaks in differentiated hysteresis loops reflect magnetostatically induced localization, which can be interpreted as creating spatially confined regions where magnons are effectively trapped. Lisiecki *et al*. [237] investigated both experimentally and numerically the dynamic properties of 1D magnonic quasicrystals structured according to the Fibonacci sequence. These structures exhibit unique SW behaviors that differ significantly from those found in periodic magnonic crystals. In Fibonacci quasicrystals composed of dipolarly coupled NiFe nanowires, the quasiperiodic sequence further enhances the complexity of the SW spectrum, yielding flat bands and localized magnonic excitations.

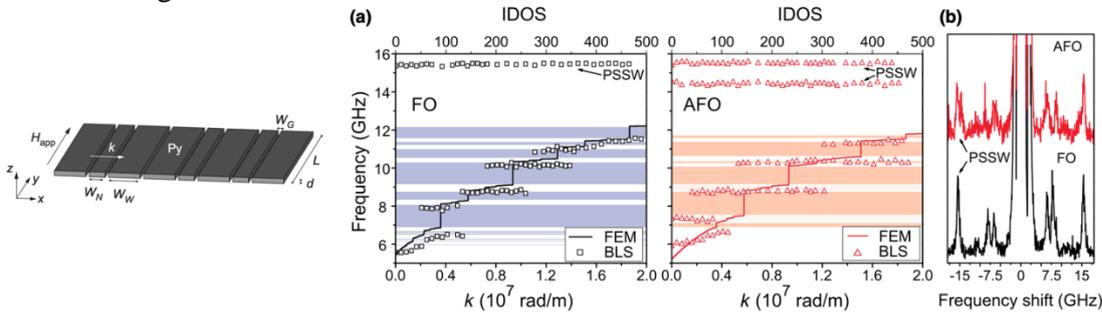

**Fig. 3.11** Left: A sketch of a Fibonacci structure consisting of NiFe nanowires of different widths separated by air gaps. The static magnetic field is applied along the nanowires. (a) Experimental SW frequencies from BLS data vs wavevector k compared with the band structure obtained from numerical calculations using the finite-element method (FEM, solid lines) shown as frequency vs. integrated density of states (IDOS) for $H_{app} = $ 12.5 mT for structures with $W_N = $ 350 nm and $W_W = $ 700 nm in ferromagnetic and antiferromagnetic order; FO and AFO, respectively. The violet (pink) regions indicate the magnonic bandgaps for FO (AFO) from the numerical calculations. (b) Typical experimental BLS spectra measured for $k = 0.41 \times 10^7$ rad/m. Adapted from Ref. [238].

Figure 3.11(a) presents the measured dependence of the SW frequency on the wavevector $k$. In both magnetization configurations, namely ferromagnetic order (FO) and antiferromagnetic order (AFO), the SW spectra display a set of discrete peaks, with only the lowest-frequency mode being dispersive. [238] This indicates that the Fibonacci array supports the propagation of long-wavelength collective SWs in both FO and AFO magnetic states. As $k$ increases, additional modes emerge within a confined wave-vector range and exhibit nearly constant frequencies. These flat modes are signatures of magnon trapping, arising from the localization of SWs in specific regions of the quasicrystalline structure. Such behavior contrasts with that observed in periodic nanowire arrays of alternating width, where the modes follow Bloch's theorem, yielding continuous dispersions with periodic frequency modulation. [239] The observed magnon traps, resulting from the combined effects of quasiperiodic geometry, dipolar interactions, and magnetization configurations, enable tunable and reprogrammable SW confinement, offering promising prospects for spatially selective control in magnonic circuits.

*3.6 Spin-wave edge and cavity modes in twisted moiré superlattices*
When two overlapping 2D lattices experience a slight angular twist or possess a minor discrepancy in lattice parameters, a moiré superlattice forms, distinguished by a novel periodic arrangement with a periodicity vastly exceeding that of the original lattice constant. The formation of such moiré superlattices has garnered significant attention due to their ability to modify the electronic, optical, and magnetic properties in various material systems. Remarkable electronic properties, including unconventional superconductivity and correlated insulating states induced by the electronic flat bands, have been observed in magic-angle twisted bilayer graphene, characterized by a precise twist angle of 1.1 degrees, the so-called "magic angle". [240,241,242] These studies have opened a new horizon dedicated to the exploration of novel superconductors. Building on these foundational insights, the principles of moiré physics have transcended their initial application in electronic systems and have been extended to other domains, such as photonics, phononics, and magnonics. [243,244,245,246]
Recent research into moiré physics within magnonics has garnered considerable interest. Achieving magnonic applications in all 2D magnetic systems, such as $CrI_3$, presents challenges in sample fabrication and detection techniques. However, artificial magnetic superlattices, such as magnonic crystals with antidot lattices, may offer a more practical approach for exploring magnonics based on moiré patterns. In a magnetic moiré superlattice, magnon flat bands can arise when two square antidot lattices are stacked at a precise twist angle, namely 3.5°, referred to as the magnonic magic angle, as shown in Fig. 3.12(a). [247] A bilayer structure of YIG is considered, which can have very low Gilbert damping in the real material system. The twist angle and interlayer exchange coupling are critical factors in forming these moiré magnon flat bands. Micromagnetic simulations reveal that a magnon flat band forms under an optimal combination of this twist angle of $\theta = 3.5°$ and the interlayer exchange coupling strength $A_{12} = 11\ \mu J/m^2$, extending across nearly half of the first magnonic BZ. The resulting zero magnon group velocity leads to a high density of states, enabling a highly localized magnon mode, namely a magnonic nanocavity, at the center of the moiré unit cell in the AB stacking region. This behavior contrasts with photonic moiré superlattices, where confinement occurs in the AA region. It is found that the magnon flat band stems from the lowest band, shaped by mode anti-crossing in the magnon dispersion due to magnon-magnon coupling. Remarkably, the magic-angle magnonic nanocavity achieves a spatial linewidth of approximately 185 nm, corresponding to the flat band wavevector distribution. The magnon intensity distribution can be calculated based on $I_0(x) = \frac{A^2 S}{\pi^2}\left|\frac{\sin\left(\frac{1}{2}\Delta k_0 x\right)}{x}\right|^2$, where $A$ is the SW amplitude, which is constant for all modes at the flat band, and $S$ is the area of one moiré unit cell. By taking the flat-band wavevector bandwidth $\Delta k_0 \approx 30$ rad/μm, the calculated magnon distribution for the nanocavity aligns well with simulation results, as shown in Fig. 3.12(b). Additionally, when a nanomagnonic waveguide is placed atop the moiré superlattice, magnon confinement extends to the waveguide, driven by interlayer exchange interactions. [248] This confinement is effectively transferred to the overlying waveguide, opening promising paths for developing future magnonic waveguides and circuits with high tunability.

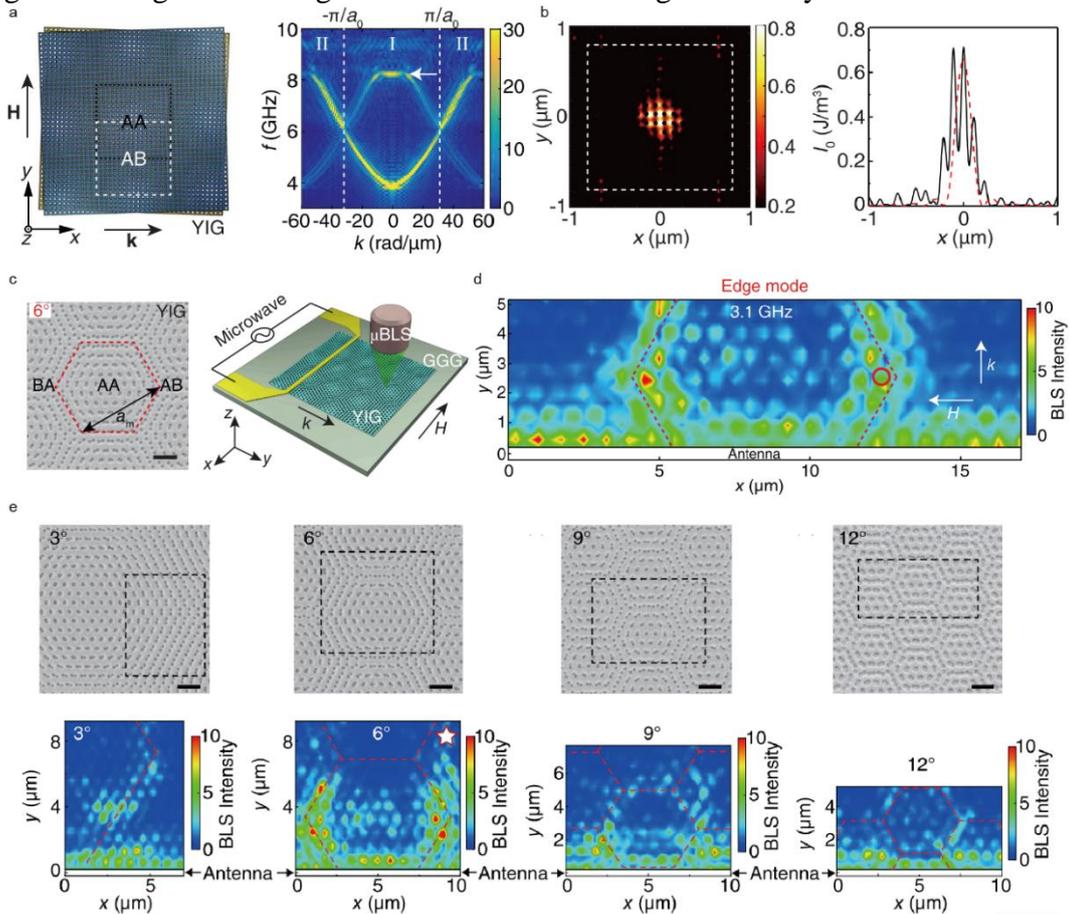

**Fig. 3.12** SW edge and cavity modes in twisted moiré superlattices. (a) Schematic of a magnetic moiré superlattice based on YIG, modelled using micromagnetic simulations. The simulated SW dispersion reveals a magnonic flat band. (b) The magnonic cavity mode arises from the magnonic flat band, with a spatial scale of approximately 185 nm. (a) and (b) are taken from Ref. [247]. (c) SEM image of a YIG-based moiré

magnonic lattice grown on a GGG substrate with a 6 ° twist angle, alongside schematics of spatially resolved SW measurements on moiré magnonic lattices using μBLS. (d) 2D SW intensity maps measured by μBLS at 3.1 GHz. (e) 2D SW intensity maps measured by μBLS at twist angles of 3°, 6°, 9°, and 12°, with an excitation frequency of 3.10 GHz and an applied magnetic field of 50 mT. (c), (d) and (e) are taken from Ref. [249].

SW edge and cavity modes have been experimentally observed in twisted magnetic moiré superlattices, as shown in Fig. 3.12(c). [249] Two antidot sublattices with a relative twist angle are merged into a single YIG thin film, forming a moiré magnonic lattice. This is different from the previous simulation structures for the magnetic bilayer moiré system. A triangular lattice is selected here to mimic the hexagonal lattice of graphene. A nanostripline is patterned on the YIG moiré superlattice to excite SWs propagating in the magnetic moiré superlattice, with an external magnetic field applied parallel to the microwave antenna. Using micro-focused BLS microscopy, two distinct SW modes are mapped in 2D within a moiré unit cell, i.e. the localized mode at the center and a propagating mode along the edges. The confined SW mode at the moiré center exhibits a slightly lower frequency than the edge mode along the boundary of the unit cell, as shown in Fig. 3.12(d). Devices with different twist angles are fabricated, and the most prominent moiré edge mode emerges at an optimal twist angle of 6° under an external magnetic field of 50 mT, as shown in Fig. 3.12(e). The magnetic field provides an additional degree of freedom for tuning the magnon edge mode beyond the twist angle. Micromagnetic simulations reproduce the propagating SWs at the moiré unit cell edges. Fast Fourier transform analysis further reveals that the edge mode arises from a magnonic bandgap at the intersection of two magnon branches. A nontrivial topology is demonstrated by analytical calculations, attributed to magnon-magnon coupling with dipolar interactions in the chiral magnonic edge mode. This experimental investigation of moiré magnon modes is crucial for advancing moiré physics in magnonics and may also enhance the development of new magnonic devices.

So far, we have discussed spin-wave confinement in artificial magnetic moiré superlattices, where the lattice constants are orders of magnitude larger than those in natural crystals. The primary experimental challenge lies in realizing confined magnons in twisted van der Waals systems, owing to difficulties in the efficient excitation and detection of magnons, as well as in sample fabrication. Instead, Wang et al. investigated magnon excitations in bilayer CrI$_3$ with a small twist angle through theoretical calculations, where lattice mismatch and interlayer rotational misalignment give rise to a long-period moiré pattern. [250] Local stacking variations between the two CrI$_3$ layers spontaneously form stacking domain walls that act as one-dimensional magnetic channels. Employing first-principles calculations in combination with linear spin-wave theory, the authors demonstrated that magnon modes in these twisted structures are spatially confined along the domain walls, exhibiting energies well below those of the bulk magnon bands. These one-dimensional magnons within the stacking domain walls are inherently more robust against external perturbations. Moreover, the local interlayer exchange coupling within the moiré unit cell can be modulated by the twist angle, allowing precise control over the magnon confinement strength and mode frequencies. The predicted moiré magnon confinement in emergent domain-wall channels open new avenues for reconfigurable, twist-tunable magnonic circuits in atomically thin magnets.

**3.7 Localized magnons with periodic interfacial Dzyaloshinskii-Moriya interaction**
The periodic modulation of the iDMI introduces a novel and versatile mechanism for engineering magnon band structures in ultrathin ferromagnetic systems. Unlike static magnetic potentials or geometric modulation, which are widely used, the spatial variation of DMI directly introduces a periodic chiral component into the SW dynamics, enabling the emergence of unconventional SW phenomena, such as flat bands, indirect gaps, and spatially localized magnons. [135] From a theoretical perspective, the role of the periodic DMI is captured by the terms $A^{xx}_{G_iG_j}$ and $A^{yy}_{G_iG_j}$ in Eqs. (8–9). These matrix elements are given by [134]

$$A^{xx}_{G_iG_j} = A^{yy}_{G_iG_j} = -i\frac{2D(G_i-G_j)}{\mu_0 M_s}\left(\frac{\mathbf{G}_i+\mathbf{G}_j}{2}+\mathbf{k}\right)\cdot\hat{z}, \tag{10}$$

where $G_i = 2\pi i/a$, with $a$ being the periodicity along the z-direction. If the DM is non-zero only in regions of width $w$, the Fourier coefficient is

$$D(G_i) = D\frac{w}{a}\,\text{sinc}\left(\frac{w}{2}G_i\right).$$

In this way, the periodic Dzyaloshinskii-Moriya interaction is adequately accounted for within the plane-wave method.

*3.7.1 Emergence of flat bands in the magnonic spectrum*
In ultrathin ferromagnetic films coupled with periodic heavy-metal (HM) stripes, characterized by high spin-orbit coupling, the interfacial DMI becomes spatially nonuniform, being active only under the HM zones. This breaks inversion symmetry locally, yielding a position-dependent antisymmetric exchange interaction. As shown by Gallardo *et al.*, the periodic DMI leads to modulated effective fields that induce a reduced local frequency underneath the HM layers. [1354] These discontinuities of the chiral effective field act as barriers for SWs, trapping them within the DMI-active or inactive regions depending on the material parameters, frequency, wave vector, and band index.

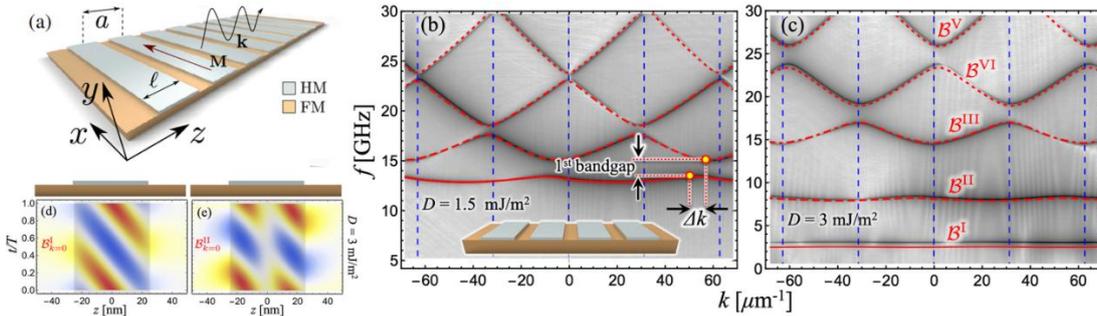

**Fig. 3.13** (a) Illustration of a chiral magnonic crystal where a ferromagnetic ultrathin film is covered with a periodic array of heavy metal wires such as Pt. SWs propagate along z, and the equilibrium magnetization points along x. (b) and (c) show the magnon band structure for a $Ni_{80}Fe_{20}$ film covered with a HM-wire array, where a = 100 nm, D = 1.5, and 3 mJ/m$^2$. The lines correspond to the theory, and the color code corresponds to the MUMAX results, where darker (lighter) colors represent intensity maxima (minima). (d) and (e) show the temporal evolution of the real part of the in-plane dynamic magnetization evaluated at D = 3 mJ/m$^2$. Adapted from Ref. [135].

One of the most significant outcomes of periodic DMI is the formation of flat magnonic bands. These arise when SWs become confined within specific spatial regions (e.g., beneath the heavy-metal stripes), suppressing their propagation and resulting in a nearly zero group velocity. Figure 3.13(a) illustrates a 1D chiral magnonic crystal formed by coupling a ferromagnet (FM) with periodically patterned HM stripes. When the SWs propagate perpendicular to both the equilibrium magnetization and the HM stripes, the resulting band structure exhibits indirect bandgaps, as shown in Fig. 3.13(b). These indirect gaps appearing in these chiral magnonic crystals are related to Bragg reflections at the BZ edges, which, due to DMI-induced nonreciprocity, no longer cancel out, thus creating indirect gaps that are regions of forbidden frequencies where the top of one band and the bottom of the next occur at different wave vectors. As the DMI strength increases, the low-frequency bands are flattened, as seen in Fig. 3.13(c) for the bands $\mathcal{B}^I$ and $\mathcal{B}^{II}$. The mode flattening is associated with the strong spatial localization of SWs beneath the HM layers. This dynamic behavior is visualized in Fig.

3.13(d–e), which shows the real part of the in-plane dynamic magnetization as a function of time. The localized mode exhibits a nontrivial temporal evolution, characterized by time-dependent maxima and minima in the dynamic magnetization amplitude. [135]

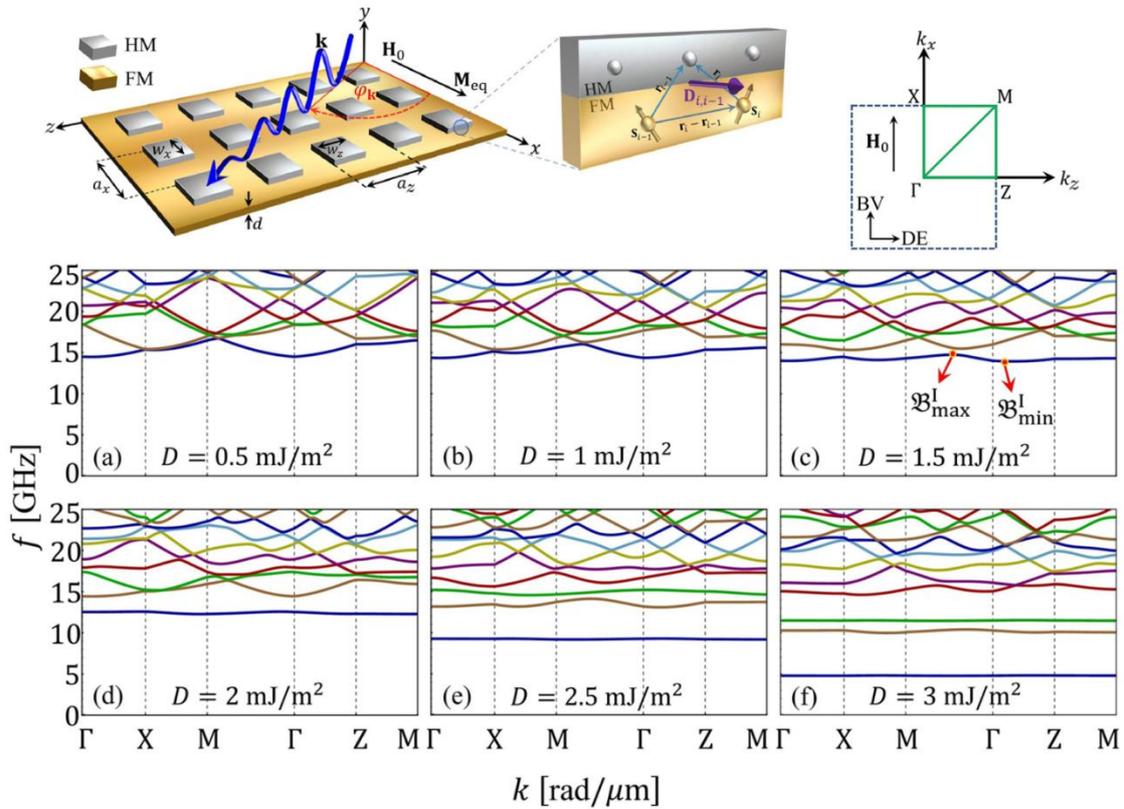

**Fig. 3.14** Top-left, Illustration of a ferromagnetic ultrathin film of thickness d in contact with a 2D periodic array of heavy-metal islands, which induce an interfacial DMI only below the array. In this 2D chiral magnonic crystal, the SWs k in the xz plane with a given angle $\varphi_k$. The equilibrium magnetization points along x due to an external magnetic field applied in the same direction. (a)-(f) Illustrate the bandstructure evolution for different values of the DM strength D. [251]

In the 1D MC, only SWs propagating perpendicular to the HM stripes experience strong localization, while those traveling along the stripe direction remain fully dispersive. Nonetheless, Flores *et al*. have shown that complete localization can be achieved by introducing a 2D periodic arrangement of the heavy metals. [251] This is shown in Fig. 3.14, where the emergence of omnidirectional flat bands reveals that SW localization occurs regardless of the propagation direction. Crucially, these dispersionless bands are not confined to specific directions but represent a global feature of the band structure. For sufficiently large DMI strength, the low-frequency modes acquire an omnidirectional flat character, confirming the robustness of chiral localization in 2D periodic DMI systems.

The localization properties of dispersionless modes can be effectively controlled through the combined effects of the DMI and perpendicular magnetic anisotropy (PMA). These two interactions act selectively, localizing specific SW branches depending on their symmetry and coupling to the chiral magnetic background. [252] When low-frequency flat bands are localized in DMI-rich zones, the resulting SW modes exhibit asymmetric intensity with respect to wave-vector direction, showing nonreciprocal behavior. In contrast, when these modes are confined beneath regions with strong PMA (but without DMI), they display reciprocal intensity profiles under wave-vector inversion. Figure 3.15(a) shows the simulated and calculated SW dispersion for a system with dominant PMA, where the lowest flat band is localized under Ruthenium layers (which exhibit strong PMA). In this case, the simulation reveals a flat band with a reciprocal simulated power-spectrum amplitude. Conversely, when DMI is the dominant interaction, the lowest flat band becomes nonreciprocal, as seen in Fig. 3.15(b), where the simulated spectrum displays a stronger amplitude for positive wave vectors.

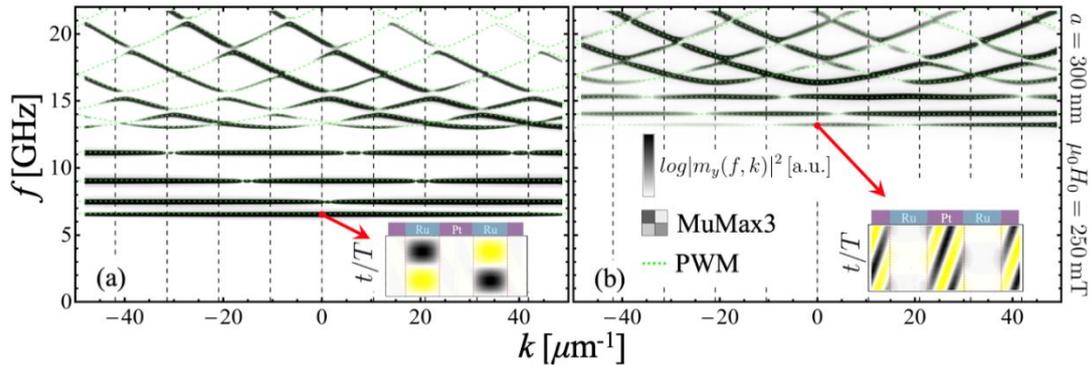

**Fig. 3.15** (a) and (b) illustrate the magnonic band when the PMA and DMI dominate, respectively. The dotted green lines represent the calculated dispersion, while the power spectrum of the MuMax3 simulations is depicted using a gray scale. The insets display the SW profiles for the lowest-frequency band evaluated at k=0 where localization occurs beneath the (a) Ru and (b) Pt regions. Adapted from Ref. [252].

Recent experiments have provided convincing evidence of flat magnonic modes in 1D chiral magnonic crystals using wave-vector-resolved BLS measurements. [253,254] The investigated systems, ferromagnetic films patterned with periodic heavy-metal stripes, exhibit spatially modulated interfacial DMI, which plays a key role in tailoring the SW band structure. As the DMI strength increases, flat bands emerge in the low-frequency range, corresponding to magnon modes strongly localized in regions with active DMI (e.g., beneath Pt stripes). These dispersionless modes exhibit zero group velocity and directional nonreciprocity, which is evident in the BLS spectra as asymmetries in the frequencies of the Stokes and anti-Stokes peaks. Figures 3.16(a–l) present the measured and simulated band structures of bicomponent chiral magnonic crystals with periodically modulated DMI, realized in Ir/Fe/Ir(Pt) and Au/Fe/Au(Pt) trilayers. The experiments by Wei et al. reveal clear signatures of mode flattening and indirect bandgaps, [254] along with a marked increase in the density of states (DOS) for low-frequency bands as the DMI intensity increases. A similar behavior was first reported by Tacchi *et al*. [2533] who studied a CoFeB film coupled with a periodic array that alternates platinum and ruthenium nanostripes. Their BLS spectra, shown in Figs. 3.16(m–o), confirm strong low-frequency peaks associated with flat modes and their dependence on the HM stripe width. Together, these results validate theoretical predictions and demonstrate that spatially engineered DMI enables robust magnon trapping without relying on geometric confinement.

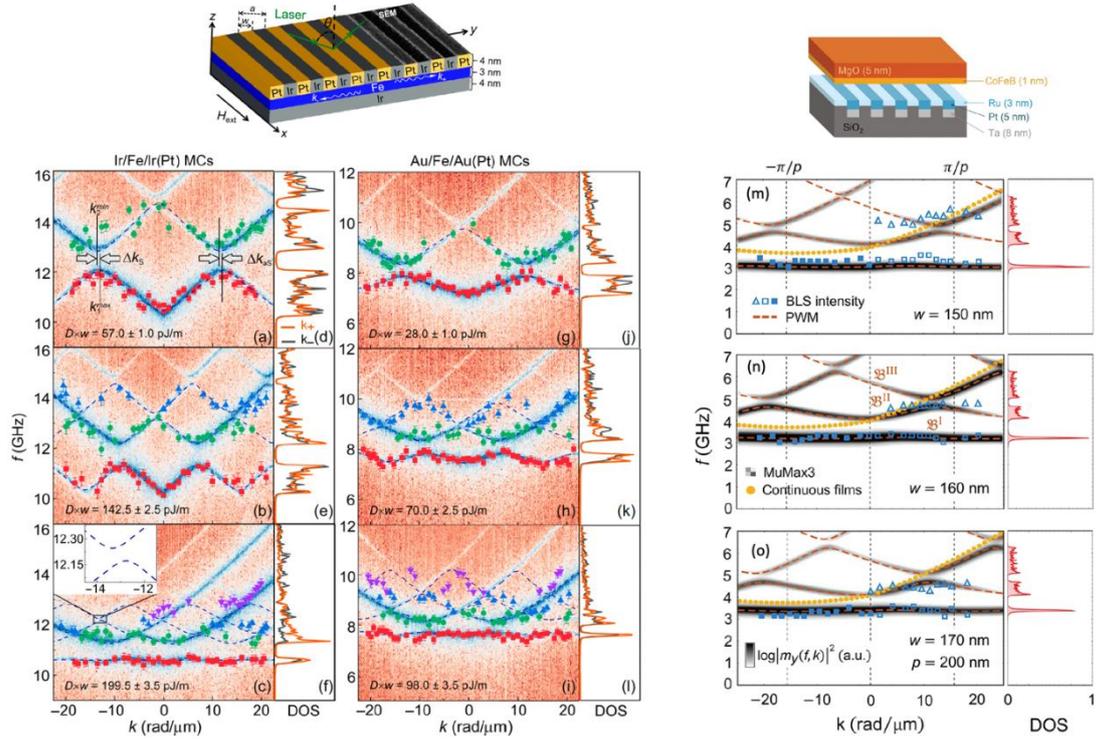

**Fig. 3.16** Band structures of SWs in chiral MCs with different periods for Ir/Fe/Ir(Pt) and Au/Fe/Au(Pt) chiral MCs under an 800 Oe magnetic field. The periods are (a) a = 250 nm, (b) 400 nm, (c) 500 nm for Ir/Fe/Ir(Pt) chiral MCs, and (g) a = 250 nm, (h) 400 nm, (i) 500 nm for Au/Fe/Au(Pt) chiral MCs. The background intensity plots are simulated using MuMax3. The solid symbols represent the dispersion branches extracted from the BLS spectra, and the dashed lines depict the FDFEM calculations. The inset in (c) shows a magnified view of the FDFEM calculation. The effective interfacial DMI energy $Dw$, with w being the Pt stripe width is labeled in the figures. (d–f), (j–l) Calculated magnonic DOS. The orange and black lines represent the DOS in the positive and negative wave-vector regions, respectively. Band structure for the chiral magnonic crystals with 200-nm period and (m) w = 150 nm, (n) 160 nm, and (o) 170 nm. The symbols correspond to the BLS data, the dashed lines to the PWM calculations, and the gray map represents the power spectrum of the out-of-plane dynamic magnetization obtained from MuMax3. (a-l) Adapted from Ref. [254]. (m-o) Adapted from Ref. [253].

The SW localization in the chiral magnonic crystal with periodic DMI (underneath Pt/CoFeB) and periodic perpendicular anisotropy (at the Ru/CoFeB regions), is evidenced in Fig. 3.17, which shows the spatial profiles of the in-plane dynamic magnetization component, $m_x$, for the first three modes evaluated at $k = \pi/p$, for a system with a period $p = 200$ nm. [253] The spatiotemporal evolution of the first bands reveals distinctive behaviors in regions characterized by the presence of interfacial DMI (Pt) and perpendicular anisotropy (Ru). In the sample with $p = 200$ nm, the (a) low-lying frequency mode is located underneath the Ru stripes, while the (b) second and (c) third bands localize underneath the Pt wires.

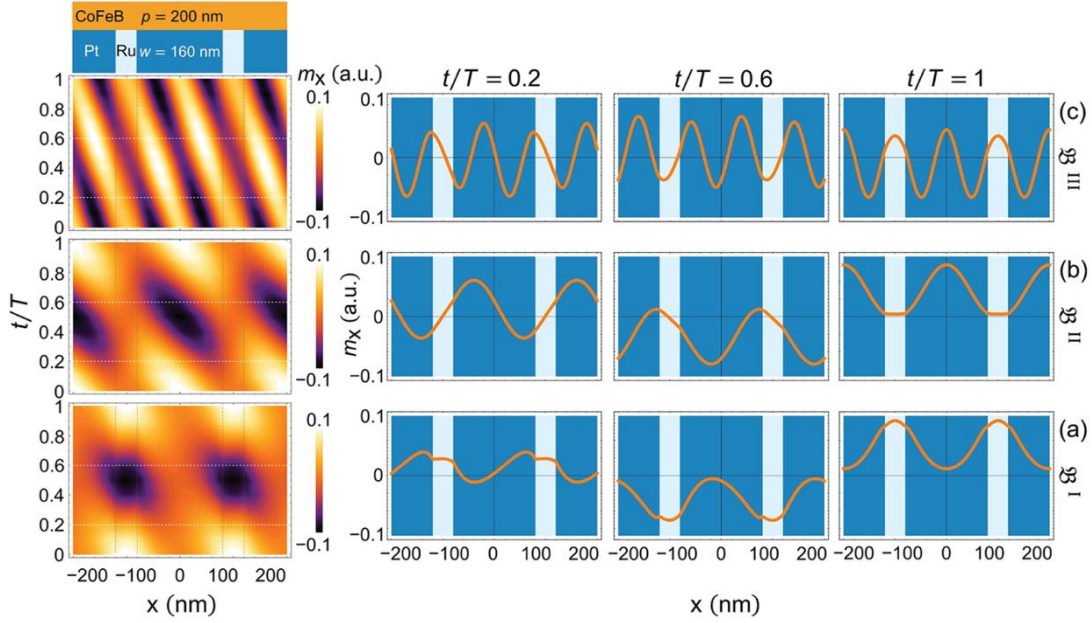

**Fig. 3.17** (left) Spatiotemporal profiles of the dynamic magnetization $m_x$ for a chiral magnonic crystal with a Pt width $w = 160$ nm and for $k = \pi/p$, which has been evaluated for the first three low-frequency bands: (a) $\mathcal{B}^{I}$, (b) $\mathcal{B}^{II}$, and (c) $\mathcal{B}^{III}$. The blue (light blue) color represents the regions with (without) interfacial DMI. The right panels show the dynamic magnetization profiles at three stages of the process, emphasized with the dashed horizontal lines, where $T$ is the oscillation period. Extracted from Ref. [253].

By varying key parameters such as the magnonic crystal period, the DMI constant, or the effective magnetization, it is possible to manipulate the features of SW localization. This is illustrated in Fig. 3.18, which presents a SW localization diagram for the low-frequency mode calculated for a sample with a 300-nm period, filling fraction $f = 0.5$, $M_s = 900$ kA/m, and $\mu_0 H_0 = 250$ mT. The color code results are obtained from the plane-wave method calculations, while the circle dots correspond to an analytical formula given by Eq. (8) in Ref. [252]. In Fig. 3.18(a), each triangle defines a pair ($D$ [mJ/m$^2$], $\mu_0 H_s$ [mT]) where $H_s$ is the strength of the perpendicular anisotropy field. In Fig. 3.18(b), the spatiotemporal profiles of the SW modes were calculated at the points $P_1$-$P_4$. For $P_1$, the modes are localized behind the Ru wires. In contrast, for points $P_2$ and $P_4$, the SW excitations are confined underneath the Pt areas. Point $P_3$ represents cases where the excitations are simultaneously beneath the Pt and Ru nanostripes. Fig. 3.18(c) depicts the magnon dispersion for a continuous film without interfacial DMI ($D = 0$) and for the anisotropy fields: $\mu_0 H_s = 1200$ mT (dashed line), $\mu_0 H_s = 613$ mT (dot-dashed line), and $\mu_0 H_s = 200$ mT (dotted line). The solid line considers the case of an extended film with $D = 2.5$ mJ/m$^2$ and $\mu_0 H_s = 0$. For the zero DMI case and $\mu_0 H_s = 613$ mT, the nominal film dispersions without DMI present a frequency minimum that coincides with the dispersion of a continuous film with zero anisotropy and $D = 2.5$ mJ/m$^2$, which means that for these particular values, SWs can be excited either underneath the Pt or Ru nanowires. Such a condition allows for determining the transition curve (circles) in Fig. 3.18(a) between localization at the CoFeB regions in contact with the Ru or Pt zones. [252]

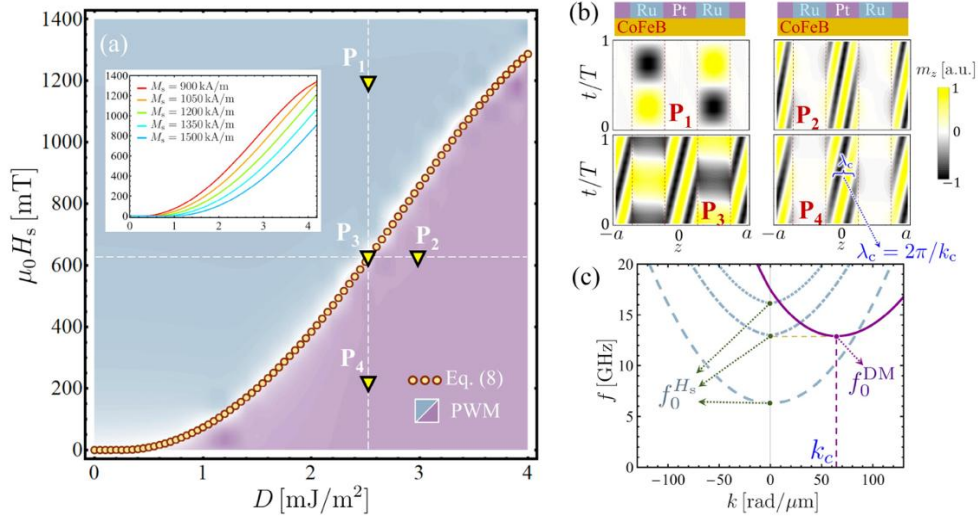

**Fig. 3.18** (a) Magnonic localization diagram for the low-frequency mode calculated for a chiral MC with periodic DMI and periodic anisotropy, where $D$ is the interfacial DMI strength and $H_s$ the perpendicular anisotropy field. (b) Spatiotemporal profiles of the first SW mode for cases $P_1$-$P_4$ showing that spatial localization depends on $D$ and $H_s$. (c) Dispersion for a continuous film with $D = 0$ and three anisotropy fields: $\mu_0 H_s = 1200$ mT (dashed line), 613 mT (dot-dashed line), and 200 mT (dotted line). The solid line represents a film with $D = 2.5$ mJ/m2 and $\mu_0 H_s = 0$. Extracted from Ref. [252].

*3.7.2 Enhancement of magnon density of states*
The (DOS) quantifies the number of available modes per unit frequency (or energy) and volume. It provides insight into the system's dynamic behavior and is crucial for understanding various thermodynamic and transport properties of materials. In electronic systems, for example, a peak in the DOS near the Fermi level enhances the probability of electronic transitions because it signifies a high concentration of available electronic states at a given energy, which is foundational to understanding, for instance, electrical conductivity and superconductivity. [255] Similarly, in magnonics, the magnon DOS determines how SW modes populate energy levels and how they respond to thermal excitations or external fields. A high magnonic DOS implies a large number of states available per frequency, which can strongly affect magnon scattering, energy localization, and even collective phenomena like magnon BEC.

Flat bands in magnonic crystals are characterized by a zero or nearly zero group velocity, indicating highly localized SW modes. As the dispersion relation $\omega(k)$ becomes flat over a finite region in reciprocal space, its derivative tends to zero, leading to a divergence or sharp enhancement in the DOS. This is analogous to the situation in photonic or electronic systems, where flat bands cause strong resonances and density peaks. As mentioned in the previous section, in recent experimental and theoretical studies, particularly those involving chiral magnonic crystals with periodically modulated interfacial DMI, flat bands have been observed and linked directly to spatial magnon localization and DOS enhancement. In Ref. [253] BLS experiments on 1D chiral magnonic crystals with a periodic interfacial DMI reveal the emergence of low-frequency flatbands, with the calculated DOS exhibiting sharp peaks at the flat band frequencies. These modes are localized underneath heavy-metal stripes and are a direct result of the DMI modulation. The work by Wei *et al*. [254] further supports the emergence of flat bands in chiral magnonic crystals, showing through both simulations and BLS measurements that increasing the DMI strength not only induces nonreciprocal dispersion but also leads to the appearance of dispersionless bands. The corresponding DOS, calculated from simulations, shows a significant enhancement at the flat-band frequencies. This enhancement grows with increasing DMI or lattice constant, demonstrating the tunability of the magnonic DOS via structural parameters.

Recent theoretical studies demonstrated that in defect-free [252] and superlattice chiral MCs, [256] flat bands give rise to pronounced peaks in the magnon density of states. When heavy-metal stripes are periodically removed or modified, constituting a superlattice, multiple flat modes with high DOS are observed within the bandgaps, often associated with localized excitations in defect regions. This is illustrated in Fig. 3.19, showing on the left panel the SW dispersions of a chiral magnonic lattice and superlattice. The right panel shows the dispersion of two groups of low-frequency flat bands, (a) $\mathcal{B}_1$-$\mathcal{B}_9$ and (b) $\mathcal{B}_{10}$-$\mathcal{B}_{18}$, with the corresponding spatial distributions of the modes (f,g) and their localization underneath the gray zones with active DMI. (c) Illustrates the higher frequency bands ($\mathcal{B}_{28}$-$\mathcal{B}_{64}$) containing the flat modes localized in the defects (green curves). Their associated DOS are displayed in (d,e), where the narrower peaks are seen for the flattest bands. Note that in (h), the first group of flat defect modes, $\mathcal{B}_{28}$-$\mathcal{B}_{30}$, is strongly localized around the defect (orange regions) and exhibits (e) very sharp DOS peaks. In contrast, the DOS peak of the higher frequency flat modes ($\mathcal{B}_{41}$, $\mathcal{B}_{52}$, $\mathcal{B}_{53}$, and $\mathcal{B}_{64}$) in (e,h) decreases notably, and the excitations are localized in both zones, with periodic DMI (gray) and periodic anisotropy (orange).

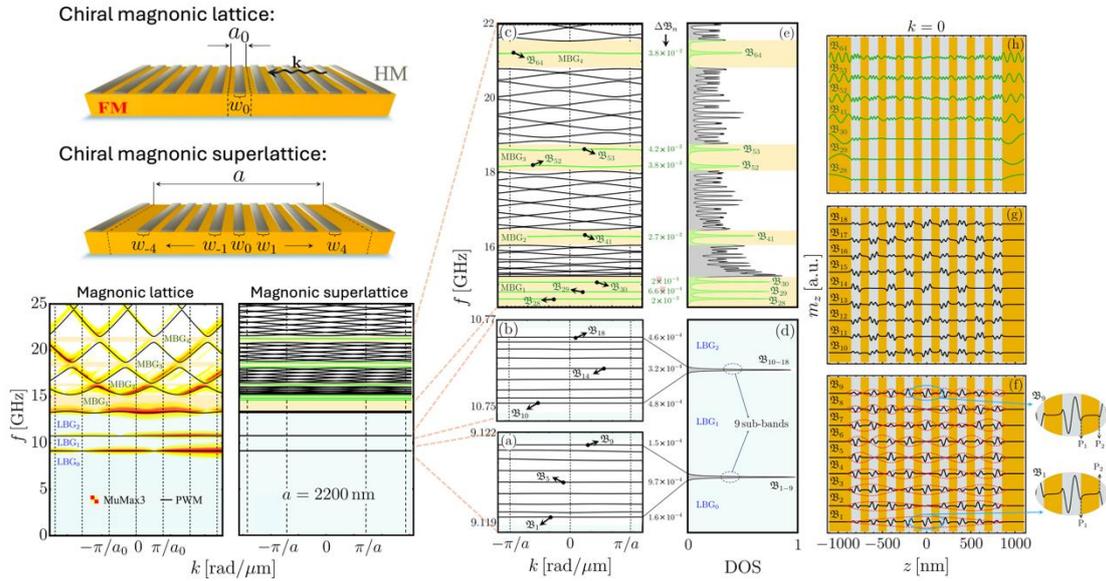

**Fig. 3.19** (left) SW band structure of defect-free and superlattice chiral magnonic crystals showing flat bands. In the superlattice, a hyperfine structure appears (a-c), which is provided by the complex supercell of period $a$. (d-e) Magnonic DOS for selected bands with pronounced peaks. (f-h) Spatial profiles of the dynamic magnetization, which demonstrate the selective localization of the SW excitations in the magnonic superlattice. Adapted from Ref. [256].

The enhancement of the magnon DOS through flat bands is a phenomenon of profound relevance, with implications that extend well beyond magnonics. In systems such as twisted bilayer graphene, the emergence of flat electronic bands leads to a dramatic increase in the electronic DOS, enabling strong electronic correlations and unusual phenomena like superconductivity. [257,258,259] Magnon flat-band formation in twisted bilayer magnonic crystals has also been observed in moiré superlattices, [247] as discussed in Section 3.6. As shown in Fig. 3.18, the interlayer twist leads to the emergence of nondispersive (flat) magnon modes in the second branch, particularly at small wave vectors. The frequency position of these modes can be tuned via the interlayer exchange coupling. As displayed in the inset of Fig. 3.18(b), the DOS exhibits a clear enhancement around 8.2 GHz, indicating the presence of a flat band.

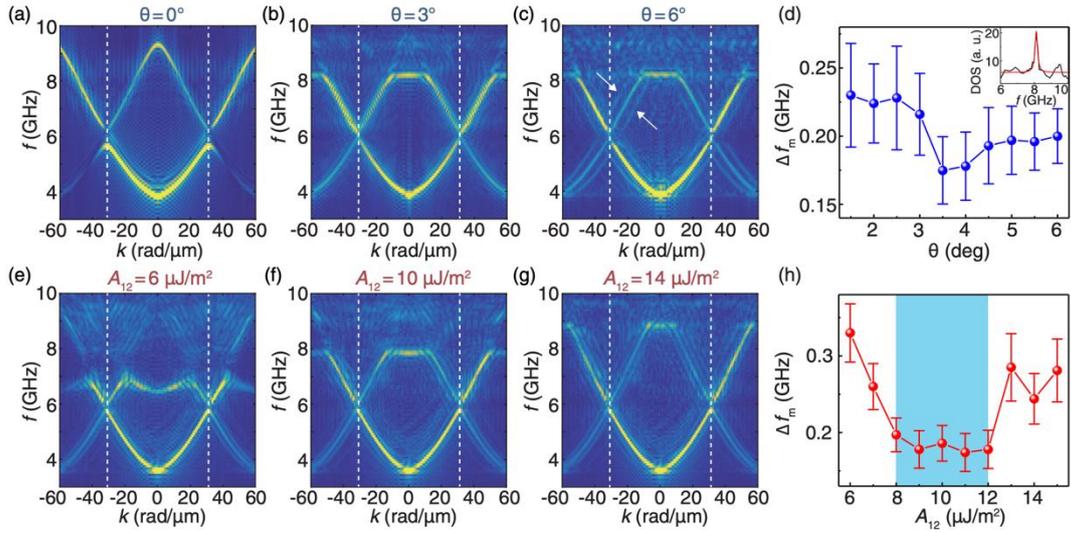

**Fig. 3.20** (a) SW dispersion on commensurate bilayer magnonic crystal, i.e., with θ = 0. (b) SW dispersion at the moiré unit cell center with θ = 3°. (c) SW dispersion at the moiré unit cell center with θ = 6°. White arrows indicate the side bands. The interlayer exchange coupling is fixed at $A_{12} = 11$ μJ/m² in (a-c). (d) Moiré flat-band bandwidth $\Delta f_m$ as a function of twist angle θ. $\Delta f_m$ is extrapolated from the linewidth of magnon DOS around 8.2 GHz shown in the inset as an example at θ = 3.5°. (e-g) Simulated SW dispersion at the moiré unit cell center (AB stacking region) with a fixed twist angle θ = 3.5° and alternating interlayer exchange coupling of (e) 6 μJ/m², (f) 6 μJ/m², and (g) 14 μJ/m². (h) Interlayer exchange coupling dependence of moiré flat-band bandwidth $\Delta f_m$ with θ = 3.5°. The light-blue shaded area highlights the region with high-quality moiré flat band. Adapted from Ref. [247].

The previous examples highlight a unifying principle: flat dispersions amplify the DOS, which in turn enhances magnon-magnon coupling and wave confinement. Therefore, an increased DOS due to flat SW bands, engineered via periodic interfacial DMI or other superlattice architectures, can give rise to a robust **magnon trapping** strategy, allowing the SW modes to be excited in a localized way within the nanostructure.

**Chapter 4: Chirality and topological effects**

*4.1 Chiral magnon edge modes*
Chirality describes the property of an object that is not identical to its mirror state, a concept widely investigated across diverse physical systems. In magnonics, chiral magnons enable the precise directional control of their propagation. From a theoretical perspective, chirality in magnonic systems originates from asymmetric spin interactions that break spatial inversion symmetry, such as the DMI, dipolar interactions, or geometrically engineered asymmetries. These chiral interactions lift the degeneracy between magnons propagating in opposite directions, giving rise to nonreciprocal behavior. Additionally, chiral edge magnon states, analogous to chiral edge states in the quantum Hall effect, may support topologically protected magnons, making them immune to backscattering.

*4.1.1 Directional magnon propagation due to dipolar interactions*
Controlling the direction of magnon propagation is crucial. Chen *et al*. [260] have experimentally found that magnon excited by Co nanowires on top of YIG can propagate unidirectionally depending on the magnetic configuration between Co and YIG, specifically in parallel and antiparallel states. Such unidirectional emission of magnons is theoretically explained by Yu *et al*. [261] as a chiral excitation due to the interlayer dynamic dipolar coupling [262] between the Co nanowire and YIG film. When the magnetizations of Co and YIG are in an antiparallel configuration, the effect is significantly enhanced as the strong interlayer magnon-magnon coupling is achieved. [263] The chirality manifested in this hybrid magnonic system is further understood and interpreted with its topological nature exhibiting tunable magnonic Chern bands. [264] As shown in Fig. 4.1, the experimental observation of unidirectional magnons may be considered as the 1D counterpart of the chiral magnon edge modes theoretically predicted by Shindou *et al*. [265] in a 2D bicomponent MC. [75, 266]

In addition, Yu et al. demonstrated that unidirectional propagating magnons, excited by a driven nanowire, can be fully trapped by a second, initially passive nanowire via a dynamical interference effect. [267] The spin waves generated by the first nanowire interact with the second one, which is shielded from direct microwave excitation, thereby exciting its magnetization and prompting it to emit its own spin waves. When these spin waves from both sources interfere destructively outside the nanowires, the pair effectively forms a magnonic cavity that confines the traveling waves, independent of the geometric phase shift arising from their separation. Unlike conventional standing-wave formation, which relies on repeated reflections, this mechanism exhibits robustness against disorder and enables nearly perfect spin and energy transfer between the wires.

Although the mechanisms for the 1D and 2D chiral magnons both stem from the dipolar interaction, the proposed 2D MC with out-of-plane magnetization is highly challenging to prepare while maintaining a low damping for both material systems of YIG and Fe. Even though one can fabricate such a device and be able to excite the magnon edge mode, it is hardly possible to spatially probe the magnon edge mode by micro-focused BLS (µ-BLS). For a perpendicularly magnetized film, the static magnetization is out-of-plane, and therefore the dynamic magnetic component is in-plane. As BLS is mostly sensitive to the out-of-plane dynamic magnetization to the first-order magneto-optical effect, [268] the chiral magnon edge mode is hardly detectable by the µ-BLS technique [269] (a more recent and comprehensive study on the mechanism of µ-BLS was provided recently by Wojewoda *et al*. [270]). As a result, there has been no experimental demonstration of the chiral magnon edge mode theoretically proposed by Shindou *et al*. [265] despite a reasonable analogy with its 1D counterpart observed by Chen *et al*. [260]. However, recent progress on the fabrication of low-damping YIG film with perpendicular anisotropy [271] provides an opportunity to overcome the challenge in the preparation of the proposed device.

Moreover, recent studies have demonstrated the extended capability to probe forward-volume exchange SW in YIG films with deeply nonlinear excitation, which provides an exceptional solution for efficient detection of chiral magnon edge mode in a perpendicularly magnetized film. [272]

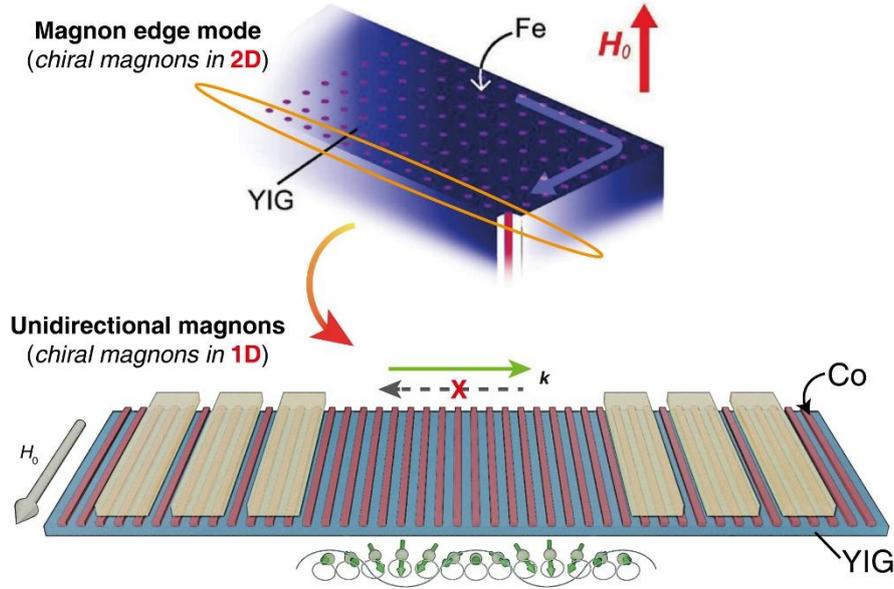

**Fig. 4.1** Schematic illustration of a bi-component magnonic crystal with Fe nanowires embedded in YIG, where a chiral magnon edge mode is theoretically predicted by Shindou *et al*. [265]. The 1D version of the chiral magnons may manifest itself as unidirectional magnons, which are experimentally observed by Chen *et al*. [260] in a hybrid magnetic nanostructure consisting of periodical Co nanowires grown on a YIG thin film. Images rearranged and reproduced from Refs. [260, 265]

Recently, Wang et al. [273] have demonstrated a fast-switching method to control the directionality of forward-volume magnon modes by simply implementing an alternating Oersted field to modify the magnon dispersion. The dipolar interactions not only lead to unidirectional excitation of coherent magnons, but can also induce nonreciprocity for thermally excited magnons, as demonstrated in a YIG/NiFe hybrid structure by the inverse spin-Hall effect [274] and also the BLS detection.[275] While NiFe nanowires may affect the magnon propagation in YIG, conversely propagating magnons in YIG can affect the magnon dynamics in magnetic nanowires due to the effect of a magnon trap; if the magnon accumulation reached some threshold level, one can even trigger the magnetization switching [276,277] in the NiFe nanowires by magnon spin torque. [278,279] The unidirectional magnons are also experimentally demonstrated in synthetic antiferromagnets with the propagating SW spectroscopy [280,281,282,283] owing to interlayer magnon-magnon coupling induced by dynamic dipolar interactions and even unidirectionality for zero-momentum magnons as recently demonstrated with BLS detection. [284] However, if the magnetization dynamics enter the nonlinear regime, [285] the interlayer three-magnon coupling [286] is triggered, where one magnon in the ferromagnetic nanowire (e.g., CoFeB) will split into two counter-propagating magnons at half of the original magnon frequency and with positive and negative wavevectors, where the unidirectionality vanishes.

*4.1.2 Chiral magnon propagation induced by the Dzyaloshinskii-Moriya interaction*

As previously discussed in Section 3, the DMI may introduce a drift in magnon dispersion [287,288,289] with an asymmetric linear term of $\omega_a = p \frac{2\gamma}{M_s} Dk$, as shown in Fig. 4.2(a). [290] Here, $p = \pm 1$ indicates the sign of the drift determined by the chiral property of $(\hat{\mathbf{n}} \times \hat{\mathbf{H}}) \cdot \hat{\mathbf{k}}$ with $\hat{\mathbf{n}}$, $\hat{\mathbf{H}}$ and $\hat{\mathbf{k}}$ denote the unit vector of film normal direction, magnetic field direction and magnon wavevector direction, respectively, as shown in Fig. 4.2(b). [291] As a direct consequence of the magnon dispersion drift, a chiral magnon group velocity arises naturally as

$$v_g^{\text{DMI}} = \left[(\hat{\mathbf{n}} \times \hat{\mathbf{H}}) \cdot \hat{\mathbf{k}}\right] \frac{2\gamma}{M_s} D \qquad (7)$$

This indicates that magnons propagating towards, e.g., +**k** direction are faster than those propagating towards the reversed directions -**k** as presented in Fig. 4.2(b). Notably, it is nontrivial to discover the DMI in magnetic oxide insulators with ultra-low damping. [292] The DMI in centrosymmetric insulating iron garnets, including YIG, had been undetected for decades despite their ubiquity in spintronic and magnonic research. The recently observed chiral SW velocities in YIG thin films provide an important measurement protocol to characterize the DMI in thin-film garnets, together with some other recent experimental studies on current-driven DW motions [293,294] that characterize the DMI in ultra-thin films of $Tm_3Fe_5O_{12}$ (TmIG) and $Tb_3Fe_5O_{12}$ (TbIG) covered with Pt. The chiral magnon transport in ultra-thin yttrium-iron-garnet films will be further elaborated in the latter part of this chapter.

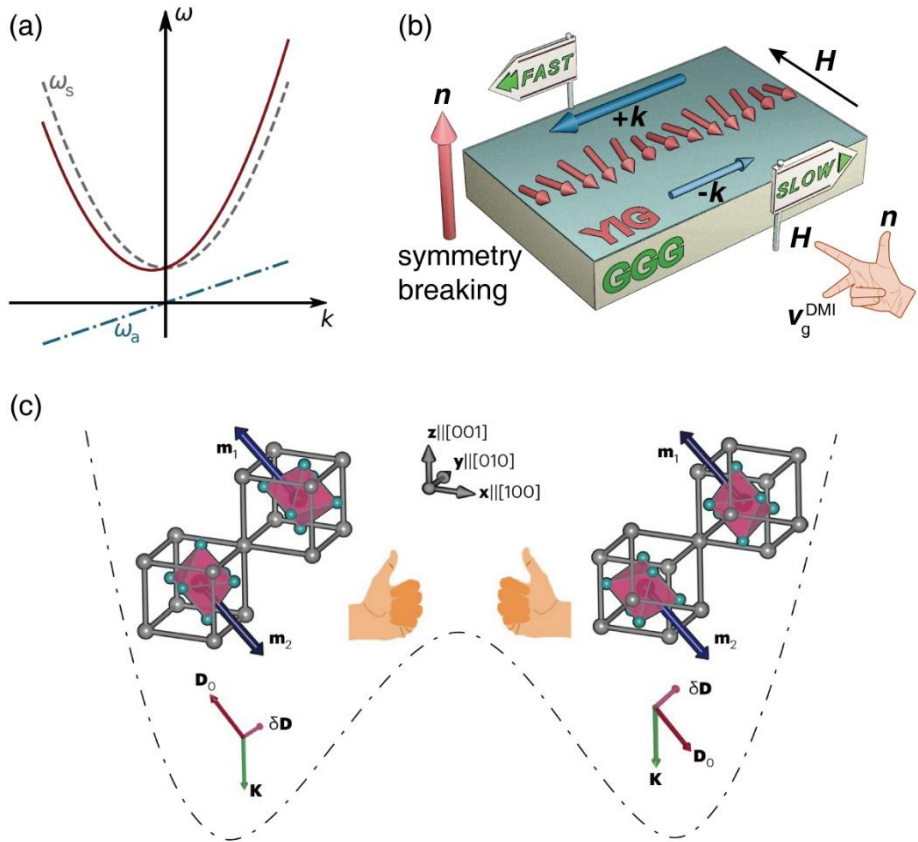

**Fig. 4.2** (a) Magnon dispersion in the presence of the DMI. The actual magnon dispersion (red solid line) consists of the symmetric parabolic part (gray dashed line) and the asymmetric linear contribution arising from the DMI (blue dash-dotted line). (b) Schematic presentation of the chiral group velocity ($v_g^{\text{DMI}}$) induced by the DMI. The sign of the chiral group velocity is determined by the cross product of the magnetic field **H** and symmetry-breaking direction **n** with respect to the magnon wavevector **k**. (c) Schematic mechanism for the chiral propagation of magnons in $BiFeO_3$ in the presence of the DMI. $\mathbf{D}_0$ represents the DMI that dictates the spin cycloidal structure, which couples with the electric polarization being controllable with an applied electric field. Images rearranged and reproduced from Refs. [290,2911,292].

Not only ferromagnets or ferrimagnets exhibit DMI. In fact, the antiferromagnets can possess DMI as well, such as $BiFeO_3$, a well-known multiferroic, namely being simultaneously antiferromagnetic and ferroelectric. [295] Recent studies by Huang *et al*. [296] have demonstrated the electric control over the magnon spin transport. The applied electric field can switch the ferroelectric polarization, and thereby modify the DMI [$D_0$ in Fig. 4.2(c)] in association with the spin cycloidal structure. [297] The DMI was originally discovered and introduced by Dzyaloshinskii [298] and Moriya [299] in order to explain the weak ferromagnetism in antiferromagnets such as hematite (α-$Fe_2O_3$). [300] Spin pumping of hematite in its easy-plane phase [301,302] is found to be enhanced by the presence of the DMI. Recent observation of magnon nonreciprocity [303,304] and alternating magnon polarization [305] in hematite may arise from its intrinsic DMI and/or dipolar interactions.

*4.1.3 Non-Hermitian magnonic effects in magnetic arrays*

Recent advances in magnonics have uncovered striking non-Hermitian phenomena in magnetic arrays with long-range, chiral interactions. Non-Hermitian physics and properties offer functionalities for next-generation devices, spanning ultra-sensitive sensors, energy funneling mechanisms, wave isolators, nonreciprocal signal amplifiers, and dissipation-driven phase transitions. Non-Hermitian topological magnonics focuses on engineering dissipation and/or gain to realize such phases and properties in magnetic systems, which unlocks remarkable capabilities, including dramatic enhancements in magnonic frequency combs, magnon amplification, quantum-grade magnetic field sensing with exceptional sensitivity, magnon accumulation, and perfect microwave absorption. Yu et al. deliver a thorough review of these advances, encompassing theoretical foundations for non-Hermitian magnonic Hamiltonians, exceptional nodal phases, and skin effects in chiral configurations. [306] Two pivotal studies from T. Yu and his co-workers explore how chirality, damping, and topology conspire to produce novel localization effects of SW modes. [307,308] In Ref. [307], authors investigate a 1D system composed of a periodic array of magnetic nanowires placed atop a thin magnetic film. The nanowires support discrete magnetic resonances (Kittel modes), and the underlying film acts as a SW medium that mediates dynamic, long-range coupling between them. The critical feature of this configuration is the chirality of the coupling; due to the relative orientation of the wire and film magnetizations, the SWs propagate asymmetrically, preferentially in one direction. This results in nonreciprocal interactions across the array. Crucially, this chirality, combined with the damping in the SW medium, leads to a striking manifestation of the non-Hermitian skin effect. Unlike conventional Hermitian systems, where eigenmodes are delocalized, non-Hermitian systems with asymmetric couplings can support eigenmodes that accumulate exponentially at a boundary—a phenomenon known as the skin effect. In the system studied here, the long-range chiral interactions naturally generate a non-Hermitian effective Hamiltonian, and damping suppresses destructive interference from counterpropagating waves, allowing robust edge-localization of all eigenmodes. The main experimental finding is that when a weak local microwave excitation (in the microtesla range) is applied to one edge of the array, a giant response is observed at the opposite edge, making it a compelling platform for microwave sensing and magnonic signal amplification. This highly non-local and nonreciprocal effect arises because all SW energy accumulates on one side due to the skin effect.

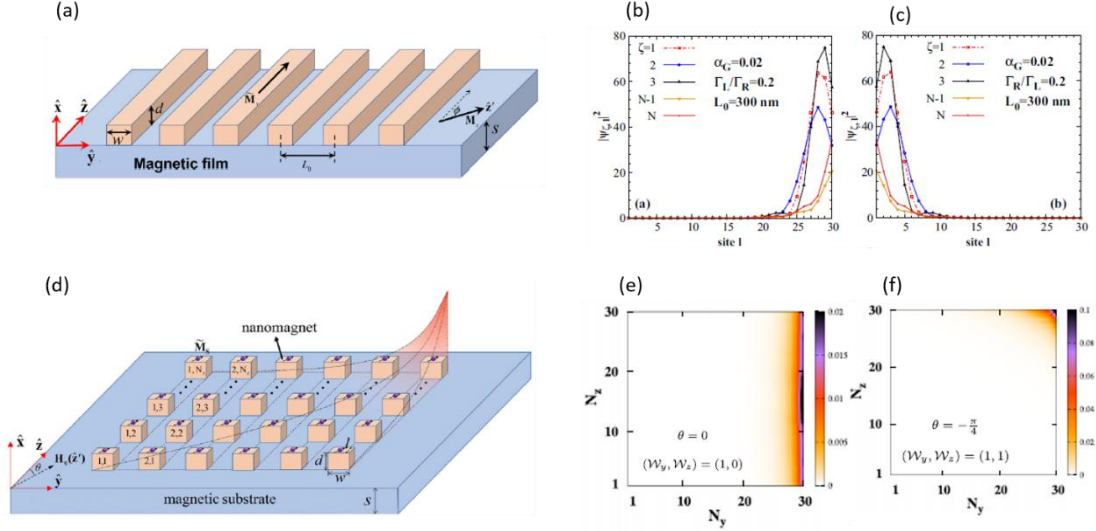

**Fig. 4.3** A periodic array of (a) 1D and (d) 2D array of nanowires and nanomagnets on top of a thin magnetic film. Distribution of normalized eigenmodes under different conditions: all the modes are localized at the edge in (b) and (c) when the coupling is chiral and film damping is strong.[307] (e,f) Edge or corner aggregations of the magnon eigenstates for two different orientations of the magnetic field $\theta = 0$ and $\theta = \pi/4$ and $W_y, W_z$ [308].

Extending the 1D work, Cai *et al*. explore a 2D array of magnetic nanomagnets placed on a magnetic film. [308] The system again features chiral, long-range dynamic couplings mediated by film SWs. However, the 2D nature introduces qualitatively richer physics: now, localization can occur not just along edges but also at corners, depending on the symmetry of the coupling and the direction of the applied magnetic field. The system's response is tunable via the direction of the in-plane magnetic field. For example, when the field is aligned with one of the array axes (say, the $x$-direction), the chiral couplings become asymmetric along $x$ but symmetric along $y$, resulting in edge localization along $x$. Rotating the field diagonally (e.g., at 45°) induces asymmetric couplings in both $x$ and $y$, driving corner localization at one of the array's four corners.

The key theoretical innovation introduced in these works is the topological winding tuple: a pair of integers $(W_y, W_z)$ that quantifies the complex spectral winding of the non-Hermitian system. Each component corresponds to a winding number of the complex eigenfrequency as one traverses the BZ zone along a specific direction (e.g., $k_y$ or $k_z$). This tuple plays a critical role in predicting the spatial localization of magnon modes:

- If $W_y \neq 0$ and $W_z \neq 0$, eigenmodes accumulate along one vertical edge.
- If $W_y = 0$ and $W_z \neq 0$, localization occurs along one horizontal edge.
- If both winding numbers are nonzero ($W_y, W_z \neq 0$) magnons localize at a single corner of the 2D array.

The tunability, combined with the predictive power of the winding tuple, allows one to program where SW energy accumulates simply by rotating the magnetic field. Importantly, the results hold even for long-range couplings that defy traditional Bloch-band-based topology. Interestingly, edge and corner skin effects can coexist in a single system, leading to high energy accumulation, and can be selected dynamically by field orientation.

*4.1.4 Chiral magnon edge modes in stripe domains*

The long-sought-after electronic edge and surface states in topological insulators [309] are promising to realize electron transport immune to backscattering from defects and along the crystal edge, offering the possibilities to construct novel electronic devices with low-dissipation, benefiting from the nontrivial topology of the materials. The topological edge and surface modes have been experimentally demonstrated in topological insulators such as $Bi_2Te_3$ at low temperatures (~1 K). [310] It hitherto remains challenging to realize such topological edge states for electrons at room temperature for potential applications. Following the relevant concept and methodology, topological edge states or edge modes for magnons [311,312,313] are proposed theoretically in analogy to those in electronic systems. Unfortunately, in most of the theoretical studies, the dipolar interactions are directly dropped out, given the argument that they are "weak" and "negligible", even though most of the experimental works in the field of magnonics study dipolar magnons such as Damon-Eshbach and backward volume modes. This leads to the predictions of novel topological magnons at high energy levels, e.g., 100 meV (~24 THz) and with atomic wavelengths (~1 nm) that are almost impossible to realize in experiments, particularly for coherent excitation and detection. Opportunities arise when Shindou *et al*. [265] studied chiral magnon edge mode based on dipolar interaction in a hybrid magnetic system, as shown in the upper panel of Fig. 4.1. As previously mentioned, it is almost unrealistic to fabricate such bicomponent magnonic device by etching and embedding magnetic nanowires into another magnetic system. However, spatially periodic noncollinear spin textures such as skyrmion lattices [314] may potentially do the same job as the proposed artificial magnonic crystal. Topological magnon edge modes were theoretically studied in antiferromagnetic skyrmion crystals, [315] considering DMI (no dipolar interactions). Recent experiments have investigated topological magnon bands in a skyrmion lattice using polarized inelastic neutron scattering. [316] More recently, Che *et al*. [317] have experimentally observed the short-wave magnons in the topological bands of a skyrmion lattice in single-crystal $Cu_2OSeO_3$. The direct experimental observation of topological magnon edge modes in magnetic textures remains challenging.

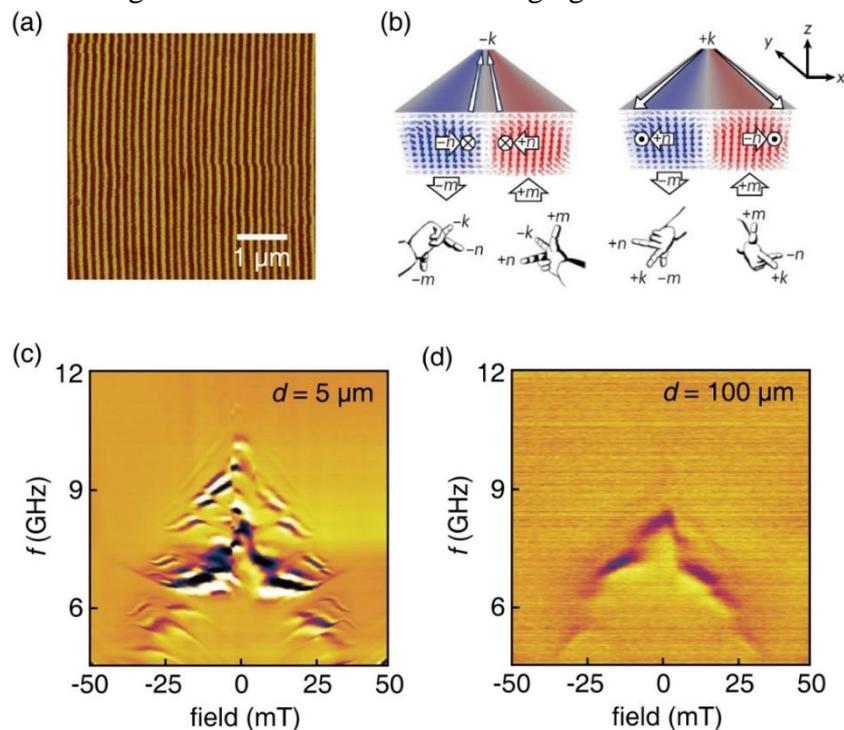

**Fig. 4.4** (a) Magnetic force microscope imaging of the periodical stripe domains in an epitaxial LSMO film in the top-view perspective. The periodicity of the stripe domains is approximately 130 nm. (b) Conceptual schematic of the magnon edge modes located within the domains stripe channels. Chirality is dictated by the three orthogonal vectors: domain magnetization (**m**), magnon wavevector (**k**), and edge normal (**n**). (c) Transmission spectra $S_{12}$ (Imaginary part) measured as a function of an in-plane magnetic field along the wavevector direction, where magnons are excited by CPW2 and detected by CPW1 with a propagation distance $d = 5$ μm. (d) Transmission spectra $S_{12}$ (Imaginary part) measured over a long propagation distance of $d = 100$ μm. The magnetic field is applied along the wavevector direction. Images reproduced and rearranged from Ref. [322].

Apart from the 2D skyrmion lattices, some materials can host 1D spin texture alignment as stripe domains. [318] The stripe domains can act as one-dimensional magnonic crystals, where the magnon band structures have been studied in Fe-N thin films. [319] As the magnetic damping of the Fe-N thin film is still relatively high (~0.01), it is difficult to achieve long-distance magnon propagation in the system. The magnetic oxide $La_{1-x}Sr_xMnO_3$ (LSMO), a perovskite well known for its colossal magnetoresistance, can also host such stripe domains as shown in Fig. 4.4(a) captured by magnetic force microscopy. By growing LSMO on $NdGaO_3$ substrate, Qin *et al*. [320] achieve an ultralow magnetic damping of LSMO thin film down to 0.0005, and explained this low damping based on a breathing Fermi surface model. The periodic stripe domains with alternating up and down magnetic domains can form a macroscopic antiferromagnetic order with an effective lattice constant around 130 nm. The collective dynamic mode of such magnetic order can be considered an effective antiferromagnetic resonance, [321] and has been observed in LSMO with a resonance frequency of approximately 10 GHz at zero magnetic field. The resonance frequency of this mode decreases with an applied in-plane magnetic field. When the external field exceeds a certain threshold value of around 100 mT, the spin textures are saturated into uniform magnetization. Consequently the resonance frequency goes up with the magnetic field as conventional ferromagnetic resonance modes. When the effective antiferromagnetic resonance mode hybridizes with the DW mode, [102,104] then we observe intricate transmission spectra [322] representing some of the features of magnon band structures around 5-10 GHz as shown in Fig. 4.4(c) with a propagation distance $d = 5$ μm. The transmission spectra show multiple anticrossing features that can be reproduced in comparison with micromagnetic simulations. The simulation results indicate that some of the hybridized modes, particularly the lower branches, exhibit a salient feature of chiral propagation as presented in Fig. 4.4(b). For a given wavevector direction (e.g., $+k$), magnons propagate along the right or left edge of the domain depending on the domain magnetization (**m**) orientation, which points up or down. When the propagation direction reverses, magnons propagate along the left edge instead of the right one in the domain with up magnetization. The chirality follows a simple cross-product relation of $\hat{\mathbf{n}} = \hat{\mathbf{k}} \times \hat{\mathbf{m}}$, where $\hat{\mathbf{n}}$ denotes the left $(+n)$ or right $(-n)$ edge, $\hat{\mathbf{k}}$ is the magnon wavevector direction. Such chirality of the magnon edge mode in the stripe domain is dominated by the dynamic dipolar interaction in the spin texture. The chirality generated by dynamic dipolar coupling can be theoretically understood in terms of generalized spin-orbit interactions [323] in low-symmetry magnetic systems. Interestingly, the chiral edge modes appear to propagate over substantially long distances of more than 100 μm. Figure 4.4(d) shows the transmission spectra measured with two coplanar waveguides placed about 100 μm apart. Although the signal is much weaker than the short-distance measurement as in Fig. 4.4(c), it is quite surprising to detect sizable transmission signals over a propagating distance as long as 100 μm in a metallic system, such as LSMO. The intrinsic edge mode nature may potentially lead to the long-distance detection, avoiding magnon backscattering from magnetic defects similar to its electronic counterpart. Future studies are imperative to understand more of the microscopic mechanism and further improve the functionality and capability of magnon edge modes in magnonic devices. Che *et al*. [324] employed scanning transmission x-ray microscopy to probe coherent magnons propagating in stripe domains and domain walls formed in Fe/Gd multilayers. The magnons are detected around 1 GHz with short wavelengths down to about 280 nm and exhibit strongly nonreciprocal propagation in the domain-wall channels.

*4.2 Magnon localization in 2D and quasi-2D materials*
Recently discovered 2D magnetic materials, such as $CrI_3$ and $Cr_2Ge_2Te_6$, with atomic-layer thickness, provide a new dimension for exploring magnonics in 2D systems. These magnons exhibit modified characteristics, where enhanced quantum fluctuations and reduced dimensionality amplify the roles of magnetic anisotropy, interlayer exchange coupling, and the reduced dipolar interaction. These 2D magnons exhibit distinct properties compared to their 3D counterparts, offering a pathway for investigating quantum magnonic phenomena and designing magnonics devices for future 2D magnonic network.

*4.2.1. Magnons in ultra-thin yttrium iron garnet films*

Traditionally, most experimental studies in magnonics focus on investigating SW propagation in mm-thick yttrium iron garnet films prepared by liquid phase epitaxy (LPE). [325] As a result, the dipolar interactions dominate the SW dispersion, particularly in the magnetostatic (low-wavevector) regime, leading to MMSW, BVMSW and forward volume modes depending on the relative orientation between the wavevector and the magnetization. [22] Only in 2014, two groups [326, 327] independently achieved high-quality growth of YIG films with a thickness down to about 10-20 nm. Recently, Wei *et al.* [328] have managed to prepare ultra-thin YIG films grown by LPE [329] and study the nonlocal magnon transport in the films with different thicknesses from 3.7 nm to 53 nm. The measurement protocol is presented in Fig. 4.5(a). Combined with their previous studies, [330] it is found that the magnon decay length is substantially enhanced for nanometer-thick films compared to the thicker (≥210 nm) films as summarized in the chart of Fig. 4.5(b), where the data for the 3.7 nm-thick film are highlighted. The optimized magnon transport in ultrathin YIG films is further explained theoretically considering a quasi-2D scenario, as 3.7 nm is merely about three lattice constants of YIG, leading to a transition from 3D to 2D. [331]

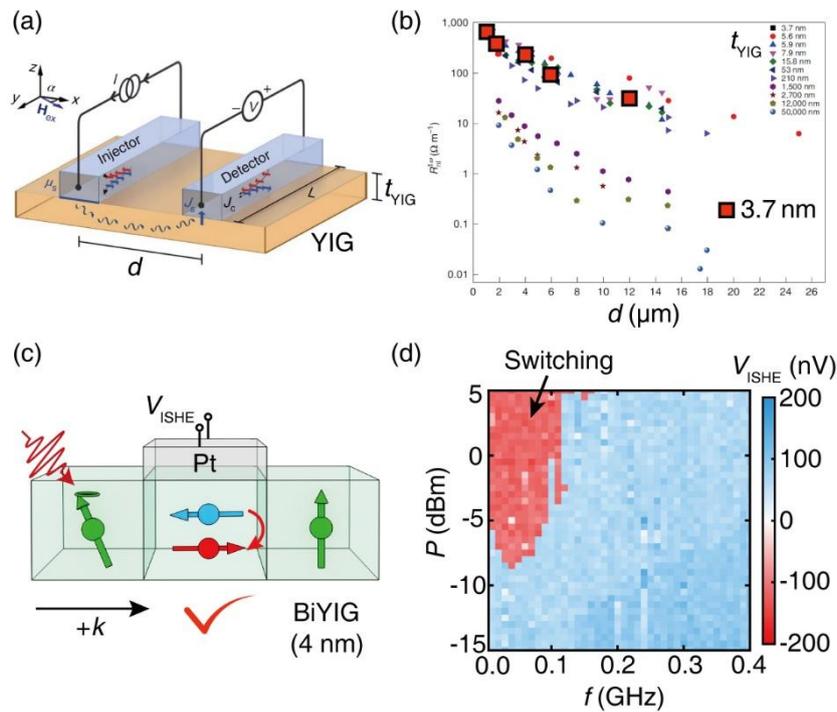

**Fig. 4.5** (a) Device layout for nonlocal magnon transport in ultrathin YIG films with two Pt bars deposited on top. The YIG films are prepared with the different thickness ($t_{YIG}$) down to 3.7 nm. The distance ($d$) between the Pt injector and detector is varied in different devices. (b) Nonlocal first-harmonic resistance measured on YIG thin films with different thickness $t_{YIG}$ from 3.7 nm to 5 μm and plotted as a function of the injector-detector distance $d$. The data points for $t_{YIG}$ = 3.7 nm are highlighted as red squares for better visibility. (c) Magnon-driven switching between different chiral spin frustration states in a 4 nm-thick BiYIG film. Magnons are coherently excited at the left side of the Pt bar and propagate towards the in-plane region to switch the local magnetization orientation beneath Pt. (d) The measured data of the switching experiment as a function of excitation frequency ($f$) and excitation power ($P$). At around 0.4 GHz, a threshold power of -8 dBm is required to trigger the switching by chiral magnon torque injected from the left side (torque injected from the right side is ineffective, not shown). Images reproduced and rearranged from Refs. [328] and [335].

Most spintronic devices favor perpendicular magnetic anisotropy (PMA) for applications. [332] By substituting yttrium by bismuth in YIG, one may prepare BiYIG films with PMA and still maintain low magnetic damping, [271] in which a record-high speed of domain walls was observed of up to 4.3 km/s. [333] Lin *et al*. [334] managed to grow BiYIG with a thickness below 10 nm while possessing low damping and PMA. Very recently, Wang *et al*. [335] leverage magnon spin torque to switch between different chiral spin frustration states in a 4 nm-thick BiYIG film as illustrated in Fig. 4.5(c). The ultrathin BiYIG film grown on a $Y_3Sc_2Ga_3O_{12}$ substrate and covered by a Pt bar on top, exhibits a sizable DMI [291,292] of about 5 $\mu J/m^2$ in the system, which leads to chiral spin frustration beneath the Pt bar probed by the nitrogen-vacancy (NV) magnetometry. [336] When magnons are injected coherently from the left side at around 0.4 GHz, the magnetization beneath the Pt bar can be switched from left to right, given the condition that the excitation power exceeds a threshold of -8 dBm, as presented in Fig. 4.5(d). If one needs to switch back the magnetization from right to left, then magnons should be injected coherently from the right side instead, indicating a chiral behavior of the magnon spin torque due to the DMI in the ultrathin BiYIG film. The magnon-driven switching of the chirally frustrated spin states stems from an intrinsic asymmetry in the system's energy landscape. To illustrate this, consider a minimal three-spin system consisting of left and right out-of-plane spins and a middle in-plane spin. Due to competing interactions such as Dzyaloshinskii-Moriya interaction, the middle spin experiences frustration in its coupling with one of the edge spins. That is, their energetically preferred orientations become mutually incompatible. This frustrated state is exceptionally sensitive to external perturbations. When a spin wave excites the OOP spin that is frustrated with respect to the IP spin, the resulting effective torque on the middle spin is significantly enhanced. This enhancement occurs because the contributions from exchange and DMI constructively interfere, yielding a large torque capable of driving the switching of the middle spin. In contrast, when the excited OOP spin shares a non-frustrated interaction with the IP spin, the exchange and DMI contributions compete destructively, leading to a markedly weaker torque that is insufficient to induce switching. Thus, the chirality-dependent switching is not externally imposed but inherently encoded in the frustrated spin configuration. The system is predisposed to respond strongly to magnons propagating from one direction while remaining largely inert to those from the opposite direction.

*4.2.2 Magnons in van der Waals magnetic materials*
The discovery of 2D materials, such as graphene, has revolutionized condensed matter physics by offering fresh insights into material properties and their potential applications. These atomically thin structures exhibit unique electronic, mechanical, and thermal properties that starkly contrast with those of their bulk counterparts, enabling unprecedented control over physical phenomena at the nanoscale. In 2017, two independent research groups identified 2D van der Waals magnets, namely $CrI_3$ and $Cr_2Ge_2Te_6$, which introduced a new class of materials with intrinsic magnetic properties in the 2D limit. [337,338] This breakthrough has heralded a new era for spintronics and magnonics, broadening the scope of both fundamental research and technological innovation. Due to the inherent 2D nature of these materials, magnons are intrinsically localized in 2D, displaying distinct differences from their 3D counterparts in bulk magnets. [339,340]
Experimental studies of 2D magnons have been conducted on $CrI_3$ using Raman spectroscopy and MOKE techniques, respectively. [341,342] Optical methods prove effective for detecting SWs in 2D materials, with the addition of monolayer $WSe_2$ for featuring strong excitonic resonance in order to enhance the sensitivity. Two magnon modes, namely the optical and acoustic modes, have been observed with energy levels of approximately 17 meV and 0.3 meV, respectively. Additionally, two degenerate optical magnon modes with opposite angular momentum and conjugate optical selection rules have been identified. Sub-terahertz magnetic resonances in bilayer $CrI_3$ have enabled the determination of anisotropies and interlayer exchange fields. Furthermore, electrostatic gating enables the tuning of the antiferromagnetic resonance in CrI$_3$ by tens of gigahertz, allowing for the control of 2D confined magnons. However, these techniques predominantly detect the antiferromagnetic resonance modes of 2D magnons in van der Waals magnets, rather than the propagating magnons, which are critical for magnonic applications in logic computing and information processing. CrSBr is an A-type antiferromagnetic van der Waals semiconductor that exhibits tightly bound excitons with significant oscillator strength and coherent magnons, attributed to its bandgap and spatial confinement. [343] As a direct-gap semiconductor down to the monolayer, CrSBr features an electronic bandgap of 1.5 eV and an excitonic bandgap of 1.34 eV. Despite the orders-of-magnitude difference in energy levels between magnons and excitons, their indirect coupling enables efficient optical detection of magnons. In CrSBr, the strong magnon-exciton coupling modulates the frequency of above-gap excitons, enabling the direct measurement of exciton frequency shifts to reflect magnon modes, providing the opportunity for studying the magnon propagation in such material. Atomically thin CrSBr flakes can be exfoliated from bulk single crystals while preserving the material properties of the bulk crystal. Using pump-probe techniques, magnons in CrSBr have been demonstrated to propagate up to 7 μm with a coherence time exceeding 5 nanoseconds, as shown in Figs. 4.6(a) and (b). By varying the external magnetic field direction relative to the crystal lattice, the rotational symmetry of the crystal can be adjusted, thereby tuning the exciton-magnon coupling. Furthermore, applying uniaxial strain modulates the magnon dispersion, resulting in the emergence of a dispersionless dark magnon band at a critical strain threshold. These confined magnons in van der Waals magnetic thin layers are further demonstrated to be mediated by long-range dipole-dipole interactions rather than short-range exchange interactions, as shown in Figs. 4.6(c) and (d). The dipolar SW packets are found to propagate in the CrSBr thin flakes with different thickness. The magnon group velocity is found to depend on the film thickness, resulted from the dipolar interaction. Additionally, Diederich *et al.* demonstrated pronounced nonlinear coupling between magnons and excitons in CrSBr, observing exciton states dressed by up to 20 magnon harmonic modes up to the frequency of sub-terahertz, due to their extreme nonlinearities, as shown in Fig. 4.6(e).[ 344 ] Leveraging the versatility of van der Waals heterostructures, these coherent 2D magnons in CrSBr offer a promising foundation for optically accessible magnonics.

Utilizing a microwave antenna, magnons can be efficiently generated in van der Waals magnets. The conventional ferromagnetic resonance technique, employing a broad coplanar waveguide or a transmission line, has been used to investigate both ferro- and antiferromagnetic resonances in corresponding van der Waals magnetic single crystals. By integrating a microwave stripline directly onto an exfoliated thin flake, coherent magnon propagation was observed in a nanometer-thick flake of the 2D van der Waals ferromagnet Fe$_5$GeTe$_2$ below its Curie temperature, as detected via X-ray microscopy and shown in Fig. 4.6(f). [345] Two distinct frequencies, 2.77 GHz and 3.84 GHz, were measured, corresponding to magnon wavelengths of 1.5 μm and 0.5 μm, respectively. This method provides simultaneous access to phase and amplitude information in both the BV and DE configurations, enabling the determination of magnon dispersion in Fe$_5$GeTe$_2$ despite the material high damping. Additionally, SW dynamics have been investigated in the antiferromagnetic van der Waals material CrPS$_4$ using all-electrical spectroscopy. [346] Two CPWs were employed for the excitation and detection of magnon propagation through CrPS$_4$, revealing strong coherent magnon-magnon coupling between the optical and acoustic magnon branches, which results in hybridized magnon modes. These modes can be precisely tuned by modulating the coupling strength due to the breaking of the rotational symmetry, offering a broad frequency tunability exceeding 10 GHz. The microwave-based generation and detection of magnons in van der Waals magnets present significant opportunities for developing energy-efficient magnonic devices operating at accessible frequencies, opening a pathway for advanced spintronic and magnonic applications.

Although experimental studies of 2D magnons have been carried out on monolayer CrI$_3$, the observed signals are more likely associated with $k = 0$ resonances or thermal magnons rather than coherent magnon modes. Observing coherent magnon propagation in such confined 2D systems therefore remains a significant challenge. An important approach to probe magnon transport in CrSBr flakes is through magnon–exciton coupling. However, as the thickness decreases, the dipolar interaction weakens, leading to a flatter magnon dispersion near $k = 0$ and this flattening results in locally confined magnons that cannot propagate efficiently. Hence, techniques capable of exciting and detecting 2D magnons with large wavevectors are highly desirable. Moreover, because of the extremely small volume of monolayer or bilayer 2D magnets, inductive detection of magnon signals becomes increasingly difficult. This underscores the need for quantum sensing techniques, such as NV magnetometry, to achieve efficient detection of 2D magnons.

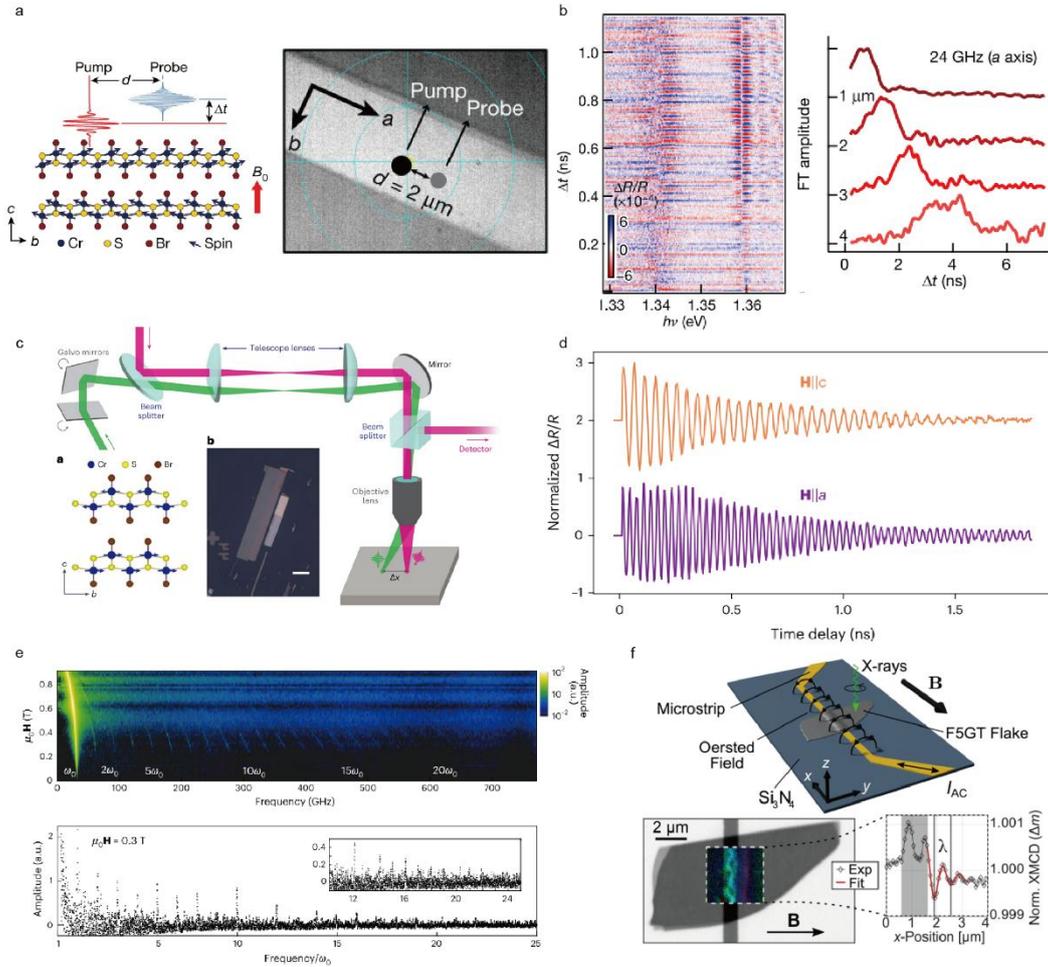

**Fig. 4.6** Excitation and detection of magnons in van der Waals magnetic materials. (a) Schematic of the pump–probe technique used to detect exciton-magnon coupling in CrSBr. (b) Transient reflectance spectra corresponding to the measurement in a (pump–probe separation $d = 2$ μm). Probe-wavelength-integrated spectra at different pump–probe separations for the 24 GHz magnon mode along the $a$-axis (from C). (a) and (b) are taken from Ref. [341]. (c) Schematic of the experimental setup with two orthogonal galvo-driven mirrors and a 4f telescope allow automated control of the pump–probe separation (taken from Ref. [342]). (d) Transient reflectance measured on CrSBr flakes under different directions of the applied magnetic field (taken from Ref. [342]). (e), Field-dependent magnon spectrum showing high-harmonic magnon modes (taken from Ref. [344]). (f) Probing magnon propagation in a Fe5GeTe2 flake using TR-STXM. Reproduced with permission from Ref. [345].

## 5. Magnon trap using external fields

The control of magnons using external methods, such as electric, thermal, and mechanical techniques, is central to effective magnon manipulation. From a theoretical perspective, magnon trapping relies on creating spatially varying effective potentials that modify the local magnon dynamics. Within the framework of the Landau–Lifshitz equation, such confinement can be described analogously to particles in quantum wells. In this chapter, we review recent advancements in trapping magnons through nanoscale microwave cavities, static and resonant fields, and thermal effects. These approaches enable the confinement of magnons within nanoscale regions, offering potential for precise control and manipulation of magnons into the quantum regime.

*5.1 Microwave and RF field trapping*

Cavity magnonics, also known as spin cavitronics, investigates the coupling between microwave photons and magnons in microwave cavities or resonators. This field has emerged as an important branch of magnonics, enabling efficient energy exchange between diverse quantum systems. In this section, we review magnon-photon coupling at the nanoscale, which differs from the conventional use of millimeter-sized YIG spheres, offering enhanced coupling efficiency at nanoscale dimensions.

*5.1.1 Efficient magnon-photon coupling in the nanoscale*

Cavity magnonics is an emerging field that explores the coupling between magnons and microwave photons confined in high-quality electromagnetic cavities. [347,348] Strong magnon-photon coupling, typically demonstrated in magnetic systems such as polished YIG spheres, provides a powerful platform for studying hybrid quantum systems, enabling coherent quantum information transfer and precise control of spin dynamics. To date, submillimeter-sized YIG spheres have been the primary focus in cavity magnonics due to their large number of spins and exceptionally low Gilbert damping. [349] According to the Dicke scaling law, the coupling strength scales as $g \propto \sqrt{N}$ where $N$ is the number of spins. As a result, when magnetic elements are miniaturized to the nanoscale-such as in integrated magnonic devices-the reduced spin number leads to significantly weaker coupling, making it challenging to achieve coherent microwave control at the nanoscale.

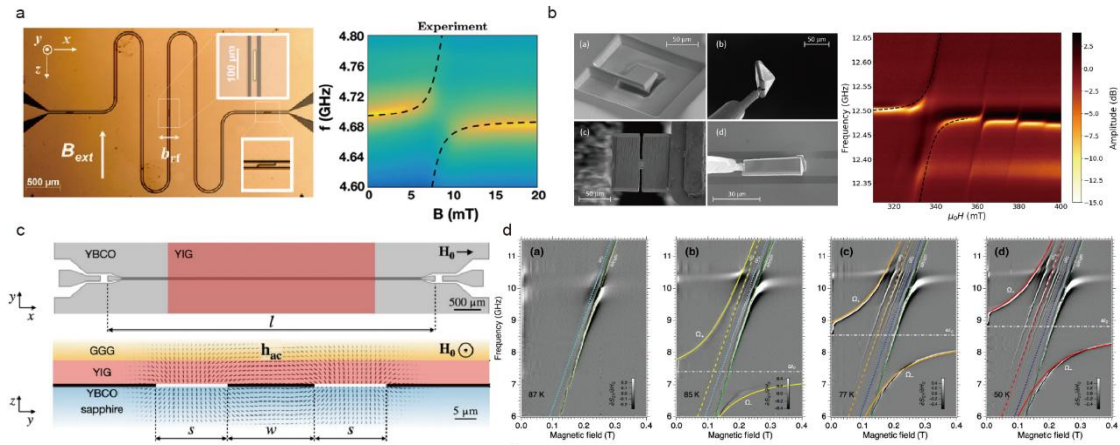

**Fig. 5.1** Nanoscale Magnon-Photon Coupling. (a) Strong magnon-photon coupling in a μm-scale Py stripe on a superconducting resonator (taken from [350]). (b) PFIB technique for precise μm-scale YIG flake placement on superconducting resonators. Multiple anticrossing gaps from confined PSSW modes. Reproduced with permission from [354]. (c) Superconducting YBCO resonator with YIG film stacked atop. (d) Temp-dependent evolution of coupled SW and resonator modes. Dash-dot lines show $\omega_c$ at each temperature. Dashed lines indicate $\omega_{\text{FMR}}$ (green) and $\omega_0$ (blue). Reproduced with permission from [357].

Co-localized magnon and photon modes in the nanoscale may lead to strong magnon-photon coupling at the single-quanta level, opening new pathways for quantum information processing and conversion schemes. In this section, we review recent progress in techniques for achieving efficient magnon–photon coupling in nanoscale magnetic devices, which is essential for quantum information transfer and control in miniaturized systems.

Superconducting resonators are microwave circuits fabricated from superconducting materials such as Al or Nb, and they exhibit ultra-low energy dissipation and extremely high-quality factors (Q-factors). These resonators are widely used in quantum science, particularly for coupling with qubits, photons, and magnons, due to their ability to confine electromagnetic fields within a well-defined, highly localized region. Strong magnon–photon coupling has been demonstrated using superconducting resonators integrated with polished YIG spheres. [349]

However, achieving scalable and miniaturized integrated hybrid quantum system has been challenging. By using lithographically defined Nb superconducting resonators, strong magnon-photon coupling has been observed in a nanometer-sized NiFe thin film stripe, as shown in Fig. 5.1 (a), [350,351] where the number of spins is in the level of $10^{13}$, much lower than the case of YIG spheres. Using a macrospin model, one could derive this hybrid quantum system, where the total Hamiltonian can be written as:

$$\widehat{H} = \hbar\omega_r\left(\hat{a}_r^\dagger \hat{a}_r + \frac{1}{2}\right) - \omega_m(B_0)\hat{S}_z + g_s(\hat{S}_+\hat{a}_r^\dagger + \hat{S}_-\hat{a}_r) \qquad (8)$$

where $\hat{a}_r^\dagger(\hat{a}_r)$ is the creation (annihilation) operator of the microwave photon modes, $\widehat{\boldsymbol{S}} = \frac{1}{2}(\hat{S}_+ + \hat{S}_-)\hat{\boldsymbol{x}} + \frac{1}{2i}(\hat{S}_+ - \hat{S}_-)\hat{\boldsymbol{y}} + \hat{S}_z\hat{\boldsymbol{z}}$ is the macrospin operator, with $\hat{S}_+(\hat{S}_-)$ raising (lowering) the z component of the macrospin. The resonance frequency of the superconducting resonator ($\omega_r$) and the macrospin ($\omega_m$), given by $\omega_r = \frac{1}{\sqrt{LC}}$ and the Kittel formula, respectively. The coupling strength between microwave photons and individual spins can be written as $g_s = g_e\mu_B b_{rf}\omega_r/\sqrt{8\hbar Z_r}$, with $Z_r = \sqrt{L/C}$ as the impedance of the $LC$ resonator. Then the eigenfrequency of the coupled system can be expressed as:

$$\omega_\pm = \omega_r + \Delta/2 \pm \sqrt{\Delta^2 + 4g^2}/2 \qquad (9)$$

where $\Delta = \omega_m(B_0) - \omega_r$ is the detuning frequency and $g = g_s\sqrt{N}$ is the magnon-photon coupling strength. Therefore, to achieve strong magnon-photon coupling while scaling down the number of spins $N$, the coupling strength of the individual spin $g_s$ must be enhanced. Two strategies can be employed to increase $g_s$. The first is to maximize $b_{rf}$ by positioning the nanomagnet as close as possible to the region of maximum RF magnetic field within the resonator. The second approach involves reducing the resonator impedance $Z_r$, which can be achieved by designing a superconducting resonator with high inductance $L$ and small capacitance $C$. By carefully optimizing these parameters, a single-spin coupling strength of $\frac{g_s}{2\pi} = 263$ Hz can be achieved in a nanometer-thick NiFe wire. A 40 μm× 2 μm× 10 nm NiFe wire was patterned on top of a superconducting resonator made of Nb, with the resonator frequency of $\omega_r/2\pi = 5.253$ GHz. The device was mounted in a cryostat at 1.5 K, and the microwave signal was measured using a vector network analyzer. With a small spin ensemble ($N = 7.3 \times 10^{10}$), a magnon-photon coupling strength of $g/2\pi = 74.5$ MHz was observed in the reflection spectra. Given the resonator mode decay rate of $\frac{\kappa_r}{2\pi} = 1.05$ MHz and the magnon mode of $\frac{\kappa_m}{2\pi} = 122$ MHz, the magnon-photon cooperativity can be obtained as $C = g^2/\kappa_r\kappa_m = 43.3$, a remarkably high value considering the small number of spins. This strong magnon-photon coupling between a Kittel mode in Py nanomagnets and a resonator mode in planar superconducting resonators highlights the significant potential for integrated hybrid quantum systems leveraging magnon physics for the nanoscale confined magnons.

Although the NiFe-based ferromagnet/superconductor hybrid device exhibits strong magnon-photon coupling with high cooperativity, the intrinsic high damping of NiFe limits its efficiency for energy-efficient magnon-photon coupling at the nanoscale. In contrast, the ultra-low damping alloy $Fe_{75}Co_{25}$, with its high saturation magnetization and spin density, is better suited for nanoscale magnon-photon coupling. [352] Haygood *et al.* demonstrated strong coupling between a patterned nanometer-thick $Fe_{75}Co_{25}$ bar and a superconducting resonator, achieving a coupling rate exceeding 700 MHz-significantly higher than that of NiFe and well within the strong coupling regime. [353] By varying the volume of the ferromagnet, they showed that the coupling rate scales linearly with the square root of the number of spins.

Due to the absence of magnon-electron scattering, magnetic insulators like YIG offer much lower magnon dissipation rates compared to metallic magnets, making them more suitable for efficient magnon-photon coupling. However, the smallest commercially available YIG spheres, with diameters around 200 μm, are likely incompatible with nanoscale systems. Planar YIG samples are an alternative, but they must be grown on gadolinium gallium garnet (GGG) substrates to achieve desirable damping rates. A key limitation is that GGG exhibits high microwave losses at low temperatures, rendering it a suboptimal substrate for fabricating planar resonators. To solve this challenge, Baity *et al.* develop a method for using Plasma Focused Ion Beam (PFIB) technology to precisely place the YIG thin flakes down to the micrometer scale in order to integrate with superconducting resonators at low temperature, as shown in Fig. 5.1 (b). [354] To achieve this, a YIG film with the thickness of 100 μm is first grown on a GGG substrate. Then a large section of the film is milled out using a wide beam column, followed by a narrower column beam that is used to mill beneath the desired section of the YIG film, creating an inverted pyramid shape. Before the sample is fully removed from the surface of the film, the finger of a nanomanipulator approaches the edge and connect the sample to the finger with the deposited Pt. After the sample is fixed to the nanomanipulator, it can be fully removed from the film and moved to a stage where finer milling can be performed to achieve the desired shape. Finally, the nanomanipulator repositions the sample, mounting it onto an NbN superconducting resonator. Strong magnon-photon coupling is observed in devices with a width of 6 μm. Based on the first anticrossing, attributed to the homogeneous magnon mode, the coupling strength is measured as $\frac{g}{2\pi} = 70$ MHz. The single-spin coupling constant is calculated as $\frac{g_s}{2\pi} = \frac{g}{2\pi\sqrt{N}} = 7.2$ Hz, and the cooperativity is determined to be $C = g^2/\kappa_r\kappa_m = 15$, with decay rates of $\frac{\kappa_r}{2\pi} = 6.4$ MHz for the resonator and $\frac{\kappa_m}{2\pi} = 40$ MHz for the magnon mode. Compared to NiFe, the magnon dissipation rate in free-standing YIG is reduced threefold. Furthermore, at least two additional anticrossings are observed at higher magnetic fields, indicating strong magnon-photon coupling in higher-order standing SW modes. The high-quality YIG fabricated via PFIB represents a significant advancement toward developing hybrid quantum devices integrating SW and superconducting components.

The strong confinement of the electromagnetic field between superconducting plates gives rise to magnon-photon hybridized modes characterized by ultrastrong magnon-photon coupling and exceptionally high cooperativity. The magnon-photon quantum vacuum comprises squeezed magnon and photon states, with the degree of squeezing adjustable over a wide range through an external magnetic field. [355] This vacuum supports a substantial ground-state population of virtual photons and magnons, which can be harnessed to generate correlated magnon and photon pairs. Furthermore, magnon-photon excitations exhibit bipartite entanglement between magnons and photons. To this end, an alternative strategy to exploit the low dissipation rate of YIG in nanoscale devices for enhanced magnon-photon coupling involves depositing YIG films onto superconducting substrates other than GGG. $YBa_2Cu_3O_7$ (YBCO), a well-known high-$T_c$ superconductor with a critical temperature near the boiling point of liquid nitrogen, serves as a promising candidate. [356,357]

Given these properties, YIG/YBCO film stacks offer a compelling platform for studying magnon-photon coupling, as shown in Fig. 5.1 (c) and (d). In the experiment, two types of YBCO antennas-CPW and resonator-are employed. Firstly, the resonance frequency of a 5 μm YIG film is characterized using the CPW antenna above and below $T_c$ to determine magnetic parameters, such as saturation magnetization at various temperatures and the corresponding magnon spectra across a frequency range. Subsequently, a half-wavelength YBCO planar resonator with a resonance frequency of $\omega_c/2\pi$=10.1 GHz at $T = 10$ K is used. The fundamental mode frequency decreases with increasing temperature up to $T_c$. By placing the YIG film on the YBCO resonator, transmission spectra are measured at different temperatures. At $T = 87$ K, which is above $T_c$, only the YIG mode is observed. As the temperature is reduced to $T = 87$ K, the superconducting resonator mode emerges, and upon crossing with the YIG mode, a significant anticrossing gap is observed. With further temperature decrease, the resonant frequency of the resonator increases, leading to an upward shift in the anticrossing gap. The maximum collective coupling strength $g = 2\pi \times 1.7$ GHz confirms the attainment of the ultrastrong coupling regime. An analytical model is proposed to describe the coupling between the superconducting resonator mode and the collective SW mode in YIG, with the eigenvalues given by:

$$\omega_\pm = 1/\sqrt{2}\sqrt{\widetilde{\omega}_c^2 + \omega_b^2 \pm \sqrt{(\widetilde{\omega}_c^2 - \omega_b^2)^2 + 16\omega_c\omega_b g^2}} \qquad (10)$$

where $\widetilde{\omega}_c = \sqrt{\omega_c(\omega_c + 4\beta)}$, and $\beta$ is the small diamagnetic term. A single magnetic mode couples to the resonator with $\omega_b = \omega_0 + \delta_{sc}$, where $\omega_0$ is the frequency of the lowest YIG SW mode. The temperature-dependent shift $\delta_{sc}$ is quantified by self-consistently incorporating the SW-induced Meissner current into the LLG equation as:

$$\delta_{sc} \approx \gamma\mu_0 M_s k_y dr \frac{1-e^{-\frac{2t}{\lambda_L}}}{(k_y\lambda_L+1)^2 - (k_y\lambda_L - 1)^2 e^{-\frac{2t}{\lambda_L}}} \qquad (11)$$

where $r$ is a dimensionless geometrical factor. Notably, the frequency-shift parameter $\delta_{sc}$ displays a distinct temperature dependence: it remains small for $T < T_c$, while for $T \ll T_c$, it saturates to a maximum value of $\frac{\delta_{sc}}{2\pi} \approx 1.1 - 1.2$ GHz. The authors replicate the experimental trends using this simple model that accounts for the interplay between SW excitations and Meissner currents, demonstrating that the evolution of the hybrid magnon-photon system is directly linked to the penetration depth of the superconductor.

Future research on strong magnon-photon coupling in the nanoscale should focus on optimizing the integration of YIG with advanced superconducting materials to enhance coupling efficiency and reduce energy losses. Exploring novel heterostructure designs, such as multilayered YIG/superconductor stacks, could enable precise control of magnon-photon interactions at various temperatures. Investigating the scalability of these systems for quantum information processing and exploring their potential in quantum sensing applications will also be critical steps forward.

*5.1.2 Localization of magnons via resonant magnetic fields*

Beyond achieving strong magnon-photon coupling, the application of resonant magnetic fields offers a promising approach to enhance magnon trapping by stabilizing localized states. By finely tuning the resonant magnetic field to align with the frequency of these localized magnons, the coupling strength between magnons and other quasiparticles can be substantially increased. This precise adjustment could introduce nonlinear effects, such as frequency mixing, which are essential for manipulating quantum states within hybrid systems. One effective method to excite the resonant magnetic field involves stimulating a magnetic medium to its resonance frequency. The dynamically generated dipolar field at resonance is significantly stronger than that observed off-resonance, enabling efficient coupling with magnons and potentially facilitating the creation of magnonic nanocavities and magnonic resonators to support localized magnon states.

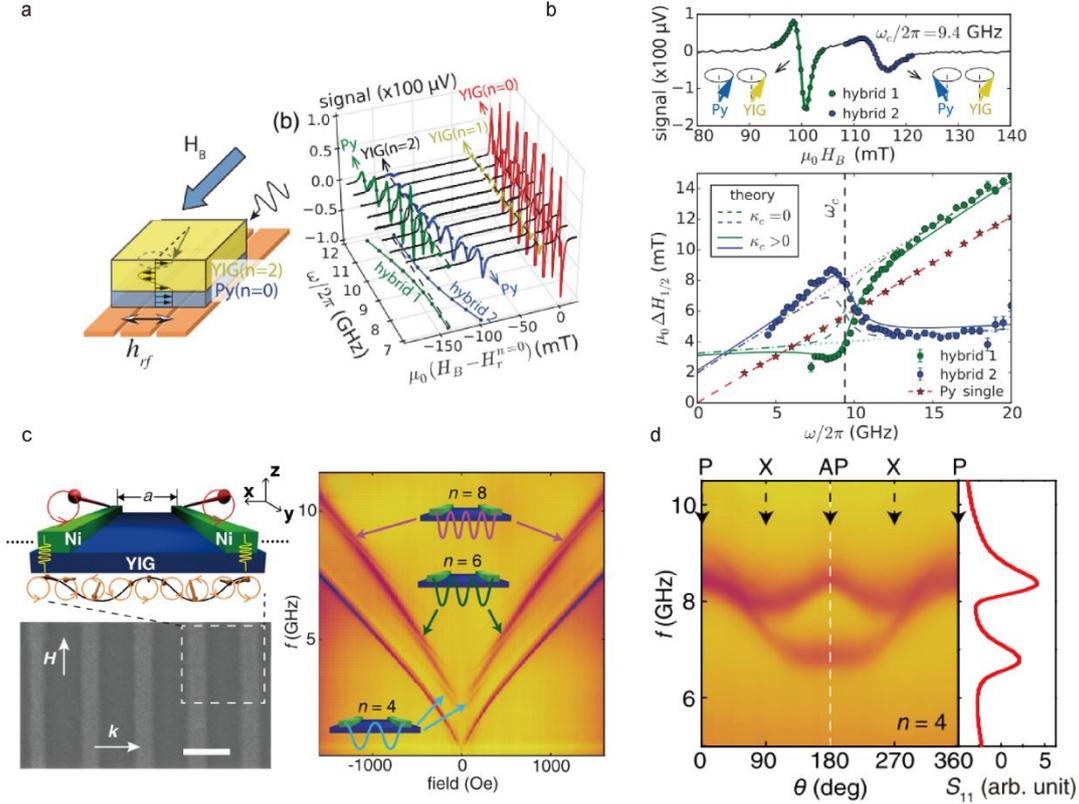

**Fig. 5.2** Localized magnons stabilized by resonant magnetic fields. (a) Excitation of PSSW modes in a YIG/Py bilayer through Py resonance. Line plots illustrate the first three SW modes of YIG alongside the uniform mode of Py. (b) Varying linewidths of the hybridized mode, comparing the $n = 2$ mode of YIG with the $n = 0$ mode of Py. (a) and (b) are taken from Ref. [360]. (c) Ni-based nanowire array on a YIG thin film, with reflection spectra showing high-order ISSW modes in YIG excited by Ni resonance. (d) Angular-dependent measurements of the $n = 4$ ISSW mode. (c) and (d) are taken from Ref. [362].

The study of magnon-magnon coupling represents a compelling frontier in nanoscale magnonic systems. [358,359,360,361,362] This phenomenon occurs when localized magnon states interact, driven by enhanced dipolar fields generated through resonant magnetic excitation. By tuning the magnetic field to resonate with multiple magnon modes simultaneously, the interaction strength between these modes can be significantly amplified, leading to collective magnon dynamics. A common approach to achieving magnon-magnon coupling involves coupling confined perpendicular standing spin wave (PSSW) modes with the FMR mode in a bilayer magnetic heterostructure.[358,359] Experimental evidence shows that high-order PSSW modes can be excited by interfacial spin torques in YIG/Co heterostructures. These spin torques couple the magnetization dynamics of YIG and Co via microwave-frequency spin currents. Anticrossing behaviors between YIG PSSWs and the Co FMR mode have been observed, involving approximately 40 distinct PSSWs with wavelengths as small as $\lambda_{PSSW} \approx 50$ nm. A theoretical model based on a modified LLG equation accounts for the combined effects of exchange, damping-like, and field-like torques localized at the YIG/Co interface. The excitation of PSSWs is suppressed by a thin aluminum oxide interlayer but persists with a copper interlayer, consistent with the proposed model. Furthermore, coherent spin pumping has been identified in a magnon-magnon hybrid system comprising a YIG/Py bilayer, as shown in Fig. 5.2 (a). [360] By reducing the thicknesses of both YIG and Py, the strong interfacial exchange coupling generates a significant anticrossing between the YIG PSSWs and the Py FMR, allowing precise modification of the linewidth through damping-like torque. The suppression and enhancement of the linewidth in these hybrid modes can be attributed to the collective damping-like torque arising from mutual spin pumping, as shown in Fig. 5.2 (b). Additionally, coherent spin pumping has been observed in a low-damping ferrimagnetic bilayer consisting of epitaxially grown GdIG and YIG, where magnon-magnon coupling is achieved. A pronounced variation in the coherent SW-mediated spin current across the magnetic compensation temperature, exhibiting a strong dependence on the relative alignment of magnetic moments. [361]

The realization of strong interlayer magnon-magnon coupling in this on-chip nanomagnonic device underscores the potential of employing resonant magnetic fields to stabilize localized magnon states within a magnonic nanocavity, as shown in Fig. 5.2 (c). [362] Periodic magnetic nanowires, fabricated from Py or Co, are patterned atop YIG thin films. These nanowire arrays serve as Bragg scattering gratings, generating in-plane standing waves characterized by large wavenumbers and nanoscale wavelengths. Within this periodic potential, SWs develop a band structure featuring gaps at the BZ boundaries, where the wave number equals $\pi/a$, with $a$ representing the unit cell length. In the presence of a strong periodic potential, the superlattice band structure becomes dispersionless, localizing magnons within each unit cell, and the band index $n$ indicates the number of nodes. When the frequency of ISSWs in YIG nears a resonance with the nanowire array, where the nanowire resonance acts as a dynamically generated resonant magnetic field, coupling occurs, resulting in an anticrossing between these two modes. Notably, significant anticrossing gaps of up to 1.58 GHz have been observed in a YIG/Co nanowire coupled system, marking the entry into a strong coupling regime with a magnon-magnon cooperativity of $C = 21$. Moreover, the coupling strength exhibits tunability across a broad range by adjusting the magnetization alignment between the nanowires and the YIG film, achieving a maximum when the magnetizations are antiparallel, as shown in Fig. 5.2 (d). This coupling is chiral, which enables the excitation of unidirectional propagation SWs, as detailed in Sec. 4. This tunability enhances the system's versatility, allowing for precise control over confined magnons at the nanoscale. Furthermore, Wang et al. recently reported ultrastrong to nearly deep-strong magnon-magnon coupling in synthetic antiferromagnets with intrinsic magnetic anisotropy. [363] Unlike most ultrastrong coupling systems, where the counter-rotating coupling strength is strictly equal to the co-rotating coupling strength, the synthetic antiferromagnet offers high tunability of these two couplings. The deep-strong magnon-magnon coupling holds promise for highly efficient enhancement of the magnon population in confined nanostructures, and thereby enabling the development of magnon-based quantum systems.

**5.2 Magnonic confinement by localized static fields and thermal control of magnons**
Nanopatterned structures on ferromagnets can modify localized static fields, structuring ferromagnetic materials on the nanoscale and reshaping the internal magnetic energy landscape, thereby modifying the allowed spin-wave eigenmodes and enabling the creation of nanoscale magnonic cavities within confined regions. Furthermore, thermal gradients can generate magnon chemical-potential imbalances, facilitating precise magnon control and guide magnon propagation.

In this section, we review YIG-based magnonic nanocavities implemented with various methods, as well as thermally guided magnons in ferromagnetic films.

*5.2.1 Magnonic confinement using localized static fields at the nanoscale*
As discussed in previous sections, YIG, a low-damping ferrimagnetic insulator, is an excellent material for magnonic studies. [323] YIG exhibits a saturation magnetization of approximately 140 kA/m, significantly lower than that of conventional ferromagnetic metals such as Py and CoFeB. When a ferromagnetic metal with carefully designed structures is patterned onto YIG films, the resulting dipolar and exchange fields influence the local internal field of YIG, leading to distinct magnetic properties compared to unpatterned regions. This enables the design of magnonic confinement and nanocavities in YIG-based magnetic heterostructures. In this section, we review recent advancements in magnonic nanocavities and nanoscopic SW guiding.

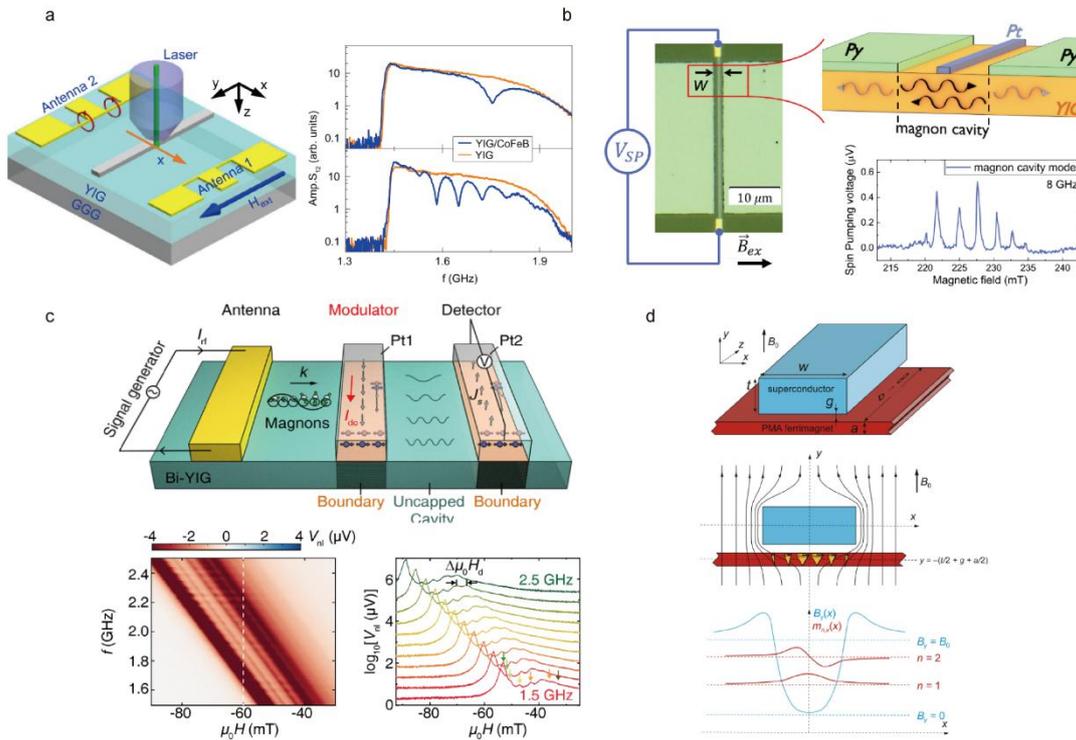

**Fig. 5.3** Localized magnon states by static magnetic fields. (a) Magnonic Fabry-Pérot interferometer enabling dynamic control of SWs beneath a CoFeB wire (taken from [364]). (b) Magnonic cavity formed by confining a YIG film between two YIG/Py bilayers. Confined BVSWs and MSSWs are detected via the ISHE (taken from [366]). (c) On-chip magnonic cavity resonator utilizing two Pt bars. Spin pumping spectra from the Pt2 nanostripe reveal multiple high-order confined magnon modes (taken from [367]). (d) Ferromagnetic film subjected to the stray field of a superconducting stripe. The internal static magnetic field is reduced in the ferrimagnetic region beneath the superconducting strip, resulting in the confinement of SW modes quantized within a quasiparabolic potential well (taken from [368]).

Magnonic resonators offer valuable insights into controlling SW propagation and localization for magnonic applications. In optics, a Fabry-Pérot interferometer, consisting of two parallel reflecting surfaces, forms an optical cavity that allows light to pass only when in resonance. This principle is widely applied in telecommunications, dichroic filters, lasers, and spectrometers for light control and wavelength measurement. Analogously, a magnonic Fabry-Pérot resonator requires two parallel reflecting interfaces within a low-loss magnetic material, enabling SWs to circulate coherently within the cavity before damping out, as shown in Fig. 5.3 (a). Qin *et al.* demonstrated a magnonic Fabry–Pérot resonator by placing a ferromagnetic wire on a low-damping YIG thin film. [364] Local dynamic dipolar coupling between the wire and YIG film creates two magnonic interfaces at the bilayer edges. At these interfaces, propagating SWs partially reflect and transmit, with their wavelengths converted. When SWs within the bilayer accumulate a total phase of $(2n + 1) \times \pi$ after two consecutive internal reflections, the incoming and circulating SWs interfere destructively, forming a magnonic gap in the SW spectra at discrete frequencies. The resonance condition is given by $(|k_2| + |k_3|)w + \varphi_0 = (2n + 1) \times \pi$, where $k_2$ and $k_3$ are the wavevectors for long- and short-wavelength SWs in the YIG/CoFeB bilayer, respectively, depending on propagation direction, $w$ is the width of the magnetic wire, $\varphi_0$ is the phase shift from two internal reflections, and $n$ is the resonance order. This drastic down conversion of SW wavelengths within the narrow bilayer enables manipulation of micrometer-scale SWs ($\lambda = 10 - 50$ μm) using single nanostripes as narrow as 270 nm. The fabrication of such magnonic resonators is compatible with other YIG-based magnonic elements, offering a straightforward approach to integrating compact magnonic and control the SW in a reconfigurable manner. In addition to utilizing the localized field provided by CoFeB to realize a magnonic cavity, a nanoscale magnonic channel confining propagating magnons can be developed within a YIG thin film through dipolar coupling with CoFeB nanostripes. [365] This approach enables the creation of a 160 nm wide waveguide, supporting long-distance magnon transport exceeding 20 μm while achieving a broad frequency range under small external magnetic fields. Furthermore, the ability to redirect SWs using stray-field-induced bends in continuous YIG films highlights a novel strategy for implementing confined magnonic integrated circuits tailored for SW computing.

In addition to creating magnonic resonators by patterning regions on YIG with localized static fields, an inverse design approach, where an unpatterned YIG region is flanked by two patterned regions, also exhibits pronounced magnonic cavity effects. Santos *et al*. developed an on-chip YIG cavity using two YIG/Py bilayers, as shown in Fig. 5.3 (b). [366] By leveraging the modified magnetic properties of the covered and uncovered YIG films, they defined distinct on-chip regions with boundaries capable of confining magnons. The exchange and dipolar interactions within the YIG/Py bilayer establish magnetically distinct regions, forming effective reflective boundaries for magnons and resulting in a robust magnonic cavity. By positioning a Pt nanowire as an inverse spin Hall effect (ISHE) detector both inside and outside the magnonic cavity, multiple spin-pumping voltage peaks are observed inside the cavity. These peaks are pronounced and comparable in magnitude to the intensity of FMR excited in the uncovered YIG film, indicating the formation of standing-wave resonance modes. The confined magnons can be attributed to backward volume (BV) and Damon-Eshbach (DE) PSSW modes confined within the magnonic nanocavity, as calculated from the dispersion relation in a YIG film. Notably, these multiple peaks are absent outside the cavity, confirming that they do not originate from PSSWs formed along the thickness direction of the YIG film. Here, the the cavity is a consequence of both the exchange and dipolar interaction in the YIG/Py bilayer. Furthermore, Wang *et al*. demonstrated a similar approach by placing two Pt nanowires on a low-damping Bi-YIG film to define the boundaries of a magnonic cavity, as shown in Fig. 5.3 (c). [367] In this configuration, the exchange interaction, which significantly changes the local anisotropy of the YIG, is the primary contributor to the formation of the magnonic nanocavity. The $n = 5$ cavity mode is observed. Spin current injection via the spin Hall effect in a Pt nanostripe disrupts the cavity boundary conditions, suppressing both cavity modes and hybridization when the system is driven beyond the damping compensation threshold.

The integration of magnonics and superconductivity offers significant potential for low-loss magnonic applications at cryogenic temperatures, laying the groundwork for quantum magnonics in quantum information processing and conversion. In superconductors, eddy currents shield internal magnetic fields, generating external stray fields that influence nearby ferromagnetic materials, such as by modulating magnetization dynamics. Kharlan *et al*. conducted a theoretical study on magnon confinement induced by the stray field of a superconducting strip in a YIG layer, as shown in Fig. 5.3 (d). [368] This stray field, tunable via an external magnetic field, impacts the magnetization dynamics within the magnetic system. The system is subjected to an external magnetic field perpendicular to the ferromagnetic Ga:YIG layer with perpendicular magnetic anisotropy (PMA), which overcomes shape anisotropy, aligning magnetization out-of-plane even without an external field. The study predicts SW confinement within a potential well formed by the static stray field, with the number of bound states and their frequencies tunable by an external magnetic field. These confined magnon modes exhibit frequencies below the FMR frequency of the ferromagnetic layer, accounting for their exponential decay outside the well region. In addition, Yu et al. explore the gating of magnons via proximity superconductors in their theoretical study. [369] The authors demonstrate that placing a superconducting strip atop a ferromagnetic insulator, such as YIG, induces chiral frequency shifts on the order of tens of GHz. This chirality stems from the constructive interference between the Oersted fields generated by the spin-wave dipolar fields and the induced eddy currents in the superconductor. A wide superconducting gate can fully reflect coherent magnons in the microwave band that propagate perpendicular to the "up" magnetization orientation, yet it transmits them upon magnetization reversal. Furthermore, by applying two superconducting strips, the authors achieve perfect trapping of magnons within a submicrometer-scale region, enabling nanoscale localization and substantial enhancement of the magnon population. These findings enable the design of superconductor-ferrimagnet hybrid magnonic devices for precise, on-demand control of magnon localization and propagation.

*5.2.2 Thermal control of magnons and magnon-phonon coupling*
The emerging field of spin caloritronics [370,371] explores the interplay between spins and heat currents, driven by physical phenomena such as the spin Seebeck effect (SSE), [372,373] anomalous Nernst effect (ANE) [374,375] and spin Nernst effect (SNE), [376] to enhance thermoelectric device performance. Heat currents can be carried by magnons, [377] and they can transfer spin angular momentum opposite to the magnetization direction. A magnetically controllable heat flow caused by a unidirectional SW current is found, and the direction of the flow can be switched by applying a magnetic field, as shown in Fig. 5.4 (a) and (b). [378] At elevated temperatures, phonons, electron-hole excitations, and magnons coexist, collectively contributing to heat transport simultaneously. Their interactions, particularly magnon-phonon coupling, play a pivotal role in thermoelectric phenomena, significantly influencing thermopower, especially at low temperatures. Magnons couple with phonons (lattice vibrations), enabling the manipulation of magnon populations through temperature gradients. This coupling allows magnons to either be directed toward or be confined within regions with varying phonon density population. In ferromagnetic metals, however, the short mean free path of magnons alone cannot explain the length scales observed in the SSE in lateral configurations. [379] Instead, the magnon-phonon coupling provides a critical mechanism for understanding the SSE in a single magnetic wire, capturing effects that a simple drift-diffusion model for the magnon-phonon system fails to account for. While magnon–phonon coupling has been revealed through thermal effects, such as the SSE, coherent magnon–phonon coupling in confined geometries has recently garnered significant attention in the field. In 2011, Weiler et al. first observed coherent magnon-phonon coupling by depositing a ferromagnetic layer atop a piezoelectric substrate. [380] Using an interdigital transducer to coherently excite phonons, the authors identified coupling between these two quasiparticles at the resonance frequency of the magnon mode. Enhancing the strength of coherent magnon-phonon coupling is crucial for efficient energy transfer between magnons and photons. However, reducing the number or relaxation rate of magnons remains challenging due to material limitations. One effective strategy is to increase the number of phonons. Recently, Hwang et al. demonstrated strong coupling between magnons and shear surface acoustic wave (SAW) phonons at room temperature. [381] The SAWs were generated using a nanostructured high-frequency acoustic cavity, and the authors' design enables tunability of the film thickness while maintaining a fixed phonon wavelength. They observed a monotonic increase in coupling strength with increasing film thickness. Strong coherent magnon–phonon hybrid quasiparticles could prove essential for realizing localized quantum states in quantum information processing.

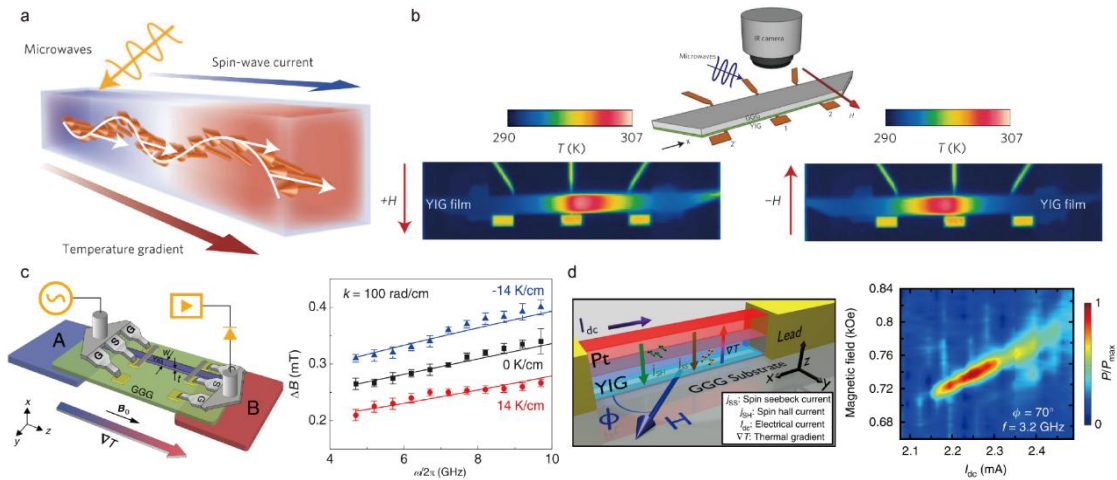

**Fig. 5.4** Thermal control of magnons. (a) Schematic of a SW heat conveyer, where heat emission creates a negative temperature gradient. (b) Temperature distribution images of excited surface SWs for opposing orientations, captured using a micro-electromechanical system infrared camera. (a) and (b) are taken from Ref. [378] (c) SW propagation under a thermal gradient, with SW linewidth plotted as a function of frequency at a fixed temperature gradient (taken from [379]). (d) SCNO based on a YIG/Pt bilayer nanowire, with microwave power output from the nanowire shown as a function of the charge current (taken from [385]).

Thermal spin torques emerge from the interplay of spin and heat currents in spin-caloritronic systems, where temperature gradients generate spin-polarized currents that exert torques on magnetization. These torques, often associated with phenomena like the SSE, enable manipulation of magnetization dynamics without external electrical currents. The first experimental evidence of thermal spin-transfer torque was observed in Co/Cu/Co spin valves positioned at the center of Cu nanowires. [382] Both the switching field and the voltage response amplitude depend on the heat current. Slonczewski predicted that spin-transfer torque driven by thermal magnons could be more efficient than conventional electrically induced torques. [383] Additionally, experiments in magnetic insulators have demonstrated thermal spin torque, showcasing magnon control via thermal gradients, as shown in Fig. 5.4 (c). [384] The linewidth of magnon transmission spectra can be broadened or narrowed by applying a thermal gradient. In YIG films, magnon damping is proportional to the temperature gradient, with its sign determined by the relative orientation of the magnetic field, wavevector, and temperature gradient. In spin torque nano-oscillators (STNOs), a spin-polarized current injected into a ferromagnetic layer generates an antidamping torque, counteracting the intrinsic damping at a critical current density. This induces sustained magnetization oscillations within a confined region, enabling the conversion of charge current into microwave signals and facilitating the excitation of SWs for magnonic applications. These confined magnon modes, spatially restricted within the nanoscale geometry, enhance the coherence and tunability of the oscillatory dynamics, making STNOs promising for high-frequency magnonic devices. Additionally, a thermal spin current, driven by a temperature gradient rather than an electrical charge current, can produce an antidamping thermal spin torque, also known as the spin Seebeck torque. This mechanism operates at ferromagnetic/heavy metal interfaces, giving rise to spin caloritronic nano-oscillators (SCNOs). [385] Safranski *et al.* fabricated nanowire devices, 350 nm wide and 15 μm long, using YIG/Pt bilayers, as shown in Fig. 5.4 (d). As YIG is an electrical insulator, the electric current and associated ohmic heating are confined to the Pt layer, creating a significant temperature gradient $\nabla T$ across the YIG/Pt interface at high current densities. This gradient drives a sharp onset of microwave emission when the DC current exceeds a critical threshold of 2.15 mA. The distinct contributions of the spin Hall effect and the spin Seebeck effect are identified through the angular dependence of the critical current in YIG/Pt nanowires. A theoretical framework explains these findings, proposing a mechanism where a single magnon in the spin current splits into two magnons, with one of them being the magnon mode resonating at the nanostructure. [386] The confined modes in a magnetic nanowire, amplified by the thermal gradient, are promising for advanced spin caloritronic devices that harness ohmic heat to drive magnon-based technologies.

## 6. Macroscopic quantum and nonlinear effects

*6.1 Magnon Bose-Einstein Condensation (BEC)*
In a typical system where magnon BEC is observed, an external source, such as microwave radiation, is used to excite magnons into higher energy states via parametric pumping. As magnons relax toward the system's ground state, they can accumulate at the minimum energy level, leading to the formation of a condensate. Once in a condensed state, magnons can exhibit coherent behavior, similar to the macroscopic quantum effects seen in superfluidity or superconductivity.

*6.1.1 Parametric pumping leading to magnon BEC*
BEC is a quantum phenomenon in which a macroscopic number of bosonic quasiparticles occupy the lowest energy state of a system, resulting in the emergence of coherent collective behavior. [387,388,389,390] While predicted initially and observed in ultracold atomic gases, [391,392] analogous condensates can form in solid-state systems, including ensembles of magnons. As bosonic quasiparticles, magnons obey Bose-Einstein statistics and can undergo condensation under nonequilibrium conditions. Unlike atomic BECs that form at ultralow temperatures, magnon BECs were first observed two decades ago in the magnetic insulator $TlCuCl_3$ [393,394,395,396] and in superfluid phases of 3He. [397,398,399,400,401,402] In 2006, magnon BEC was observed by Demokritov *et al.* in a YIG film at room temperature through parametric pumping, which injects magnons at a given frequency and wavevector. [403] When the magnon population exceeds a critical threshold, interactions among them lead to a redistribution toward the lowest-energy magnon state, resulting in a spontaneous accumulation of phase-coherent magnons. [404,405,406] Parametric spin pumping involves the nonlinear interaction between the microwave magnetic field, applied parallel to the static magnetization, and the spin system, resulting in the splitting of microwave photons into pairs of magnons. These magnons are created with frequencies equal to one-half of the pump frequency and possess equal and opposite wavevectors, thereby conserving both energy and momentum. [407,408,409,410,411] This process, known as parallel parametric pumping, efficiently injects magnons at specific wavevectors that lie within the magnon dispersion relation, enabling a rapid increase in magnon population. [412] Once the magnon density surpasses a critical threshold, interactions among magnons lead to a redistribution toward the system's lowest-energy state, culminating in the formation of a magnon BEC, as has been experimentally observed in YIG films.[ 412,413,414,415,416,417,418,419,420,421,422,423,424,425] A typical setup for excitation and observation of magnons is shown in Fig. 6.1. [413]

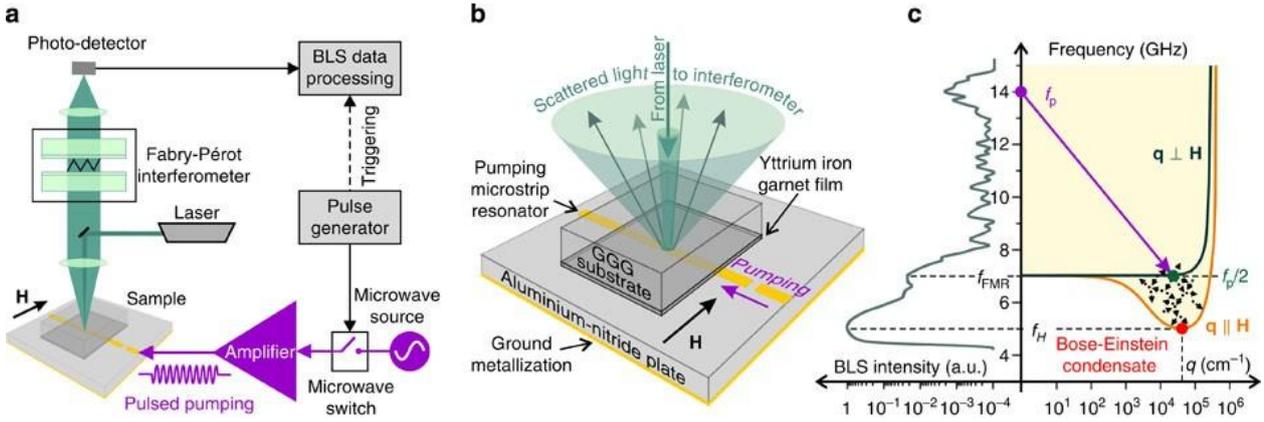

**Fig. 6.1** (a,b) A YIG film is placed above a microstrip pump resonator prepared on top of a metallized plate. A bias magnetic field of 1,735 Oe is applied in the film plane across the resonator, which is driven by microwave pulses with a carrier frequency of $f_p = 14$ GHz. A probing laser beam is focused onto the microstrip. The scattered light is directed to a Fabry–Pérot interferometer for analysis. (c) The measured frequency spectrum of the inelastically scattered light (left) is presented with the magnon spectrum (right). The purple arrow illustrates the creation of primary magnons (green dot) by parametric pumping (purple dot). The dashed arrows schematically represent the thermalization of the parametrically injected magnons. The high-intensity BLS peak is associated with the BEC at the lowest-energy state (red dot). The second BLS intensity maximum at frequency $f_p/2$ is created by the parametrically injected magnons, while the third at frequency $f_p$ results from non-resonant magnetization precession directly forced by the pumping field. Adapted from Ref. [413].

*6.1.2 Magnon BEC in nanoconduits*

The realization of magnon BEC in confined geometries has opened new avenues for studying macroscopic quantum phenomena in magnetic systems. In particular, nanoconduits and quasi-one-dimensional traps have proven to be versatile platforms for the creation, manipulation, and study of magnon condensates. Unlike atomic BECs, magnon condensates are formed by quasiparticles with finite lifetimes and non-conserved particle number, requiring continuous or pulsed pumping to sustain the condensate state. Nevertheless, under suitable conditions of low temperature and long magnon coherence time, a well-defined magnon chemical potential can be established, leading to condensation.

Autti *et al*. [399] demonstrated the condensation of magnons in a three-dimensional magnetic trap formed in superfluid $^3$He-B. The trap emerges from the interplay between the magnetic field and the spatially inhomogeneous orbital texture of the superfluid order parameter. At low excitation densities, the trap exhibits a harmonic confinement potential; however, as the magnon population increases, spin-orbit interactions modify the trap into a box-like potential with nearly impenetrable walls. This self-trapping behavior resembles the formation of electron bubbles in liquid helium and represents a rare example of a self-localized bosonic field in condensed matter. Furthermore, the authors reported the formation of condensates not only in the ground state but also in excited states of the trap, establishing the coexistence of multiple BECs and enabling energy-resolved control over magnon dynamics.

In parallel, Mohseni *et al*. [417] achieved magnon BEC in YIG-based waveguides via parametric pumping. Their experiments reveal the confinement and spatial propagation of magnon condensates in micrometer-scale conduits, showing that nonlinear spin-wave interactions govern the condensate's propagation. Figure 6.2 illustrates the behavior of parametrically excited magnons in a narrow YIG conduit under microwave pumping. Panel (a) shows the setup, where magnons are generated at half the pumping frequency ($f_p/2 = 2.1$ GHz) via a dynamic Oersted field from a microstrip antenna. At moderate current ($I_{rf} = 12$ mA), magnons are primarily excited in the first spin-wave mode, as seen in the spectrum (panel c) and the dispersion map (panel d). As the pumping current increases to 15 mA, nonlinear four-magnon scattering becomes significant, broadening the spectrum (panel e) and redistributing magnons in momentum space (panel f). Two types of scattering are observed: frequency-conserving events involving parametrically injected magnons, and frequency-non-conserving processes that affect the whole magnon population. At even higher current ($I_{rf} = 20$ mA), magnons are distributed more broadly across the band (panel g). When the pumping is turned off, the system relaxes, and a clear accumulation of magnons at the band minima is observed (panel h), signaling the formation of a magnon Bose-Einstein condensate. [417]

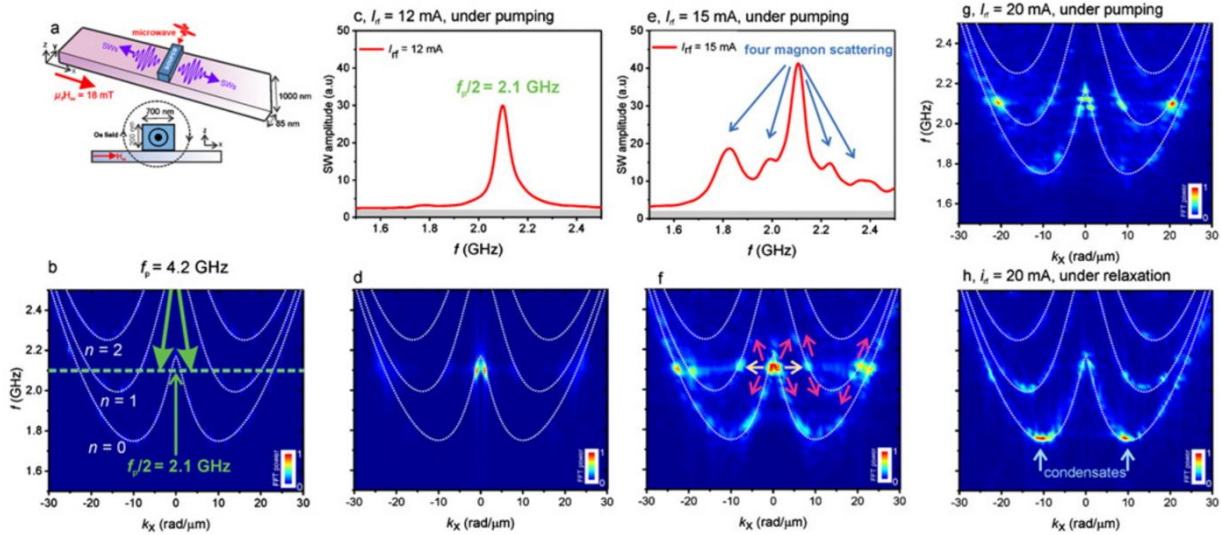

**Fig. 6.2** (a) Schematic view of the sample under study. (b) Magnon band structure near the band bottom obtained from numerical simulations (color plot) and analytical calculations (dotted lines). The first three width modes (n = 0, 1, 2) are present in the band structure. (c)–(f): frequency spectra and the magnon band structures during pumping when microwave currents of Irf = 12 mA and Irf = 15 mA are injected into the stripline, respectively. Thermal magnons are distinguished via gray color at (c)–(e). (g) and (h): magnon band structure during pumping and relaxation, when the microwave current is Irf = 20 mA. Magnon condensation at the band bottom is visible during relaxation. Reprinted with permission from Ref. [417].

It has been demonstrated that BEC of magnons is spatially stable due to a repulsive interaction between magnons, which disagrees with earlier hypotheses of an attractive interaction that would lead to a collapse of the condensate. Borisenko *et al*. provide experimental evidence for this repulsion and propose a mechanism for its origin, [418] solving a long-standing problem regarding the stability of magnon BECs. The study used a YIG film at room temperature to create a magnon BEC through microwave parametric pumping [Fig. 6.3(a)]. To analyze the condensate's spatial stability, a nonuniform magnetic field was prepared with a direct current flowing through a control line, forming a potential hill (an increase in the field), or a potential well (a reduction in the magnetic field) as shown in Fig. 6.3(b). Then the spatial and temporal dynamics of the magnon BEC density were measured using micro-focused BLS, allowing one to conclude that, when a potential well is created, the magnon density formed an intense peak at the center of the well. In contrast, a potential hill provoked a reduction in magnon density at its center. Such behavior, particularly the limited increase in density in the potential well [see Fig. 6.3(c,d)], is inconsistent with the assumption of attractive magnon-magnon interaction, which would have led to a collapse of the condensate. After the microwave pumping was turned off, the magnon BEC's density profile in the potential well began to narrow over time, instead of broadening, which was attributed to the repulsive interaction. The decay rate of the condensate was also slower at the center of the well than at the edges, which supports the idea that magnons from the surrounding areas were moving back towards the center under the influence of the potential gradient. [418] Such behavior is supported by a model based on the Gross-Pitaevskii equation, which includes a term for magnon-magnon interaction that accurately reproduces the experimental density profiles only when the interaction is assumed to be repulsive. The origin of such repulsive interaction relies on the dipolar field that appears in response to any local increase in magnon density (and decrease in the equilibrium magnetization). This field effectively increases the potential energy per magnon, thus counteracting the accumulation of magnons and stabilizing the condensate. [418]

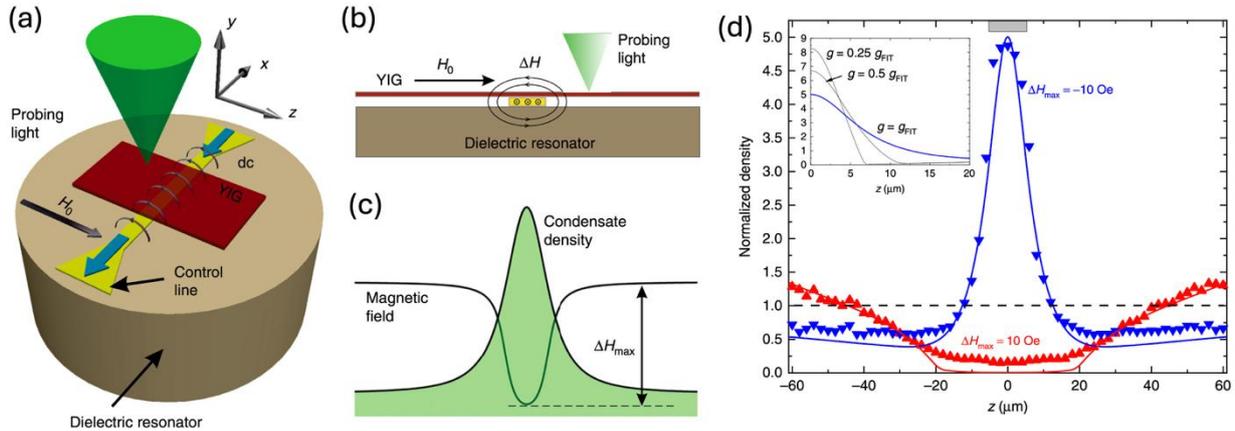

**Fig. 6.3** Spatial stability of a magnon BEC. (a) Illustration of the experimental setup. (b) A DC current creates a localized Oersted field that is used to create a potential well (or hill). (c) Schematic representation of the interplay between the inhomogeneous magnetic field and the condensate density. (d) Direct measurements of the magnon density for a potential well (blue) and a potential hill (red). Adapted from [418].

Together, these works establish that nanoconduits, magnetic field inhomogeneities, and magnetic textures provide controllable, low-dimensional environments for magnon BEC. They highlight how magnon self-interactions, trap geometry, and confinement potential can be tuned to achieve desired condensation conditions, offering a promising pathway toward practical implementations of coherent magnonic systems at the nanoscale.

*6.2 Spin-Torque Induced Magnon Trapping*
Spin-transfer torque (STT) is a fundamental mechanism in spintronics that enables the control of magnetization dynamics through the flow of spin-polarized electric currents. When a spin-polarized current passes through a ferromagnetic material, the angular momentum carried by the conduction electrons can be transferred to the local magnetization, exerting a torque that modifies its orientation or excites coherent magnetic oscillations. This phenomenon was independently predicted by Slonczewski and Berger in the 1990s [426,427] and has since become central to the operation of spin-torque nano-oscillators, magnetic random-access memory (STT-MRAM), and magnonic devices. [428,429,430] In magnonics, STT provides a powerful means of generating, amplifying, and localizing spin waves (magnons) without the need for external microwave fields. When combined with additional effects, such as the spin Hall effect and the Oersted field, STT can give rise to rich nonlinear dynamics including self-localized wave packets, auto-oscillations, and even BEC of magnons. [431] These current-driven excitations offer an electrically tunable pathway to confining magnons in nanostructures, enabling advanced functionalities in low-power, nanoscale information processing.

The following subsections describe two key manifestations of spin-torque-driven magnon localization: 6.2.1 localization induced by electric currents through ferromagnetic layers, and 6.2.2 the emergence of bullet modes—nonlinear, self-localized wave packets with distinct spectral and spatial properties.

*6.2.1 Localization achieved through electric currents in ferromagnetic layers*

The application of electric currents in ferromagnetic layers can lead to the dynamic localization of magnons through the combined effects of spin-transfer torques and the inhomogeneous magnetic fields generated by the current. In particular, when a direct current flows through a heavy-metal/ferromagnet bilayer (e.g., Pt/NiFe), it produces a spin current via the spin Hall effect that exerts a spin-orbit torque on the magnetization of the ferromagnetic layer. Simultaneously, the current generates an Oersted field whose spatial profile can form a localized magnetic potential well or hill. This localized field modifies the internal magnetic environment in such a way that magnons are energetically confined to the region where the Oersted field is minimized. [4322] Recent experiments have demonstrated that this current-induced field gradient can trap spin-wave modes within nanogaps, effectively suppressing magnon radiation losses and stabilizing localized auto-oscillations. The confinement is highly sensitive to the stacking order of the layers. For instance, as shown in Fig. 6.4, in NiFe/Pt structures, the Oersted field forms a dip at the center of the device acting as a trap, while reversing the layer order (Pt/NiFe) produces a potential hill, which instead promotes magnon leakage. [432] The trapped magnons exhibit reduced linewidths and lower threshold currents, making this mechanism highly attractive for the development of spin-torque nano-oscillators and other magnonic devices.

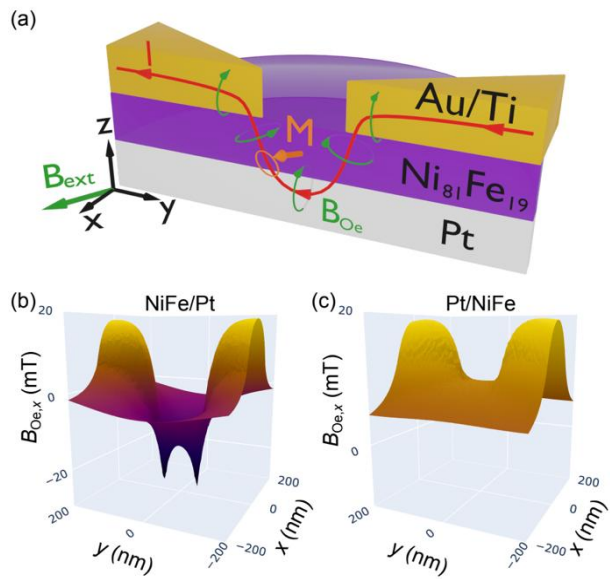

**Fig. 6.4** (a) Cross-sectional illustration of a $Ni_{81}Fe_{19}$/Pt spin torque nano-oscillator. (b,c) Calculated Oersted field in the $Ni_{81}Fe_{19}$ layer for the NiFe/Pt and Pt/NiFe nano-oscillators, respectively. The input direct current is $I = -20$ mA. Adapted from [432].

Beyond enabling localized auto-oscillatory states, spin-torque excitation can also generate coherent spin waves that propagate from the active region and serve as an efficient interaction channel between spatially separated oscillators. A recent example is provided by pairs of nano-constriction spin-Hall oscillators, [433] in which the emitted propagating magnons mediate mutual synchronization between the oscillators at separations of hundreds of nanometers. Importantly, the synchronization is not only robust but also highly tunable: the synchronized state can be continuously adjusted by varying the drive current, which modifies the spin-wave wavelength and, hence, the phase accumulated between the oscillators. It can also be shifted or controlled by the applied magnetic field and its orientation. These concepts have recently been extended from oscillator pairs to large two-dimensional arrays of spin-torque nano-oscillators, in which propagating spin waves enable global phase and frequency locking across the entire network.[434] In particular, spin-wave-assisted synchronization has been demonstrated in engineered magnetic films that promote controlled nearest-neighbor interactions and well-defined phase shifts induced by propagation, enabling robust and scalable synchronization even in the presence of thermal fluctuations. Such architectures provide additional flexibility by allowing or suppressing next-nearest-neighbor interactions. These capabilities—long-range magnonic coupling together with current/field controllability—represent a key advantage of spin-wave-mediated synchronization for prospective applications in oscillator networks and spin-wave logic-based devices.

*6.2.2 Bullet modes and their implications for wave localization in nanoscale systems*

Bullet modes are nonlinear, self-localized spin-wave excitations that arise in ferromagnetic systems when subjected to intense spin-polarized currents. [435,436] These excitations are typically observed in nanocontact spin-torque oscillators, where a direct current injected through a nanoscale contact transfers spin angular momentum to the local magnetization, inducing precessional motion via STT. When the current exceeds a critical threshold, the excitation enters a nonlinear regime and collapses into a highly localized, soliton-like structure referred to as a spin-wave bullet. [436] The bullet mode is spatially confined to a region much smaller than the excitation area and exhibits a frequency lower than the FMR of the surrounding medium. [437] The development of bullet mode theory was based on early analytical work on current-induced spin-wave excitations, including linear modes in perpendicularly magnetized nanocontacts, [438] and its nonlinear generalizations. [439] These efforts culminated in the analytic description of the bullet in in-plane magnetized systems and its subsequent extension to obliquely magnetized contacts.[440] Further studies confirmed the existence and stability of these modes in nanocontact geometries, [441] supporting their crucial role in the functionality of spin-torque nano-oscillators.

Bullet modes not only provide localized sources of coherent microwave radiation but also function as dynamic magnon traps, being confined within regions of strong localization. Their nonlinear nature, combined with high amplitude, robust spatial confinement, and sustained stability under continuous current injection, makes them particularly well-suited for applications requiring both spin-wave localization and dynamic trapping at the nanoscale.

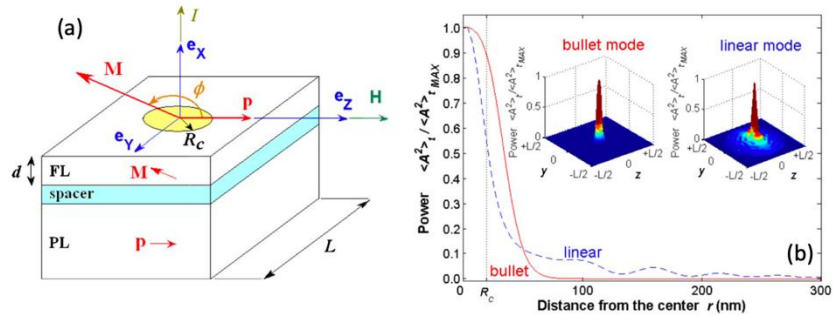

**Fig. 6.5** (a) Sketch of the point-contact device structure with the coordinate system used in the simulations. (b) Dependence of the numerically calculated normalized squared amplitude $\langle A^2 \rangle / \langle A^2 \rangle_{max}$ for the bullet mode (solid line) and the linear propagating mode (dashed line) on the distance $r$ from the nano-contact center. The dotted vertical line indicates the position of the nanocontact radius $r = R_c$. Insets show the dependence of the mode power on the coordinates in the $y$-$z$ plane for both bullet and linear propagating modes. A bias current $I = 12$ mA was used. Adapted from [436].

Figure 6.5(a) illustrates a sketch of the point-contact trilayer structure, where the spin-polarized current is injected into an in-plane magnetized NiFe free layer (FL). [436] The bias current $I$ is applied within the circular nanocontact area of radius $R_C$, which defines the excitation region. The simulated spatial profiles of both a linearly propagating spin-wave mode and a nonlinear self-localized bullet mode are depicted in Figure 6.5 (b). The normalized squared amplitude is plotted as a function of the radial distance $r$ from the center of the nanocontact. The simulations correspond to a bias current $I = 12$ mA, a regime in which both modes coexist. The results reveal a contrast between the spatial profiles of the two modes. The bullet mode exhibits an exponential decay away from the nanocontact, indicative of its evanescent nature and strong localization. Specifically, at a distance $r = 4R_c$, the bullet's amplitude has decayed by three orders of magnitude relative to its maximum at the contact center. In comparison, the linear spin-wave mode maintains a significantly broader profile, with an amplitude that remains two orders of magnitude larger than that of the bullet at the same distance. This contrast underscores the fundamentally different propagation characteristics of the two modes: while linear spin waves radiate energy outward, the bullet mode remains tightly confined near the excitation source.

The confinement properties of bullet modes can not only minimize the coupling between neighboring elements in densely packed magnonic circuits but also enhance the efficiency of localized signal generation. These characteristics make bullet modes compelling candidates for use as building blocks in future magnonic logic devices, spin-wave-based communication systems, and neuromorphic computing architectures.

## 7. Application and outlook of magnon confinement and trapping

Magnon confinement and trapping are emerging as key strategies for enabling functional magnonic devices, especially as magnons are increasingly recognized as a platform for information processing beyond the limitations of conventional electronics. By localizing SWs in space or time, magnon trapping enables refined control over SW dynamics, supporting essential operations such as buffering, logic gating, filtering, and resonant coupling. Continued advances in materials, geometrical design, and control mechanisms are expected to unlock new opportunities in both fundamental research and applied magnonic technologies.

Across these diverse platforms, several **key mechanisms** have been used to achieve magnon confinement and trapping. The principal ones are shown in the following figure

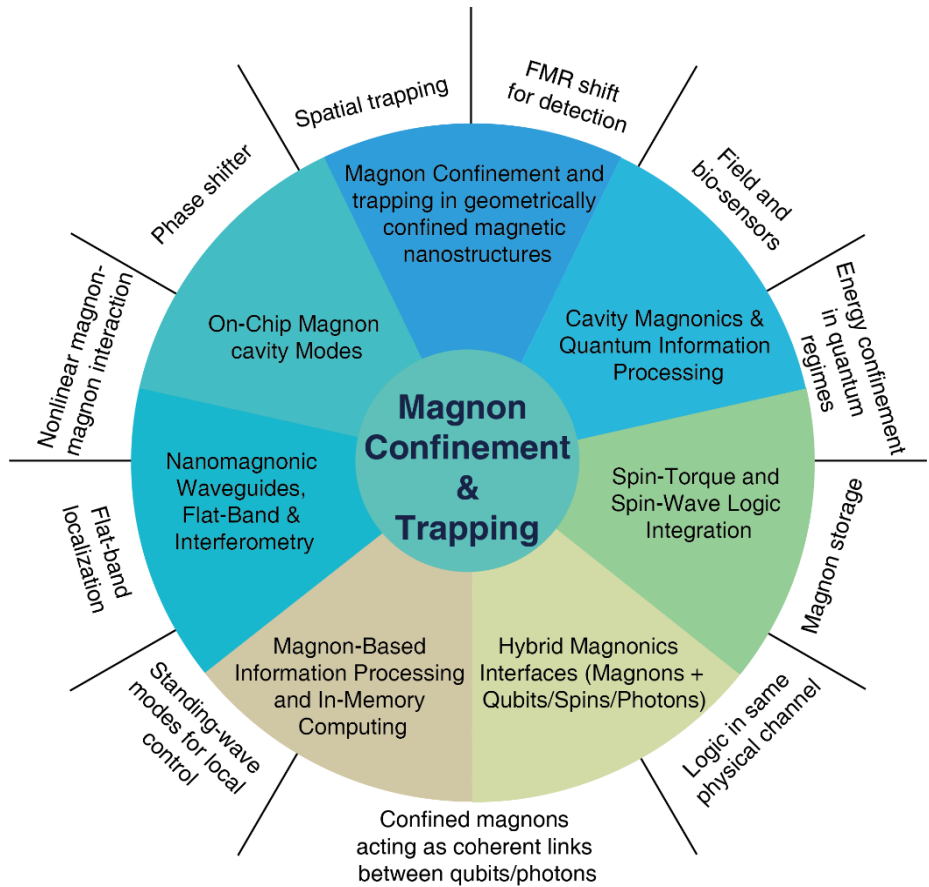

**Fig. 7.1** Illustration of the principal application of magnon confinement and trapping.

*7.1 Magnon Confinement and trapping in geometrically confined magnetic nanostructures*

The confinement and trapping of magnons in geometrically engineered magnetic nanostructures—such as stripes, dots, and antidots—play a crucial role in tailoring SW dynamics for applications in magnonic computing, signal processing, and sensing. Defined by lithographic patterning or self-assembly, these structures offer scalable, high-precision control over SW localization and propagation through boundary conditions, shape anisotropy, and interference effects.

Magnetic nanowires constitute the fundamental building blocks of magnonic waveguides, where SWs are laterally confined by the physical edges of the structure. [442,443] In layered nanowire architecture, novel level of funtional reconfigurability, not accessible in single-layer nanowire arrays, are also introduced. [444] Guided SWs can also be created in graded magnonic nanostripes (see Sec. 3). Experimental studies using BLS and time-resolved Kerr microscopy have revealed that mode interference and mode conversion at bends, notches, or junctions, create potential wells or barriers for magnons offering possibilities for wave-based logic architectures. A recent study on low-loss magnonic waveguides utilizing YIG thin films with silicon ion implantation reveals SW decay lengths surpassing 100 μm in submicrometer waveguides. [445] Moreover, the SW dispersion can be continuously tuned through localized ion implantation, distinguishing these waveguides from conventional etched magnonic structures. A large-scale magnonic network comprising 198 crossings shows great promise for developing wafer-scale magnonic integrated circuits for future industrial applications.

Directional couplers are foundational components in magnonic circuits, enabling the controlled transfer of SW energy between two parallel waveguides through dipolar or exchange interactions. As magnonics advances toward functional logic and signal processing devices, two main classes of directional couplers have emerged: planar and vertical**.** While both architectures exploit similar underlying physical principles, their geometry, scalability, coupling efficiency, and integration potential differ significantly. At the heart of most magnonic directional couplers lies the geometric confinement of SWs within narrow waveguides of low-damping magnetic materials such as YIG or NiFe. These confined structures quantize the SW modes laterally (and vertically in layered geometries), resulting in a discrete set of propagating eigenmodes.

Experimental demonstrations by Wang *et al.* [446] and Sadovnikov *et al.* [447] have shown how planar couplers can implement magnonic logic elements, including frequency-dependent half-adders and multiplexers. However, their efficiency is limited by the weaker lateral dipolar coupling and the need for tight waveguide spacing, which restricts integration density. Vertical directional couplers feature stacked magnetic layers, separated by thin dielectric or non-magnetic spacers. This architecture provides confinement in both lateral and vertical directions, resulting in stronger field overlap and thus shorter coupling lengths. Advantages include higher coupling efficiency due to improved modal overlap, smaller device footprint for high-density integration, dynamic tunability via external magnetic fields, enabling field-controlled coupling activation. Recent theoretical work by Szulc *et al.* [448] has demonstrated how vertical couplers allow on-demand coupling by tuning the internal magnetic configuration of one waveguide layer—an approach directly enabled by the confinement of SW modes within each layer. Future magnonic technologies will likely rely on hybrid coupler designs that combine the accessibility of planar geometries with the performance advantages of vertical confinement.

In magnetic nanodot and antidot lattices, edge modes play a crucial role in determining their sensitivity and functionality as magnetic field sensors. Unlike extended bulk modes that propagate through the continuous regions of the film, edge modes are highly localized—typically within tens of nanometers from the antidot edges—and their amplitude decays rapidly away from these boundaries. The localized nature of these modes makes them inherently more sensitive to any nearby variations in the magnetic field, making them promising candidates for biosensing, chemical sensing, and magnetic field imaging. [447,448] They are also compatible with wafer-scale fabrication processes, making them suitable for integration into miniaturized lab-on-chip systems. Unlike superconducting magnetometers (e.g. SQUIDs), which require cryogenic conditions, dot and antidot-based sensors operate at room temperature and are compatible with standard CMOS processing.

Beyond sensing, confined magnetic elements and magnonic crystals are also essential in the design and realization of magnonic Fabry–Pérot resonators as programmable phase shifters for SW computing, [449] control elements in nonreciprocal SW diode, [450] isolator [451] and circulator based on Dzyaloshinskii–Moriya interaction (DMI). [204]

*7.2 On-Chip Magnon Cavity Modes*

YIG is widely recognized as a model system in magnonics due to its exceptionally low magnetic damping and high SW coherence. Traditionally used in bulk or millimeter-scale microwave devices, YIG has recently gained renewed attention through advances in thin-film growth and nanofabrication, which have enabled the development of on-chip YIG cavities—microscale structures designed to confine and resonate magnon modes with high precision. These miniaturized systems combine the superior intrinsic properties of YIG with compatibility for chip-level integration, opening new avenues for coherent magnonic devices, microwave processing, and hybrid quantum systems.

The core physical principle behind on-chip YIG cavities is the confinement of SWs, or magnons, within a well-defined spatial region. In these cavities, which are often patterned from nanometer-thick YIG films (typically 20–100 nm), confinement is achieved either through geometric patterning, such as forming micrometer-scale disks or rectangular stripes, or through magnetic contrast, for instance by overlaying selected regions with metallic layers like NiFe. This confinement results in a discrete spectrum of resonant magnon modes, analogous to eigenmodes in optical cavities. Importantly, the Gilbert damping constant of YIG remains among the lowest of any known magnetic material ($\alpha \sim 10^{-5}$), which allows magnons to retain their coherence over long timescales, resulting in sharp resonance peaks and high-quality factors (Q).

Recent studies have demonstrated the quantization and spatial confinement of SW modes in on-chip YIG cavities with remarkable clarity. For example, Santos *et al.* [366] patterned nano-structured YIG/Py bilayers to define magnonic cavities where SWs were trapped between high-damping boundaries created by Py. Using spin pumping detected through Pt nanostrips, they observed distinct resonant modes within the cavity, confirming both lateral confinement and mode selectivity. Beyond passive confinement, recent work has also shown that magnon modes in on-chip cavities can be actively tuned and hybridized. In a 2025 study, for instance, Wang *et al.* [367] observed mode hybridization in Bi:YIG/Pt cavities and showed that the coupling strength could be modulated by applying a spin current via the spin-Hall effect. This created a platform for current-controlled tunable resonators, where resonance frequencies and coupling efficiencies could be dynamically reconfigured in real time.

The combination of high coherence, tunability, and miniaturization positions on-chip YIG cavities as promising components in a range of applications. In magnonic logic and computing, they can function as frequency-selective elements, buffering or gating SW signals with minimal energy dissipation. Their discrete modes can also serve as filters, delay lines, or oscillators, supporting signal processing in the microwave frequency range. Because of their sharp resonance features and sensitivity to local magnetic environments, YIG cavities are also well-suited to magnetic sensing applications. For example, shifts in resonance frequency can indicate the presence of magnetic nanoparticles, stray fields, or mechanical strain, making these cavities candidates for compact field sensors and biosensing platforms.

Despite the remarkable advantages of YIG, several challenges remain. The fabrication of high-quality nanoscale YIG films with preserved low damping is nontrivial, and integration with existing CMOS or superconducting platforms requires careful material and interface engineering. Furthermore, engineering cavity geometries that support specific mode profiles or allow selective coupling to other systems (e.g., qubits or photons) remains an active area of research. Nevertheless, ongoing developments in sputtering, pulsed laser deposition, and focused ion-beam milling are rapidly advancing the quality and reproducibility of on-chip YIG cavities.

*7.3 Cavity Magnonics & Quantum Information Processing*

Cavity magnonics is an emerging field that investigates the strong coupling between magnons and confined electromagnetic modes within resonant microwave cavities. This hybrid platform harnesses the advantages of both magnons and photons, enabling novel functionalities for quantum technologies. The confinement of magnons in high-quality magnetic materials such as YIG spheres or films allows for strong photon–magnon coupling. This strong coupling regime is critical for implementing quantum memory and quantum state transfer protocols, where information encoded in microwave photons can be coherently transferred to magnonic states and vice versa. Notable early demonstrations include the pioneering works by Huebl *et al.* [452] and Zhang *et al.* [453], which established coherent coupling between YIG magnons and cavity photons. Beyond strong coupling between both quasiparticles, cavity magnonics also enables the generation of squeezed-magnon frequency combs, which provide a resource for continuous-variable quantum computing, quantum-enhanced sensing, interferometry, and microwave-photon transduction. The nonlinear magnon dynamics in these systems can be harnessed for squeezing and entanglement generation, as discussed in recent theoretical and experimental works. [454] These developments open pathways toward quantum networks utilizing magnonic nodes as coherent interfaces. Additionally, the hybridization of magnons with superconducting qubits and spin ensembles paves the way for interfacing disparate quantum systems, combining long coherence times of magnons with fast qubit operations. Reviews by Tabuchi *et al.* [455] and Lachance-Quirion *et al.* [456] provide comprehensive insights into the integration of cavity magnonics with quantum computing platforms.

Overall, cavity magnonics stands at the forefront of developing quantum magnonic devices for quantum information processing, and nanoscale co-localization of magnon and photon modes may further enable strong magnon-photon coupling at the single-quanta level, paving the way for novel quantum information conversion approaches.

*7.4 Nanomagnonic Waveguides, Flat-Band & Interferometry*
A promising direction for further enhancing control over SW dynamics lies in the engineering of artificial magnetic superlattices, particularly magnetic Moiré patterns. By twisting two periodic magnetic layers [247] or modulating anisotropy or Dzyaloshinskii-Moriya interactions, [2533] Moiré superlattices can be designed to exhibit flat magnon bands—a regime where the magnon group velocity approaches zero, leading to spatially localized magnon modes. This phenomenon is analogous to the emergence of flat electronic bands in twisted bilayer graphene, [241,242] but applied in the magnonic domain.

Furthermore, when a magnonic waveguide is placed atop such a Moiré-patterned substrate, the flat-band character of the substrate is imprinted onto the waveguide, resulting not only in the localization of the SWs in specific regions but also in their amplification. This is a critical advancement: the enhanced local density of magnon states in the flat-band regime increases the nonlinear magnon-magnon interactions, which can counteract intrinsic damping, thereby prolonging the propagation or even enabling self-sustained oscillations.

Theoretical proposals also describe magnonic Michelson interferometers and other guided structures that confine magnons in controlled potential paths, analogous to optical waveguides, with studies quantifying confinement factor, bend-loss, mode count, and eigenvalue spectra. [457]

*7.5 Spin Torque and Spin Wave Logic Integration*
Magnon confinement and trapping play an important role in the integration of spin-torque-based components with SW logic architectures—an emerging paradigm aimed at realizing fast, reconfigurable, and energy-efficient computing systems. The ability to spatially localize magnons within well-defined regions enables a new class of hybrid devices where localized magnetic excitations can interact directly with spin currents generated by spin-transfer torque (STT) or spin-orbit torque (SOT) mechanisms. [458, 459] This spatial control is not merely beneficial—it is essential for facilitating efficient coupling between spin-torque devices and magnonic logic components. Micro-focus BLS experiment performed for in-plane magnetized elliptical STNO nanocontact on an extended NiFe film enabled to construct two-dimensional maps of the intensity of confined SW emission from the nano-oscillator. [458] In addition, dipole-field-induced localization of magnons gives rise to stable, tunable oscillation modes with narrow linewidths. Another key contribution explored early designs for magnetoelectric spin-wave amplifiers, [460,461] where strain-mediated coupling in multiferroic heterostructures could dynamically modulate the amplitude of SWs confined within specific regions, enabling logic functions through interference and phase manipulation.

In these hybrid systems, confined magnons serve as both the computational medium and the coupling interface between different device modalities. Spin-torque devices can locally inject or detect magnons within confined regions, allowing them to function as logic transducers, memory elements, or oscillatory signal sources. [462] The confined nature of the modes ensures that their spectral and spatial characteristics can be finely tuned, enabling deterministic control over logic behavior. This synergy is especially important for overcoming the traditional incompatibility between global, wave-based information transport and localized, current-driven logic operations.

*7.6 Magnon-Based Information Processing and In-Memory Computing*
SWs confined in YIG-based SW buses provide a robust mechanism for transmitting angular momentum without associated charge transport, enabling ultralow-energy signal propagation. Due to the exceptionally low Gilbert damping of YIG, these confined magnons can propagate over millimeter-scale distances with minimal loss, making them ideal candidates for information carriers in next-generation computing architectures. Li *et al.* [463] provided direct evidence of SW-driven switching, reinforcing the feasibility of using magnons for nonvolatile logic and memory operations. Furthermore, Baumgaertl *et al.* demonstrate the reversal of nanomagnets using propagating magnons in YIG thin films, enabling nonvolatile magnon memory. [464] The Py nanowires can be reversed by SWs propagating through an underlying SW bus. This facilitates the storage of charge-free angular momentum flow after transmission over macroscopic distances. These breakthroughs support the development of compute-in-memory architectures, which aim to bypass the von Neumann bottleneck by co-integrating processing and storage functionalities within the same magnonic platform. [465]

A key requirement for enabling such systems is the ability to store, delay, and retrieve SW signals with high fidelity. This is where SW confinement and magnon trapping become essential. Structures such as standing-wave resonators, closed-loop ring waveguides, and localized cavity modes can act as temporary storage elements—retaining the amplitude and phase of SW over defined time intervals. For example, Vogt *et al.* [466] demonstrated the use of standing SW modes in YIG microstructures to realize tunable magnonic delay lines, a foundational building block for phase-coherent signal buffering and synchronization.

Beyond simply delaying signals, SW traps can serve as elementary memory cells, storing information in the spatial profile or spectral properties of the SW mode. The performance of such traps is greatly enhanced when implemented in materials with ultra-low magnetic damping (e.g., YIG), and even more so when the system presents non-trivial topology. Topological magnonics introduces chiral edge states that are immune to disorder-induced backscattering, thereby ensuring robust signal retention even in the presence of material imperfections. This makes topologically protected confinement particularly appealing for scalable memory arrays. [265]

In summary, SW confinement and trapping are central to the realization of magnon-based memory and processing technologies. By leveraging well-engineered trapping potentials and exploiting the material advantages of YIG and other magnetic systems, it becomes possible to construct scalable, low-loss, and nonvolatile components that are well-suited for a new era of beyond-CMOS, energy-efficient computation. In neuromorphic architectures, confined SW modes can embody oscillatory neural states: they support spiking behavior, phase based encoding, and even associative memory through nonlinear resonance dynamics. Arrays of magnonic resonators can mimic networks of neurons and synapses; coupling strength between nodes is governed by the spatial overlap of localized magnon modes and external stimuli, allowing for programmable connectivity.

*7.7 Hybrid Magnonics Interfaces (Magnons + Qubits / Spins / Photons)*

Until recently, magnon-mediated spin-spin coupling remained largely theoretical. This changed with a landmark experimental study, which demonstrated magnon-mediated interactions between nitrogen-vacancy (NV) centers in diamond through MSSWs in YIG. [467] Using high-resolution relaxometry and spectroscopy, the authors measured self-energy shifts and non-local relaxation enhancements in NV spins, induced by thermal magnons propagating across a YIG substrate. Remarkably, this coupling was observed at room temperature, marking a significant milestone toward practical on-chip quantum communication between spin qubits using magnons as an intermediate bus. On the theoretical front, recent proposals have focused on hybridizing magnons with cavity photons and solid-state qubits, often employing Kerr nonlinearity and squeezing mechanisms to enhance coupling strengths. A notable example is the study from Si-Tong Jin *et al.*, [468] which designs a hybrid interface consisting of a YIG nanosphere embedded in a superconducting coplanar waveguide resonator, coupled to a single NV center. By driving the magnon mode into a squeezed state using a Kerr interaction, the system enters a regime where the cavity and magnon modes hybridize into two polaritonic branches—one of which becomes nearly dark (low frequency) and selectively couples to the qubit. This approach allows for exponentially enhanced, long-distance qubit–qubit coupling, mediated by the low-lying polaritonic branch, with the added advantage of tunable interaction strength. Such schemes promise a scalable architecture for entangling spin qubits, with magnons acting as an engineered quantum mediator that can be switched, squeezed, and steered through photonic interfaces.

In parallel, researchers have proposed designs to break symmetry in magnon–qubit coupling, enabling directional or chiral interactions. Ren *et al.* describe a configuration where a superconducting qubit is coupled to a one-dimensional cavity array with a controlled phase-difference. [469] This arrangement leads to chiral coupling—information flows preferentially in one direction, a property crucial for implementing nonreciprocal magnonic quantum links. Such directional coupling enables building blocks for quantum routers, delay lines, or circulators, where magnons transmit quantum information with minimal back-action or reflection. This represents a conceptual advance: magnons are no longer passive carriers but become active, programmable channels within a larger network.

One additional major future direction lies in the development of reconfigurable magnonic architectures, where magnon traps can be dynamically tuned via external stimuli—such as electric fields, strain, spin currents, or temperature. By integrating ferroelectric or multiferroic materials with magnetic layers, researchers aim to create devices where trapping potentials can be electrically written and erased, allowing for field-programmable logic circuits. These systems would enable real-time control of SW propagation paths, bandwidths, and interference conditions without physical restructuring of the device. In addition, the use of magnonic crystals is promising for future magnonic confinement applications. A seminal example is the magnon transistor developed by Chumak *et al.* [470], in which a magnonic crystal confines gate magnons through spectral bandgap tuning. These trapped magnons then interact nonlinearly with propagating source magnons, enabling the creation of a three-terminal all-magnon logic element. The localization in this system is static, determined by the patterned structure, but it marked an important proof-of-concept for confinement-enhanced magnonic control.

Another promising direction involves the use of topological magnetic textures—such as skyrmions, chiral domain walls, and artificial spin ices—to localize magnons in robust, disorder-insensitive states. These topological traps may support protected SW modes immune to backscattering or structural imperfections, enabling reliable operation even in noisy or imperfect environments. For instance, interaction-stabilized-topological magnon insulators on honeycomb ferromagnets yield chiral edge modes via magnon–magnon interactions that break certain symmetries, providing topologically protected transport along boundaries. [471,472] Similarly, skyrmion-crystal lattices exhibit tunable topological transitions by altering the external magnetic field, one can open or close magnonic bandgaps and switch between phases with and without chiral edge magnonic states. [473] Additionally, nonlinear magnon trapping—where high-amplitude magnons alter their own confinement potential via interactions—could enable self-focusing wavepackets, bistable states, or nonlinear logic elements. As magnon amplitude increases within a localized mode, these interactions can produce self-focusing wavepackets, bistable behavior, or even nonlinear threshold logic elements. This approach treats magnon–magnon interactions not as perturbations to be minimized, but as active resources for computation. Classical nonlinear magnonics literature details phenomena such as parametric pumping, bullet modes, bistability, and frequency comb generation, all of which arise from the Landau–Lifshitz–Gilbert nonlinearity. [474]

**Declaration of competing interest**

The authors declare that they have no known competing financial interests or personal relationships that could have appeared to influence the work reported in this paper.

**Acknowledgments**

We wish to acknowledge the support by the National Key Research and Development Program of China Grants No. 2022YFA1402801, the NSF China under No. 12525406, No. 12474104 and No.

12550004. G. Gubbiotti and H. Yu acknowledge the financial support of the bilateral agreement CNR/National Natural Science Foundation (CHINA), under project "Spin-orbit interaction based on topological insulator/ferromagnet heterostructures" (2024-2025). G. G. also acknowledge the European Union— Next Generation EU under the Italian Ministry of University and Research (MUR) National Innovation Ecosystem grant ECS00000041—VITALITY. CUP: B43C22000470005. R. A. Gallardo and P. Landeros aknowledgeacknowledge financial support from Fondecyt Grants No. 1241589 and No. 1250803, and Basal Program for Centers of Excellence, Grant No. CIA250002 (CEDENNA).